\documentclass[12pt]{article}
\usepackage[utf8]{inputenc}

\usepackage{latexsym,bm}
\usepackage{amsmath}
\usepackage{authblk}
\usepackage{amsthm}
\usepackage{booktabs}
\usepackage{epsfig}
\usepackage{graphicx}
\usepackage{amsfonts}
\usepackage{color}
\usepackage{amssymb}
\usepackage{multirow}
\usepackage{algorithm}
\usepackage{comment}
\usepackage{algpseudocode}
\usepackage{indentfirst}
\usepackage[left=0.74in, right=0.74in, top=0.95in, bottom=0.95in]{geometry}
\usepackage{setspace}
\doublespace
\usepackage{caption}
\usepackage{caption}
\usepackage{subcaption}

\usepackage{natbib}
\usepackage{newfile}

\usepackage{xr-hyper}

\usepackage[]{hyperref}  

\makeatletter
\newcommand*{\addFileDependency}[1]{
  \typeout{(#1)}
  \@addtofilelist{#1}
  \IfFileExists{#1}{}{\typeout{No file #1.}}
}
\makeatother

\newtheorem{theorem}{Theorem}
\newtheorem{example}{Example}
\newtheorem{remark}{Remark}
\newtheorem{lemma}{Lemma}

\newtheorem{corollary}{Corollary}

\newtheorem{assumption}{Assumption}

\newcommand{\X} {{\bf X}}
\newcommand{\Y} {{\bf Y}}
\newcommand{\Z} {{\bf Z}}
\newcommand{\K} {{\bf K}}
\newcommand{\B} {{\bf B}}
\newcommand{\bE} {{\mathbb E}}
\newcommand{\bP} {{\mathbb P}}
\newcommand{\bQ} {{\mathbb Q}}

\def\H{{ \mathbf{H} }}

\def\K{{ \mathbf{K} }}

\def\P{{ \mathbf{P} }}

\def\bbeta{{\boldsymbol{\beta}}}
\def\balpha{{\boldsymbol{\alpha}}}
\DeclareMathOperator*{\argmax}{arg\,max}

\begin{document}
	\title{A General Framework for Powerful Confounder Adjustment in Omics Association Studies}
\author[1]{Asmita Roy}
\author[2]{Jun Chen}
\author[1]{Xianyang Zhang\thanks{Roy and Zhang acknowledge partial support from NSF DMS-1811747 and NSF DMS-2113359.
Chen and Zhang acknowledge partial support from NIH 1R01GM144351-01 and NIH 1R21HG011662.
Address correspondence to Xianyang Zhang
(zhangxiany@stat.tamu.edu).}}
\date{}
\affil[1]{Texas A\&M University}
\affil[2]{Mayo Clinic}

\maketitle
\textbf{Abstract} Genomic data are subject to various sources of confounding, such as demographic variables, biological heterogeneity, and batch effects. To identify genomic features associated with a variable of interest in the presence of confounders, the traditional approach involves fitting a confounder-adjusted regression model to each genomic feature, followed by multiplicity correction. 
This study shows that the traditional approach was sub-optimal and proposes a new two-dimensional false discovery rate control framework (2dFDR+) that provides significant power improvement over the conventional method and applies to a wide range of settings. 2dFDR+ uses marginal independence test statistics as auxiliary information to filter out less promising features, and FDR control is performed based on conditional independence test statistics in the remaining features. 2dFDR+ provides (asymptotically) valid inference from samples in settings where the conditional distribution of the genomic variables given the covariate of interest and the confounders is arbitrary and completely unknown. To achieve this goal, our method requires the conditional distribution of the covariate given the confounders to be known or can be estimated from the data. We develop a new procedure to simultaneously select the two cutoff values for the marginal and conditional independence test statistics. 2dFDR+ is proved to offer asymptotic FDR control and dominate the power of the traditional procedure. Promising finite sample performance is demonstrated via extensive simulations and real data applications.
\\
\strut \textbf{Keywords:} Conditional Independence Testing, Confounding Factor, False Discovery Rate, Genomic Data Analysis, Multiple Testing.

\doublespace		

\section{Introduction}
One central theme of genomic data analysis is identifying genomic features associated with a variable of interest, such as disease status. The associated features are subject to further replication and validation. The validated features could then be followed up for a more in-depth mechanistic study or be used as biomarkers for disease prevention, diagnosis, and prognosis if they have sufficient predictive power \citep{majewski2011taming,ziegler2012personalized}. Due to the constraint of clinical sample collection,  the variable of interest is often correlated with other variables, which may confound the associations of interest. One example is the identification of microbiome biomarkers for endometrial cancer based on a comparison between benign and malignant tumor samples \citep{walsh2019postmenopause}. Patients with benign tumors are usually much younger than those with malignant tumors since the progression to malignancy requires multiple genomic events. Age has also been known to be associated with the female 
reproduction tract microbiome. Therefore, age is a confounding factor, and we need to control it if the aim is to identify cancer-related microbiome biomarkers reliably. 
Other common sources of confounding in genomic data analysis include
environmental changes \citep{fare2003effects,gasch2000genomic}, cell mixtures \citep{liang2014grasping}, technical variation or batch effects \citep{leek2010tackling,lazar2013batch} and surgical manipulation \citep{lin2006influence}. Controlling the confounders could significantly increase the rate of successful validation, reduce the overall cost and shorten the time from discovery to clinical tests. However, due to a substantial multiple testing burden, confounder adjustment exacerbates the already low statistical power for genome-scale association tests. If no confounder adjustment is performed, we are faced with a severely inflated type I error, with the extent of inflation depending on the number and strength of associations with the confounder. Increasing the statistical power of a confounded association study while controlling for the false positives is a statistical topic of critical importance. Surprisingly, few statistical efforts have been made on this important topic. 


The traditional way of confounder adjustment for high-dimensional association tests is to adjust for confounders for each genomic feature and correct the individual association $p$-values for multiple testing using false discovery rate (FDR) control \citep{benjamini1995controlling,storey2002direct}. This procedure has been a standard statistical practice for genomic association analysis to maintain the correct type I error rate level. However, in practice, confounders may affect only a subset of omics features \citep{lu2004gene,glass2013gene,gershoni2017landscape}, and adjusting confounders for every omics feature will be an over-adjustment, leading to substantial power loss. To rescue the power, one naive idea is first to test the dependence between the confounder and each omics feature. If the dependence is not statistically significant, we exclude the confounder in the model. Although this strategy substantially improves the power, it suffers from the so-called selection bias \citep{efron2011tweedie,fithian2014optimal}, which inflates the type I error if the significance cutoff is not properly chosen to reflect the selection effect from the first step.


In a recent work, \cite{yi20212dfdr} made a significant step toward solving this problem. The authors proposed a two-dimensional false discovery rate control procedure (2dFDR) based on linear models with the measurement of the omics feature as the outcome. 
The 2dFDR procedure proceeds in two dimensions. The first dimension uses the unadjusted statistic (from fitting the unadjusted linear models to each omics feature)
to screen out a large number of irrelevant features (noises) that are not likely to be associated with the covariate of interest or the confounder. In the second dimension, the procedure uses the adjusted statistic (from fitting the confounder-adjusted linear models to each omics feature) to identify the true signals within the remaining features and control the FDR at the desired level. Although the unadjusted statistic is biased and captures the effects from both the covariate of interest and the confounder, it can be leveraged to increase the signal density and reduce the multiple testing burden in the second dimension. At a high level, the idea of using the unadjusted statistics is similar to the use of marginal utilities in variable screening \citep{fan2008sure}. However, the 2dFDR procedure takes into account the selection effect from the first dimension and thus provides asymptotically valid inference. 

In this work, we propose a general framework 2dFDR+ for integrating confounder adjustment into multiple testing. The 2dFDR+ framework significantly extends the scope and applicability of the original 2dFDR in the following aspects:
\begin{enumerate}

    \item The new framework relaxes the linear model assumption in \cite{yi20212dfdr}. Indeed, the conditional distribution of the omics variables given the variable of interest and the confounders can be arbitrary and completely unknown. Thus, the new framework can be applied to different types of outcomes such as continuous/binary/count outcomes.
    
    \item We allow the joint use of any conditional and marginal independence test statistics in 2dFDR+. The marginal independence test statistic serves as an auxiliary statistic to screen out noise and improve power.
    
    \item 2dFDR+ does not require a case-by-case study to derive the joint (asymptotic) distribution of the conditional and marginal independence test statistics. It provides a unified approach to approximate their joint distribution under the null by modeling the conditional distribution of the covariate given the confounders.
    
    \item \cite{yi20212dfdr} focuses on the case of univariate covariate while the covariate of interest can be multivariate within our framework.
    
    \item 2dFDR+ allows different ways of estimating the conditional distribution of the covariate given the confounders. Examples include residual permutation/bootstrap, model-based simulation, Markov Chain Monte Carlo (MCMC), and conditional generative adversarial networks.
    
    
    \item Due to explicit modeling of the relationship between the variable of interest and confounders, the new method provides more reliable FDR control, especially when the confounding effect is strong.
\end{enumerate}

In theory, we prove that 2dFDR+ provides asymptotic FDR control (see Theorem \ref{thm-main}). By design, 2dFDR+ (using both the conditional and marginal independence test statistics) is guaranteed to deliver at least as many rejections as the corresponding 1d procedure (using only the conditional independence test statistics). The reason is that 2dFDR+ searches over a larger rejection region (2d versus 1d). 
By setting the cutoff for the marginal independence test statistics to be zero, our method would reduce to the 1d procedure. This observation suggests that the optimal cutoff for the 1d procedure is in the feasible set of cutoffs (that control the FDR estimate at the desired level) for the 2d procedure. The flexibility of choosing an additional cutoff leads to more rejections. See the detailed discussions in Section \ref{sec:4}.

A unique feature of 2dFDR+ is that it lets the data decide the usefulness/informativeness of the auxiliary statistic. If the auxiliary statistic provides helpful information, 2dFDR+ has significant power gain. When the auxiliary statistic is not informative, it typically reduces to the corresponding 1d procedure (without using the auxiliary statistic) in terms of performance. Extensive numerical studies show no harm in including the auxiliary statistic except for some very special cases (see the discussions in Remark \ref{rm:power}). Interestingly, when the FDR is controlled at level $q$, we can show that in the worst-case scenario, the asymptotic power loss for 2dFDR+ compared to the 1d procedure is at most $q$, see Section \ref{sec:4}.

Finally, it is worth mentioning another line of research on estimating latent confounding factors. Principal component analysis was first suggested by \cite{alter2000singular} to estimate the latent confounding factors. More recently, a variety of methods have been proposed for confounder adjustment in similar statistical settings, see, e.g., \cite{friguet2009factor,gagnon2012using,leek2007capturing,leek2008general}. The theoretical properties of some of these approaches were recently studied by \cite{wang2017confounder}. Although we assume that the confounders are fully observed in our framework, conceptually, our method can also be coupled with the above techniques for confounder adjustment when the confounders are unobserved. 

The rest of the article is organized as follows. Section \ref{sec:2} describes the problem setups and a two-dimensional (2d) rejection region based on a primary statistic for testing the conditional independence between the omics feature and the covariate of interest given the confounders and an auxiliary statistic for testing the marginal independence between the omics feature and covariate. Section \ref{sec:3} introduces an oracle FDR-controlling procedure, where the conditional distribution of the covariate given the confounders is assumed to be known. We prove asymptotic FDR control for the oracle procedure in Section \ref{sec:4}. We discuss several ways of estimating the conditional distribution in Section \ref{sec:5}. We review some nonparametric conditional independence tests and discuss their use in our framework in Section \ref{sec:6}. Sections \ref{sec:7} and \ref{sec:8} are devoted to numerical studies and real data analysis respectively. Section \ref{sec:9} concludes.

\section{Problem Statement and 2d Rejection Region}\label{sec:2}
We formulate the feature selection problem by allowing the omics variables to depend on the covariate of interest and confounders arbitrarily. To state the problem and the procedure carefully, suppose we have $n$ i.i.d. samples $\{(\X_i,\Y_i,\Z_i)\}^{n}_{i=1}$  with $\Y_i=(Y_{i,1},\dots,Y_{i,m})^\top$ from a population, each of the form $(\X,\Y,\Z)$, where $\X\in\mathbb{R}^p$, $\Y=(Y_1,\dots,Y_m)^\top\in\mathbb{R}^m$ and $\Z=(Z_1,\dots,Z_d)^\top\in\mathbb{R}^d$. Here $\Y$ represents a vector of omics features, $\X$ is the covariate of interest, and $\Z$ denotes the set of confounders. We aim to discover as many as possible omics features $Y_i$ that are dependent of $\X$ conditionally on the confounders $\Z$. We formulate this as the problem of testing 
\begin{align*}
H_{0,j}: Y_j \perp\!\!\!\perp \X | \Z \quad \text{against} \quad  H_{1,j}: Y_j \not\!\perp\!\!\!\perp \X | \Z
\end{align*}
for $1\leq j\leq m$. To tackle this problem, one must adjust for the confounders and the multiplicity in testing. The burden from both adjustments could lead to potential power loss, especially when the confounding effect is strong. 

Our idea to resolve this issue is to use two statistics jointly, namely a primary statistic for testing the conditional independence specified in $H_{0,j}$ and an auxiliary statistic for testing the marginal independence $Y_j \perp\!\!\!\perp \X$, for deciding whether or not to reject $H_{0,j}$.
The purpose of using the auxiliary statistic is to enrich signals, reduce the multiple testing burden and thus enhance the multiple testing power. As marginal dependence does not necessarily imply conditional dependence (e.g., $Y_j$ and $\X$ are both functions of $\Z$), the use of auxiliary statistics could lead to selection bias and requires proper adjustment in the selection of cut-off values. One of our goals is to carefully design a way to simultaneously select the cut-off values for the primary statistic and the auxiliary statistic to control the FDR at the desired level.

As a motivation, we consider $m$ independent generalized linear models:
\begin{align*}
& f(Y_j|\X,\Z,\balpha_j,\bbeta_j,\phi_j)=\exp\left\{\frac{\theta_j Y_j -b(\theta_j)}{\phi_j}+c(Y_j,\phi_j)\right\},\\
& g(\bE[Y_j])=g(b'(\theta_j))=\X^\top\balpha_j+\Z^\top\bbeta_j,
\end{align*}
for $1\leq j\leq m$, where $g$ is a known link function, $\theta_j$ is the canonical parameter, $\phi_j$ is the dispersion parameter, $c(Y_j,\phi_j)$ is some function of $(Y_j,\phi_j)$ and $\balpha_j\in\mathbb{R}^p,\bbeta_j\in\mathbb{R}^d$ are the coefficients associated with the covariate of interest and confounders respectively. Under the above model, there are four different categories to consider
\begin{itemize}
  \item[]A. Associated with both the covariate of interest and confounders: $\balpha_j\neq \mathbf{0},\bbeta_j\neq \mathbf{0};$
  \item[]B. Solely associated with the covariate of interest: $\balpha_j\neq \mathbf{0},\bbeta_j=\mathbf{0}$;
  \item[]C. Solely associated with the confounders: $\balpha_j=\mathbf{0},\bbeta_j\neq \mathbf{0}$;
  \item[]D. Not associated with either the covariate of interest or confounders: $\balpha_j=\mathbf{0},\bbeta_j=\mathbf{0}.$
\end{itemize}
We note that (i) $\balpha_j = \mathbf{0}$ if and only if $Y_j \perp\!\!\!\perp \X | \Z$; (ii) when $\bbeta_j=\mathbf{0}$ (Categories B and D), testing the conditional independence boils down to testing the marginal independence $Y_j \perp\!\!\!\perp \X$. In a model-free setting, these four categories can be described as (A) $Y_j \not\perp\!\!\!\perp \X | \Z$ and $Y_j \not\perp\!\!\!\perp \Z | \X$; (B) $Y_j \perp\!\!\!\perp \Z | \X$ and $Y_j \not\perp\!\!\!\perp \X$; (C) $Y_j \perp\!\!\!\perp \X | \Z$ and $Y_j \not\perp\!\!\!\perp \Z$; (D) $Y_j \perp\!\!\!\perp (\X,\Z)$.
As a way to enrich signals, we use a marginal independence test to screen out the omics features in Category D and further use a conditional independence test to pick out the true signals from Categories A and B. More precisely, we let $T_j^C$ and $T_j^M$ be two test statistics computed based on the samples $\{(\X_i,\Y_i,\Z_i)\}^{n}_{i=1}$ for testing the conditional independence $Y_j \perp\!\!\!\perp \X | \Z$ and the marginal independence $Y_j \perp\!\!\!\perp \X$, respectively. Throughout the discussions below, we assume that a large positive value of $T_j^M$ ($T_j^C$) provides evidence against marginal (conditional) independence. 
The readers are referred to Section \ref{sec:6} for some examples of conditional and unconditional independence tests. Given the thresholds, $t_1,t_2\geq0$, the two-dimensional (2d) procedure can be described as follows. 
\begin{itemize}
    \item[] Dimension 1. Use the marginal independence test statistics to determine a preliminary set of features $\mathcal{D}_1=\left\{1\leq j \leq m: T_j^{M}\geq t_1\right\}$.
    \item[] Dimension 2. Reject $H_{0,j}$ for $T_j^{C}\geq t_2$ and $j \in \mathcal{D}_1$. As a result, the final set of discoveries is given by $\mathcal{D}_2=\left\{1\leq j\leq m: T_j^{M}\geq t_1,T_j^{C}\geq t_2\right\}$.
\end{itemize}
Although marginal dependence does not imply conditional dependence, it can be leveraged to increase the signal density and reduce multiple testing burden in the second dimension. More precisely, the usefulness of the marginal dependence test is due to
\begin{enumerate}
    \item The marginal dependence test statistics screen out a large number of noises in Category D and thus ease the multiple testing burden in the second dimension;
    \item The marginal dependence test statistics are more effective in detecting signals from Category B as the conditional dependence test causes over-adjustment, reducing the signal strength. 
\end{enumerate}
We illustrate the rational behind 
2dFDR+ through the following example. A detailed description of the method is provided in the next section.

\begin{example}
{\rm 
Consider the following data generating process:
\begin{align}\label{eq-logistic}
Y_j \sim \text{Bernoulli}(p_j),\quad \log\left(\frac{p_j}{1-p_j}\right)= \alpha_j X + \beta_j Z,
\end{align}
where $X=(\rho Z+\epsilon)/\sqrt{\rho^2+1}$ with $Z$ and $\epsilon$ being independently generated from $N(0,1)$. 
Here $\rho\in \{0.1, 0.5, 1 \}$ represents weak (+), medium (++) and strong (+++) confounding level. We generate $\alpha_j$ and $\beta_j$ independently from the mixture distribution:
  \begin{equation*}
    0.15\times\text{Unif}(-0.7, -0.5) + 0.15\times\text{Unif}(0.5, 0.7) + 0.7\times\delta_0
\end{equation*}
where $\delta_0$ denotes a point mass at 0. We let $T_j^C$ be the t-statistic for testing $\tilde{H}_{0,j}:\alpha_j=0$ under the logistic model (\ref{eq-logistic}), and let $T_j^M$ be the t-statistic for testing $\tilde{H}_{0,j}$ under the reduced model by forcing $\beta_j=0$ in model (\ref{eq-logistic}). 
In Figure \ref{fig-1}, we plot the marginal ($T^M$) statistic against the conditional ($T^C$) statistic for various confounded scenarios. The standard approach performs (one-dimensional) FDR control based on the conditional statistic ($T^C$) only (we refer it as 1dFDR). When the correlation between the variable of interest and the confounder (denoted as $\text{cor}(X, Z)=\rho/\sqrt{\rho^2+1}\in\{0.1,0.45,0.71\}$) is high, the signals (green) and noises (red) overlap much on $T^C$. To achieve the desired FDR level, 1dFDR requires a high cutoff (black line).  For 2dFDR+, it first uses $T^M$ to exclude a large number of irrelevant features (horizontal blue line). Next, a  lower cutoff (vertical blue line) is used to achieve the same FDR level. As a result, it achieves significant power improvement, and the improvement increases with the correlation between the variable of interest and the confounder.

\begin{figure}[H]
    \centering
    \includegraphics[scale=0.42]{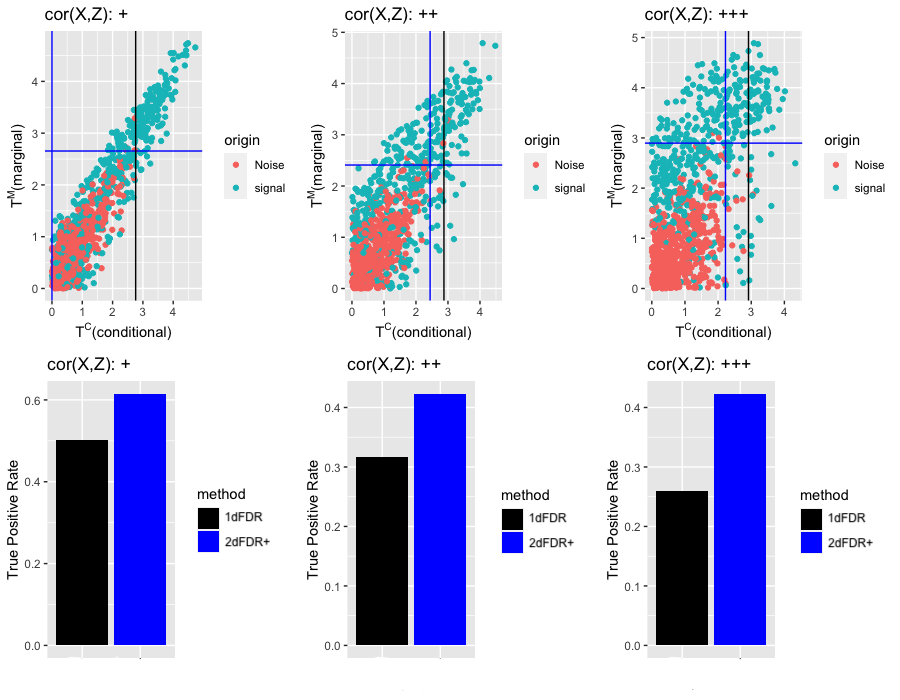}
    \caption{Illustration of 2dFDR+ using simulated datasets. The three panels in the first row denote the decision boundaries for 1dFDR (black line) and 2dFDR+ (blue lines) at the $5\%$ FDR level for three degrees of confounding. 1dFDR relies on the conditional statistic ($T^C$) only (one dimension) while 2dFDR+ is based on both the marginal and conditional statistics ($T^C$ and $T^M$), i.e., it uses two dimensions. $T^M$ is used to screen out a large number of irrelevant features (blue horizontal line), followed by a less stringent cutoff of $T^C$ that achieves higher power while keeping the FDR controlled. The figures in the second row illustrate the power difference between the two methods. When the correlation is low $(\rho= 0.1)$ using 2dFDR+ provides little improvement over 1dFDR. When the correlation is higher (“++,” “+++,” $\rho = 0.5, 1$), the signals (green) and noises (red) are more difficult to separate on $T^C$. By using $T^M$, 2dFDR+ excludes a large number of noises without losing many signals. The signal density on $T^C$ is enriched, leading to significant power gain. }
    \label{fig-1}
\end{figure}
}
\end{example}

\section{Oracle Procedure}\label{sec:3}
We introduce an oracle FDR-controlling procedure, where we assume that the conditional distribution of $\X$ given $\Z$, denoted by $P_{\X|\Z}$ below, is known. Section \ref{sec:5} introduces several ways of estimating this conditional distribution from the observations.

\subsection{Estimating the false discovery proportion}
Our goal here is to develop a principled way of finding the cutoff values $(t_1,t_2)$ such that the FDR is controlled at a desired level while the number of rejections is as large as possible. Let $\mathcal{M}_0=\{1\leq j\leq m: H_{0,j} \text{ is true}\}$ and $m_0=|\mathcal{M}_0|$ be the set and the number of true null hypotheses respectively. Write $\widetilde{\X}=(\X_1,\dots,\X_n)^\top$, $\widetilde{\Y}_j=(Y_{1,j},\dots,Y_{n,j})^\top$ and $\widetilde{\Z}=(\Z_1,\dots,\Z_n)^\top.$
Based on the 2d rejection region, the false discovery proportion (FDP) is given by
\begin{align}\label{eq-fdp-1}
\text{FDP}(t_1,t_2)=\frac{\sum_{j\in \mathcal{M}_0}\mathbf{1}\{T_j^{M}\geq t_1,T_j^{C}\geq t_2\}}{1\vee\sum^{m}_{j=1}\mathbf{1}\{T_j^{M}\geq t_1,T_j^{C}\geq t_2\}},    
\end{align}
where $a\vee b=\max\{a,b\}$ for $a,b\in\mathbb{R}$. Note that the FDP is zero when no rejection is made. We replace the numerator in the definition of $\text{FDP}(t_1,t_2)$ by its conditional expectation with respect to $\widetilde{\X}$ given $\widetilde{\Y}_j$ and $\widetilde{\Z}$, which leads to the following approximate upper bound on the FDP:
\begin{align}
\text{FDP}(t_1,t_2)\approx & \frac{\sum_{j\in \mathcal{M}_0}\bP_0(T_j^{M}\geq t_1,T_j^{C}\geq t_2|\widetilde{\Y}_j,\widetilde{\Z})}{1\vee\sum^{m}_{j=1}\mathbf{1}\{T_j^{M}\geq t_1,T_j^{C}\geq t_2\}} \nonumber
\\ \leq & \frac{\sum_{j=1}^m \bP_0(T_j^{M}\geq t_1,T_j^{C}\geq t_2|\widetilde{\Y}_j,\widetilde{\Z})}{1\vee\sum^{m}_{j=1}\mathbf{1}\{T_j^{M}\geq t_1,T_j^{C}\geq t_2\}}:=\text{FDP}_{\text{oracle}}(t_1,t_2),\label{fdp-oracle}
\end{align}
where $\bP_0(\cdot|\widetilde{\Y}_j,\widetilde{\Z})$ denotes the conditional probability under the null hypothesis $H_{0,j}$. The upper bound $\text{FDP}_{\text{oracle}}(t_1,t_2)$ relies on the conditional distribution $P_{\X|\Z}$. To find a feasible conservative estimator of the FDP, it remains to estimate the conditional probabilities in the numerator of $\text{FDP}_{\text{oracle}}(t_1,t_2)$. To this end, we write $T_j^M=T_j^{M}(\widetilde{\X},\widetilde{\Y}_j)$ and $T_j^C=T_j^{C}(\widetilde{\X},\widetilde{\Y}_j,\widetilde{\Z})$ to emphasize their dependence on the samples. As $\bP(\widetilde{\X}|\widetilde{\Y}_j,\widetilde{\Z})=\bP(\widetilde{\X}|\widetilde{\Z})$ under $H_{0,j}$ and $\bP(\widetilde{\X}|\widetilde{\Z})=\prod^{n}_{i=1}
\bP_{\X|\Z}(\X_i|\Z_i)$, we have under $H_{0,j}$ that
\begin{align*}
&\bP_0(T_j^{M}(\widetilde{\X},\widetilde{\Y}_j)\geq t_1,T_j^{C}(\widetilde{\X},\widetilde{\Y}_j,\widetilde{\Z})\geq t_2|\widetilde{\Y}_j,\widetilde{\Z})
\\=& \bE\left[\mathbf{1}\{T_j^{M}(\widetilde{\mathbf{X}},\widetilde{\Y}_j)\geq t_1,T_j^{C}(\widetilde{\mathbf{X}},\widetilde{\Y}_j,\widetilde{\Z})\geq t_2\}|\widetilde{\Y}_j,\widetilde{\Z}\right]
\\=&\int \mathbf{1}\{T_j^{M}(\widetilde{\mathbf{x}},\widetilde{\Y}_j)\geq t_1,T_j^{C}(\widetilde{\mathbf{x}},\widetilde{\Y}_j,\widetilde{\Z})\geq t_2\}d \prod^{n}_{i=1}\bP_{\X|\Z}(\mathbf{x}_i|\Z_i),
\end{align*}
where $\widetilde{\mathbf{x}}=(\mathbf{x}_1,\dots,\mathbf{x}_n)^\top$ with $\mathbf{x}_i\in\mathbb{R}^p$, which can be calculated once we know the conditional distribution $P_{\X|\Z}$. One way to approximate $P_0(T_j^{M}\geq t_1,T_j^{C}\geq t_2|\widetilde{\Y}_j,\widetilde{\Z})$ is via Monte Carlo simulation. Specifically, we generate 
$$\X_{i,b}\sim^{\text{ind}} P_{\X|\Z}(\cdot|\Z_i),\quad i=1,2,\dots,n,~~b=1,2,\dots,B.$$
Denote by $T_{j,b}^M$ and $T_{j,b}^C$ the marginal and conditional independence test statistics computed based on $(\widetilde{\X}_b,\widetilde{\Y}_j,\widetilde{\Z})$ with $\widetilde{\X}_b=(\X_{1,b},\dots,\X_{n,b})^\top$ respectively.
We propose to estimate $P_0(T_j^{M}\geq t_1,T_j^{C}\geq t_2|\widetilde{\Y}_j,\widetilde{\Z})$ by
\begin{align*}
\bar{F}_{j,B}(t_1,t_2):=\frac{1}{B+1}\sum^{B}_{b=0}\mathbf{1}\{T_{j,b}^{M}\geq t_1,T_{j,b}^{C}\geq t_2\}   
\end{align*}
with $(T_{j,0}^{M},T_{j,0}^{C})=(T_{j}^M,T_j^C)$. Hence a conservative estimate for the FDP is given by
\begin{align*}
\widetilde{\text{FDP}}(t_1,t_2)=\frac{\sum_{j=1}^m \bar{F}_{j,B}(t_1,t_2)}{1\vee\sum^{m}_{j=1}\mathbf{1}\{T_j^{M}\geq t_1,T_j^{C}\geq t_2\}}.    
\end{align*}

\subsection{Finding the optimal cut-off}\label{sec:opt-cut}
We now introduce a greedy approach to select the cut-offs.
For a desired FDR level $q$, we first define
$$\mathcal{F}_q=\{(t_1,t_2)\in \mathbb{R}^+\times\mathbb{R}^+:\widetilde{\text{FDP}}(t_1,t_2) \leq q\}$$
as the feasible set that contains all the cut-off values controlling the FDP estimate at the level $q$.
We then select the optimal cut-off as the one delivering the most number of rejections from the feasible set:
$$(t_1^*,t_2^*)=\argmax_{(t_1,t_2)\in\mathcal{F}_q}\sum^{m}_{j=1}\mathbf{1}\{T_j^{M}\geq t_{1},T_j^{C}\geq t_{2}\}.$$
Finally, we reject all the hypotheses $H_{0,j}$ such that
$$T_j^M\geq t_{1}^*~\text{ and }~T_j^C\geq t_{2}^*.$$ 
\begin{remark}
{\rm 
In the supplement, we describe a variant of the 2d procedure (2d-FWER+) to control the family-wise error rate (FWER). Simulation studies suggest that 2d-FWER+ provides reliable FWER control in finite sample.
}
\end{remark}

\subsection{Estimating the null proportion}
Following the idea in \cite{storey2002direct}, we can further improve the power of our method by estimating the proportion of null hypotheses. As a motivation, we suppose $T_j^C$ follows the mixture distribution
$$\pi_0\mathbb{P}_0+(1-\pi_0)\mathbb{P}_1,$$
where $\pi_0$ represents the null proportion, $\mathbb{P}_0$ and $\mathbb{P}_1$ denote the distributions under the null and alternative, respectively.
Under this two-group mixture model, we have
\begin{align*}
\mathbb{P}(T_j^C\leq \lambda)=\pi_0 \mathbb{P}_0(T_j^C\leq \lambda)+(1-\pi_0)\mathbb{P}_1(T_j^C\leq \lambda)\geq \pi_0 \mathbb{P}_0(T_j^C\leq \lambda),
\end{align*}
which implies that
\begin{align*}
\frac{\sum^{m}_{j=1}\mathbf{1}\{T_j^C\leq \lambda\}}{\sum^{m}_{j=1}\mathbb{P}_0(T_j^C\leq \lambda)}\approx \frac{\sum^{m}_{j=1}\mathbb{P}(T_j^C\leq \lambda)}{\sum^{m}_{j=1}\mathbb{P}_0(T_j^C\leq \lambda)}\geq \pi_0,
\end{align*}
where the approximation is due to the law of large numbers. Therefore, we propose to estimate the null proportion $\pi_0$ by
\begin{align*}
\widehat{\pi}_0(\lambda)=1\wedge\frac{\sum^{m}_{j=1}\mathbf{1}\{T_j^C\leq \lambda\}}{\sum^{m}_{j=1}F_{j,B}(\lambda)},\quad \text{where } 
F_{j,B}(\lambda):=\frac{1}{B+1}\sum^{B}_{b=0}\mathbf{1}\{T_{j,b}^{C}\leq \lambda\}.
\end{align*}
We can then implement the 2dFDR+ based on the following estimate of the FDP:
\begin{align*}
\widetilde{\text{FDP}}_\lambda(t_1,t_2)=\frac{\widehat{\pi}_0(\lambda)\sum_{j=1}^m \bar{F}_{j,B}(t_1,t_2)}{1\vee\sum^{m}_{j=1}\mathbf{1}\{T_j^{M}\geq t_1,T_j^{C}\geq t_2\}},   
\end{align*}
which can be regarded as John Storey's version of the 2dFDR+ procedure.

\section{FDR Control and Power Analysis}\label{sec:4}
We first show that under the global null, a version of the 2dFDR+ procedure provides finite sample FDR control (or equivalently FWER control). The key to the proof relies on the symmetry of the statistics $\{(T^{M}_{j,b},T^C_{j,b}):j=1,2,\dots,m\}$ across the index $b$. Let $\{(t_1(s),t_2(s))\in\mathbb{R}^+\times \mathbb{R}^+:1\leq s\leq \mathcal{S}\}$ be a sequence of thresholds such that $t_1(s)\leq t_1(s')$ and $t_2(s)\leq t_2(s')$ for $1\leq s<s'\leq \mathcal{S}$. Let 
$V^b(s)=\sum^{m}_{j=1}\mathbf{1}\{T_{j,b}^M\geq t_1(s),T_{j,b}^C\geq t_2(s)\}$ for $0\leq b\leq B$. 
Define
\begin{align*}
s^*=\min\left\{1\leq s\leq \mathcal{S}:\frac{(B+1)^{-1}\sum^{B}_{b=0}V^b(s)}{1\vee V^0(s)}\leq q\right\}.    
\end{align*}
Then we reject any hypothesis such that $T_{j,0}^M\geq t_1(s^*)$ and $T_{j,0}^C\geq t_2(s^*)$.
\begin{theorem}\label{thm:finite}
Under the global null, the above 2dFDR+ procedure provides finite sample FDR control or equivalently FWER control.
\end{theorem}

Under general setting, the symmetry among $(T^{M}_{j,b},T^C_{j,b})_{j=1}^m$ no longer holds and the finite sample FDR control is not guaranteed. Fortunately, we manage to show that 2dFDR+ provides asymptotic FDR control as $n,m$ both diverge to infinity. To achieve this goal, we impose the following assumptions. 
\begin{assumption}\label{as1}
Conditional on $(\X,\Z)$, $Y_j$'s are independent across $1\leq j\leq m$. Moreover, for $j\in\mathcal{M}_0$, $Y_j$'s are independent conditional on $\Z$.
\end{assumption}
Assumption \ref{as1} requires that the marginal models of $Y_j$ conditional on $\X$ and $\Z$ are independent across $1\leq j\leq m$. For instance, consider the model
\begin{equation}\label{eq-model}
\begin{split}
& Y_j=u_j(\X) + v_j(\Z) + \epsilon_j, \quad j\notin\mathcal{M}_0,\\
& Y_j=v_j(\Z) + \epsilon_j, \quad j\in\mathcal{M}_0,
\end{split}
\end{equation}
where $u_j(\cdot)$ and $v_j(\cdot)$ are some functions defined on $\mathbb{R}^p$ and $\mathbb{R}^d$ respectively. In this case, Assumption \ref{as1} is fulfilled provided that $\epsilon_j$'s are independent across $j.$ 

\begin{assumption}\label{as2}
Recall that $m_0$ denotes the number of true null hypotheses. Suppose $m_0/m\rightarrow \pi_0\in (0,1)$ and there exist two continuous bivariate functions $\widetilde{V}(\cdot,\cdot)$ and $\widetilde{S}(\cdot,\cdot)$ defined on $\mathbb{R}^+\times \mathbb{R}^+$ such that
\begin{align*}
& \left|\frac{1}{m_0}\sum_{j\in\mathcal{M}_0} P(T_j^M\geq t_1,T_j^C\geq t_2|\widetilde{\X},\widetilde{\Z})-\widetilde{V}(t_1,t_2)\right|\rightarrow^p 0,
\\ & \left|\frac{1}{m}\sum_{j=1}^m P(T_j^M\geq t_1,T_j^C\geq t_2|\widetilde{\X},\widetilde{\Z})-\widetilde{S}(t_1,t_2)\right|\rightarrow^p 0,
\end{align*}
for any fixed $t_1,t_2\geq 0$.
\end{assumption}
Assumption \ref{as2} is a high-level condition. We justify this assumption under model (\ref{eq-model}) in Section \ref{as2-example}. Our next assumption is similar to the requirement in Theorem 4 of \cite{storey2004strong}, which ensures the existence of cut-off values to control the FDR at level $q$.
It reduces the search region for the optimal cut-offs to a rectangle of the form $[0,t_{0,1}]\times [0,t_{0,2}]$.
\begin{assumption}\label{as3}
Assume that there exist $t_{0,1}$ and $t_{0,2}$ such that, 
$$\frac{\pi_0 \widetilde{V}(t_{0,1},0)+u_1}{\widetilde{S}(t_{0,1},0)}\leq q'<q,\quad \frac{\pi_0\widetilde{V}(0,t_{0,2})+u_2}{\widetilde{S}(0,t_{0,2})}\leq q''<q,$$
and $\widetilde{S}(t_{0,1},t_{0,2})>c>0,$ where $\pi_0$ is defined in Assumption \ref{as2}, $u_1=\limsup m^{-1}\sum_{j\in\mathcal{M}_1} \bP_0(T_j^{M}\geq t_{0,1}|\widetilde{\Y}_j,\widetilde{\Z})$ and 
$u_2=\limsup m^{-1}\sum_{j\in\mathcal{M}_1} \bP_0(T_j^{C}\geq t_{0,2}|\widetilde{\Y}_j,\widetilde{\Z})$
with $\mathcal{M}_1=\{1\leq j\leq p:H_{0,j} \text{ is non-null}\}$.

\end{assumption}
To state the main theorem, we recall that
\begin{align*}
(t_1^*,t_2^*)=\argmax_{(t_1,t_2)\in\mathbb{R}^+\times\mathbb{R}^+:\widetilde{\text{FDP}}(t_1,t_2)(t_1,t_2)\leq q}\sum^{m}_{j=1}\mathbf{1}\{T_j^{M}\geq t_{1},T_j^{C}\geq t_{2}\}.    
\end{align*}
The theorem below establishes the asymptotic FDR control of the 2dFDR+ procedure.
\begin{theorem}\label{thm-main}
Under Assumptions \ref{as1}-\ref{as3} and as $B\rightarrow +\infty$, 
\begin{align*}
\limsup_{n,m\rightarrow +\infty}\bE\left[\text{FDP}(t_1^*,t_2^*)\right]\leq q,
\end{align*}
where $\text{FDP}(t_1,t_2)$ is defined in (\ref{eq-fdp-1}).
\end{theorem}

We now turn to the power analysis of the oracle 2dFDR+ procedure. We argue that 2dFDR+ is, in general, more powerful than the corresponding 1d procedure based on the conditional independence statistics alone. Assume without loss of generality that $T_j^M$ takes non-negative values. The intuition is that for $t_1=0$, the first dimension does not screen out any null hypothesis and only the second dimension is effective in identifying signals. In this case, 2dFDR+ reduces to the corresponding 1d procedure, where we reject $H_{0,j}$ if $T_j^C\geq t^*$ with $t^*$ being the solution to the following problem
\begin{align*}
\max_t \sum_{j=1}^m\left\{T_j^C\geq t\right\} \quad \text{subject to} \quad \frac{\sum_{j=1}^m \bar{F}_{j,B}(0,t)}{1\vee\sum^{m}_{j=1}\mathbf{1}\{T_j^{C}\geq t\}}\leq q.
\end{align*}
Clearly, $(0,t^*)$ is in the feasible set $\mathcal{F}_q$ of the optimization problem in Section \ref{sec:opt-cut}. Therefore, we have $\sum^{m}_{j=1}\mathbf{1}\{T_j^{M}\geq t_{1}^*,T_j^{C}\geq t_{2}^*\}\geq \sum_{j=1}^m\{T_j^C\geq t^*\}$. In other words, 2dFDR+ is guaranteed to
deliver at least as many rejections as the corresponding 1d procedure does. 

Define $\text{FP}_{\text{2d}}$ and $\text{FP}_{\text{1d}}$ as the number of false positives for 2dFDR+ and the associated 1d procedure respectively. Similarly, we let $\text{TP}_{\text{2d}}$ and $\text{TP}_{\text{1d}}$ be the number of true positives. Suppose 
Assumptions \ref{as1}-\ref{as3} hold and both procedures make rejections (i.e., $\text{FP}+\text{TP}>0$). In addition, assume
\begin{align}\label{eq-fdr-0}
\frac{\text{FP}_{\text{1d}}}{\text{FP}_{\text{1d}}+\text{TP}_{\text{1d}}}=q_1,\quad  \frac{\text{FP}_{\text{2d}}}{\text{FP}_{\text{2d}}+\text{TP}_{\text{2d}}}=q_2, 
\end{align}
for some $0\leq q_1,q_2\leq 1.$ As 2dFDR+ makes more rejections, i.e., $\text{FP}_{\text{1d}}+\text{TP}_{\text{1d}}\leq \text{FP}_{\text{2d}}+\text{TP}_{\text{2d}}$, we must have
\begin{align*}
\text{TP}_{\text{2d}} \geq    \frac{1-q_2}{1-q_1}\text{TP}_{\text{1d}}.
\end{align*}
When $q_2\leq q_1$, $\text{TP}_{\text{2d}}\geq \text{TP}_{\text{1d}}$, i.e., 2dFDR+ makes more true rejections. In general, we have the following lower bound on the number of true positives for 2dFDR+.
\begin{corollary}\label{cor-1}
Under Assumptions \ref{as1}-\ref{as3} and as $B\rightarrow +\infty$, we have for any $\epsilon>0$,
\begin{equation}\label{eq-worse}
P(\text{TP}_{\text{2d}} \geq    (1-q-\epsilon)\text{TP}_{\text{1d}})\rightarrow 1.
\end{equation}
\end{corollary}
As $\epsilon$ can be arbitrarily small, (\ref{eq-worse}) suggests that with the FDR controlled at level $q$, 2dFDR+ asymptotically achieves at least $100(1-q)\%$ true rejections of the 1d procedure in the worst-case scenario. For instance, with $q=5\%$, the power loss compared to the 1d procedure is at most 5\%. We refer the readers to Section \ref{sec:s-power} of the supplement for more discussions on the asymptotic power of 2dFDR+.

\begin{remark}\label{rm:power}
{\rm 
Since 2dFDR+ depends on the marginal independence statistic to filter features, when the confounder and variable of interest have opposite effects on the feature with similar magnitude, they will cancel out each other’s effect, and the feature could be excluded erroneously in the first dimension. The optimal cutoff of $T^M_j$ is thus determined based on the tradeoff between power reduction due to erroneously excluding these relevant features in the first dimension and power increase due to reducing the multiple testing burden and increasing the signal density in the second dimension. If the true signals can only be revealed after adjusting for the confounder, for example, when the true and confounding signals co-locate with opposite directions, the marginal independence test statistics will not be informative. In this case, the best cutoff on $T^M_j$ should be 0 and 2dFDR+ is then reduced to the 1d procedure. In finite samples, it may not always be possible to reduce 2dFDR+ to the 1d procedure exactly. 
Nevertheless, as argued above, the power loss is relatively moderate even in the worst-case scenario.}
\end{remark}

\section{Estimating the Conditional Distributions}\label{sec:5}
As the conditional distribution $P_{\X|\Z}$ is seldom known, we need to estimate it from the data. There are indeed several ways of generating samples from
$P_{\X|\Z}$.
Examples include classical methods such as residual permutation \citep{winkler2014permutation} and parametric bootstrap \citep{davison1997bootstrap} as well as modern approaches such as conditional generative adversarial network (conditional GAN) \citep{mirza2014conditional,zhou2022deep}. In the following subsections, we shall describe the residual permutation, residual bootstrap, and parametric bootstrap in more detail. Compared to the conditional GAN, these procedures are more suitable for omics applications, given the limited sample sizes in many omics association studies. 

\subsection{Residual permutation and residual bootstrap}
When $\X$ is a continuous random vector, we can model the relationship between $\widetilde{\X}\in\mathbb{R}^{n\times p}$ and $\widetilde{\Z}\in\mathbb{R}^{n\times d}$ through a multivariate linear regression model given by
\begin{align}\label{eq-ML}
\widetilde{\X}=\widetilde{\Z}\mathbf{B}+\mathbf{E},     
\end{align}
where $\mathbf{B}\in\mathbb{R}^{d\times p}$ is the matrix of coefficients and $\mathbf{E}\in \mathbb{R}^{n\times p}$ is the error matrix. Consider the following strategy to generate samples from $P_{\X|\Z}$.
\begin{enumerate}
    \item[Step 1:]Fitting regression model. Fit the multivariate linear regression model in (\ref{eq-ML}). Let $\widehat{\mathbf{E}}=\widetilde{\X}-\widetilde{\Z}\widehat{\mathbf{B}}$ be the residuals from the fitted model, where $\widehat{\mathbf{B}}$ is the least squares estimate of $\mathbf{B}$.

    \item[Step 2:]Residual permutation. Permute the rows of the residual matrix $\widehat{\mathbf{E}}$ and denote the resulting matrix by $\widehat{\mathbf{E}}^*$.
    Let $\widetilde{\X}_b=(\X_{1,b},\dots,\X_{n,b})^\top=\widetilde{\Z}\widehat{\mathbf{B}}+\widehat{\mathbf{E}}^*.$

    \item[Step 2$'$:]Residual bootstrap. Let $\widehat{\mathbf{E}}^{**}$ be a $n\times p$ matrix whose rows are sampled with replacement from those of $\widehat{\mathbf{E}}$. Let $\widetilde{\X}_b=(\X_{1,b},\dots,\X_{n,b})^\top=\widetilde{\Z}\widehat{\mathbf{B}}+\widehat{\mathbf{E}}^{**}.$
\end{enumerate}

\begin{remark}\label{rm}
{\rm To allow nonlinearity, we can replace $Z_i$ by $(g_1(Z_i),\dots,g_{d'}(Z_i))\in\mathbb{R}^{d'}$ 
for some transformations $(g_1,\dots,g_{d'})$ in the multivariate regression model.}
\end{remark}

\subsection{Parametric bootstrap}
Suppose the conditional distribution of $\X$ given $\Z$ takes the parametric form of $P_{\X|\Z}(\mathbf{X}_i|\mathbf{Z}_i;\theta)$, where $\theta\in \Theta\subseteq \mathbb{R}^r$ is an unknown parameter. It is natural to estimate the parameter by maximizing the conditional log-likelihood
\begin{align*}
\widehat{\theta}=\argmax_{\theta\in\Theta}\sum^{n}_{i=1}\log P_{\X|\Z}(\mathbf{X}_i|\mathbf{Z}_i;\theta).
\end{align*}
Then we can generate $\X_{i,b}$ from the estimated likelihood $P_{\X|\Z}(\mathbf{X}_i|\mathbf{Z}_i;\widehat{\theta})$. For example, suppose $\X$ is a Bernoulli random variable with the
conditional success probability given by 
$\{1+\exp(-\Z_i^\top\theta)\}^{-1}.$
Then we can sample $\X_{i,b}$ from the Bernoulli distribution with success probability $\{1+\exp(-\Z_i^\top\widehat{\theta})\}^{-1},$ where $\widehat{\theta}$ is an estimate of $\theta$ by fitting a logistic model to the data with $\widetilde{\X}$ being the binary response and $\widetilde{\Z}$ being the covariates.

\section{Independence Tests}\label{sec:6}
We review some parametric and nonparametric unconditional/conditional independence tests and discuss their use within our framework. In Section \ref{sec:m-b}, we focus on the model-based (parametric) independence tests. In Sections \ref{sec:rv}-\ref{sec:hsic}, we consider two types of nonparametric independence tests targeting linear and nonlinear dependence respectively. These three types of independence tests will all be implemented in our numerical studies.

\subsection{Model-based statistics}\label{sec:m-b}
Suppose the conditional likelihood of $Y_j$ given $\X$ and $\Z$ has the form of
\begin{align}\label{m-full}
P_{Y_j|\X,\Z}(Y_j|\X^\top\balpha_j+\Z^\top\bbeta_j).    
\end{align}
The log-likelihood function based on the observations is given by
\begin{align*}
L_{n,j}(\balpha_j,\bbeta_j)=\sum^{n}_{i=1}\log P_{Y_{j}|\X,\Z}(Y_{i,j}|\X^\top_i\balpha_j+\Z^\top_i\bbeta_j).   
\end{align*}
In this case, testing $H_{0,j}$ is equivalent to testing whether $\balpha_j$ is zero. Thus we let $T_{j}^C$ be a statistic for testing $\balpha_j=0$ under the model (\ref{m-full}). Examples include the Wald test and the likelihood-ratio test. To test the marginal independence, we consider the reduced model $P_{Y_j|\X,\Z}(Y_j|\X^\top\balpha_j)$ by forcing $\bbeta_j=0$ in (\ref{m-full}). Under the reduced model, we let $T^M_j$ be a statistic for testing $\balpha_j=0$, which can be viewed as testing the marginal independence $Y_j \perp\!\!\!\perp \X$.
When $P_{Y_j|\X,\Z}$ is the likelihood function associated with a linear model with Gaussian error, we can let $T_j^C$ and $T_j^M$ be the adjusted and unadjusted z-statistics considered in \cite{yi20212dfdr}.
In this sense, the statistics in \cite{yi20212dfdr} fall into our framework.

\subsection{Nonparametric dependence metrics}
Nonparametric dependence testing, aiming to determine whether two random vectors are dependent without specifying the exact parametric forms of the distributions, is one of the fundamental problems in statistics. Classical metrics or test statistics for dependence testing include the RV coefficient, rank correlation coefficient, and nonparametric Cra\'{m}r-von Mises type statistics. Modern approaches are built on distance and kernel embedding, which can detect non-linear and non-monotone dependence. Notable examples include the distance covariance \citep{szekely2007measuring}, Hilbert-Schmidt independence criterion (HSIC) \citep{gretton2005measuring,gretton2007kernel} and the sign distance covariance \citep{bergsma2014consistent}. Below we shall review the RV coefficient and HSIC and discuss their conditional versions for testing the conditional independence.

\subsubsection{RV coefficients}\label{sec:rv}
Pearson correlation and partial correlation coefficients are perhaps the most commonly used nonparametric dependence metrics for measuring marginal and conditional dependence. Here we describe the RV coefficient and its conditional version as multivariate generalizations of the squared Pearson correlation coefficient and the squared partial correlation coefficient for detecting linear and conditional linear dependence.

For two random vectors $\mathbf{U}$ and $\mathbf{V}$, we let $\Sigma_{\mathbf{U},\mathbf{V}}$ be the covariance matrix between $\mathbf{U}$ and $\mathbf{V}$. The RV coefficient between $\X$ and $Y_j$ is defined as 
\begin{align*}
\text{RV}(\X,Y_j)=\frac{\text{tr}(\boldsymbol{\Sigma}_{\X,Y_j}\boldsymbol{\Sigma}_{Y_j,\X})}{\sqrt{\text{tr}(\Sigma_{\X,\X}^2)\text{tr}(\Sigma_{Y_j,Y_j}^2)}}. 
\end{align*}
To estimate the RV coefficient, we simply replace the covariance matrices in the definition above with the sample covariance matrices.

To introduce the conditional version of the RV coefficient, we let $\mathbf{e}_{\X}$ and $e_{Y_j}$ be the residuals by regressing $\X$ and $Y_j$ on $\Z$ respectively. The conditional RV coefficient is defined as 
$$\text{cRV}(\X,Y_j|\Z)=\text{cRV}(\mathbf{e}_{\X},e_{Y_j}).$$
Similar to Remark \ref{rm}, to account for the nonlinear dependence of $\X$ and $Y_j$ on $\Z$, we can replace $\Z$ by certain basis function transform on it, e.g., spline transformation.

\subsubsection{Hilbert-Schmidt independence criterion}\label{sec:hsic}
Hilbert-Schmidt Independence Criterion (HSIC) was introduced as a kernel-based independence measure by \cite{gretton2005measuring,gretton2007kernel}. Let $k_p(\cdot,\cdot)$ be a reproducing kernel Hilbert space (RKHS) kernel defined on $\mathbb{R}^p \times \mathbb{R}^p$. Commonly used kernels in this context include the Gaussian kernel and the Laplacian kernel. The HSIC for quantifying the strength of dependence between $\X$ and $Y_j$ can be defined as
\begin{align*}
\text{HSIC}(\X,Y_j)=\bE[k_p(\X,\X')k_1(Y_j,Y_j')] +\bE[k_p(\X,\X')]\bE[k_1(Y_j,Y_j')]-2\bE[k_p(\X,\X')k_1(Y_j,Y_j'')]       
\end{align*}
where $(\X',Y_j')$ and $(\X'',Y_j'')$ are independent copies of $(\X,Y_j)$. When $k_p$ and $k_1$ are characteristic kernels \citep{sriperumbudur2011universality}, HSIC completely characterizes the dependence in the sense that $\X$ and $Y_j$ are independent if and only if $\text{HSIC}(\X,Y_j)=0.$
To estimate the HSIC, define $\K_\X=(k_{\X,ab})_{a,b=1}^{n}$ with $k_{\X,ab}=k_p(\X_a,\X_b)$ and $\K_{Y_j}=(k_{Y_j,ab})_{a,b=1}^{n}$ with $k_{Y_j,ab}=k_1(Y_{a,j},Y_{b,j})$. Let $\H=\mathbf{I}-n^{-1}\mathbf{1}\mathbf{1}^\top$ with $\mathbf{1}$ being the $n$-dimensional vector of all ones. Set $\widetilde{\K}_\X=\H \K_\X \H$
and $\widetilde{\K}_{Y_j}=\H \K_{Y_j} \H$. The sample HSIC is defined as 
\begin{align*}
\widehat{\text{HSIC}}(\X,Y_j)=\frac{1}{n}\text{Tr}(\widetilde{\K}_\X \widetilde{\K}_{Y_j}),    
\end{align*}
which has been shown to be a consistent estimator, see \cite{gretton2005measuring}.

A conditional version of the HSIC (cHSIC) for measuring and testing conditional dependence was proposed in \cite{zhang2012kernel}. Here we describe the construction of their statistic. Let $\K_{\X,\Z}=(k_{\X,\Z,ab})^{n}_{a,b=1}$ with $k_{\X,\Z,ab}=k_{p+d}((\X_a,\Z_a),(\X_b,\Z_b))$ and define $\K_{Y_j,\Z}$ in a similar way. Denote by $\widetilde{\K}_{\X,\Z}=\H \K_{\X,\Z} \H$ and $\widetilde{\K}_{Y_j,\Z}=\H \K_{Y_j,\Z}\H$ the centered versions of $\K_{\X,\Z}$ and $\K_{Y_j,\Z}$ respectively. Further define
\begin{align*}
&\widetilde{\K}_{\X\Z|\Z}=\epsilon^2(\widetilde{\K}_{\X\Z}+\epsilon\mathbf{I})^{-1}\widetilde{\K}_{\X\Z}(\widetilde{\K}_{\X\Z}+\epsilon\mathbf{I})^{-1}, \\
&\widetilde{\K}_{Y_j\Z|\Z}=\epsilon^2(\widetilde{\K}_{Y_j\Z}+\epsilon\mathbf{I})^{-1}\widetilde{\K}_{Y_j\Z}(\widetilde{\K}_{Y_j\Z}+\epsilon\mathbf{I})^{-1},
\end{align*}
for some small positive constant $\epsilon$.
The sample cHSIC is given by
\begin{align*}
\widehat{\text{cHSIC}}(\X,Y_j|\Z)=\frac{1}{n}\text{Tr}(\widetilde{\K}_{\X\Z|\Z}\widetilde{\K}_{Y_j\Z|\Z}).
\end{align*}
We refer the readers to \cite{zhang2012kernel} for more detailed properties about the cHSIC. 


\section{Numerical Studies}\label{sec:7}
\subsection{Simulation setting}
We conduct comprehensive simulations to evaluate the performance of 2dFDR+ and compare it to competing methods. Throughout the simulations, we control the following three factors, namely the degree of confounding ($\rho$, which determines the strength of association between $\X$ and $\Z$), the signal strength (distributions of $\alpha_j$ and $\beta_j$) and the signal density (proportion of non-zero elements in $\{\alpha_j\}$ and $\{\beta_j\}$). Specifically, we generate $\alpha_j$ and $\beta_j$ independently over $j$ from the mixture distribution
\begin{equation*}
    \frac{\pi}{2}\times U(-l-0.2, -l) + \frac{\pi}{2}\times U(l, l+0.2) + (1-\pi)\times \delta_0
\end{equation*}
where $\pi\in(0,1)$ and $\delta_0$ denotes a point mass at 0. For each factor, we consider three different scenarios:

\begin{enumerate}
    \item Degree of confounding: $\rho\in \{0.1,1,1.5\}$ roughly corresponds to weak (+), medium (++) and strong(+++) confounding respectively. See Section \ref{sec:sim-model} for the role of $\rho$ in each simulated model.

    \item Signal density: $\pi\in\{5\%, 10\%, 20\%\}$ represents low, medium and high signal density respectively.
    \item Signal effect: $l\in\{0.2, 0.3, 0.4\}$ represents weak, moderate and strong effect respectively.
\end{enumerate}
We report the empirical FDR and power  averaged over 100 simulation runs for all possible combinations of the three factors.

\subsection{Competing methods}\label{sec:7.2}
We compare the finite sample performance of the following seven methods.
\begin{enumerate}
\item MS-1dFDR: The 1d procedure based on the t-statistics for testing $\boldsymbol{\alpha}_j=0$ under the full model (see the detailed descriptions of each data generating model in Section \ref{sec:sim-model}). The 1d procedure is essentially the same as the 2dFDR+ procedure, except that instead of a two-dimensional rejection region, we are searching for a cutoff along a single dimension, namely that of the conditional statistic. The FDP is estimated using the resampled $\mathbf{X}_{i,b}$ (from the conditional distribution of $\mathbf{X}$ given $\mathbf{Z}$), but only the conditional statistic is used for estimating the number of false rejections, as opposed to both in the oracle procedure. The statistics used in this 1d procedure is the model-based statistic, i.e., the z-statistic (or t-statistic, depending on the model) corresponding to the coefficient of $\mathbf{X}$ for a full model fit. 
\item RV-1dFDR: The 1d procedure based on the conditional RV coefficient. To account for the potential non-linearity in the underlying relationship between $\X$ and $\Z$ (and similarly, $\Y$ and $\Z$), the residuals obtained from a cubic spline regression of $\X$ on $\Z$ (and similarly, $\Y$ on $\Z$) have been used in the calculation of the conditional RV coefficient.

\item HSIC-1dFDR: The 1d procedure based on the cHSIC described in Section \ref{sec:hsic}. 

\item 2dFDR: The 2dFDR procedure proposed in \citet{yi20212dfdr}, which is based on linear models with the measurement of the omics feature as the outcome and the covariate of interest and confounders as the predictors. 

\item MS-2dFDR+: The proposed 2dFDR+ procedure with $T_j^C$ and $T_j^M$ being the t-statistics for testing $\boldsymbol{\alpha}_j=0$ under the full model and reduced model as described in Section \ref{sec:m-b}. 

\item RV-2dFDR+: The proposed 2dFDR+ procedure with $T^C_j=\widehat{\text{cRV}}(\X,Y_j|\Z)$ and $T^M_j=\widehat{\text{RV}}(\X,Y_j)$ which denote the sample estimates of the conditional and the unconditional RV coefficients respectively. As before, to account for the potential non-linearity in the underlying relationship between $\X$ and $\Z$ (and similarly, $\Y$ and $\Z$), the residuals obtained from a cubic spline regression of $\X$ on $\Z$ (and similarly, $\Y$ on $\Z$) have been used in the calculation of the conditional RV coefficient.

\item HSIC-2dFDR+: The proposed 2dFDR+ procedure with $T_j^C=\widehat{\text{cHSIC}}(\X,Y_j|\Z)$ and $T_j^M=\widehat{\text{HSIC}}(\X,Y_j)$, where we set $\epsilon= 0.001$ and used the Gaussian kernel with the bandwidth parameter chosen using the median heuristic \citep{Garreau2017}. 
\end{enumerate}

The 1d procedure can be viewed as a special
case of the corresponding 2d procedure by forcing the cutoff of the auxiliary statistic to be zero. As the 2d procedure is searching over a larger rejection region (by allowing the cutoff of the auxiliary statistic to be greater than zero and meanwhile lowering the cutoff for the primary statistic), the proposed 2d procedure is guaranteed to make more rejections in finite sample. 


\subsection{Data generating processes}\label{sec:sim-model-1}
To examine the performance of the above methods under different settings, we consider the following data generation scenarios. As $\X$ and $\Z$ are univariate in all cases we denote them by $X$ and $Z$. The detailed models are provided in Section \ref{sec:sim-model}. Throughout the simulations, we set the sample size $n$ to be 100, and the number of hypotheses $m$ (i.e., the number of features) to be 1000. 

\begin{itemize}
    \item[A.] \textit{Linear/nonlinear models with continuous $X$ and $Z$}. Consider the additive model
    \begin{equation}\label{sim-1}
    Y_j = \alpha_j f(X) + \beta_j g(Z) + \epsilon_j, \quad \epsilon_j\sim N(0,1), \quad j = 1,\dots,m.
\end{equation}

Here $X$ and $Z$ are associated with each other through the following model:
\begin{equation}\label{sim-2}
    X \sim N(\rho  h(Z) , 1), \quad Z \sim N( 0 , 1),
\end{equation}
where $\rho$ controls the degree of confounding and $h:\mathbb{R}\rightarrow \mathbb{R}$ is a possibility nonlinear function. We rescale $X$ to dissociate any possible entanglement between signal strength and the degree of confounding. This type of simulation setup has been used in Models 1-4 to explore the effect of the relations among $X$, $Y_j$, and $Z$ on the FDR and power. 
The empirical FDR and power of RV-1dFDR, HSIC-1dFDR, 2dFDR, RV-2dFDR+ and HSIC-2dFDR+ are summarized in Figures \ref{fig:5}-\ref{fig:3} and again in Figures \ref{fig:6}-\ref{fig:7} in the supplementary material. MS-1dFDR and MS-2dFDR+ have not been included under this scenario because the statistics associated with these procedures are directly proportional to the statistics in RV-1dFDR and RV-2dFDR+, respectively. 

\item[B.] \textit{Linear/nonlinear models with discrete $X$ and continuous $Z$.} In particular, we consider the functional form in (\ref{sim-1}) and generate
$$X\sim \text{Bernoulli}\left(\frac{e^{ \rho Z}}{1 + e^{\rho Z}}\right),$$
where $Z \sim N(0,1)$. Models 5-7 explore this setup. In this case, we generate $X_{i,b}$ through a fitted logistic regression model using $Z$ as the predictor. 
We report the FDR and power for MS-1dFDR, RV-1dFDR, 2dFDR, MS-2dFDR+ and RV-2dFDR+ as described in Section \ref{sec:7.2} in Figures \ref{fig:9}-\ref{fig:11} in the supplementary material. HSIC-2dFDR+ and HSIC-1dFDR are not used in this data generating setup because for binary variables, HSIC is not efficient and the bandwidth parameter is not well-defined. 

\item[C.] \textit{Linear models with discrete $X$ and $Z$.} 
We consider the linear model 
$$Y_j = \alpha_j X + \beta_j Z + \epsilon_j,$$
where 
$$X\sim \text{Bernoulli}\left(\frac{e^{ \rho Z}}{1 + e^{\rho Z}}\right)~~\text{ and }~~Z \sim \text{Bernoulli}(0.7).$$ 
The results for MS-1dFDR, RV-1dFDR, MS-2dFDR+ and RV-2dFDR+ are reported in Figure \ref{fig:12}. 
\item[D.] \textit{Binary response.} The following logistic regression model has been considered:
\begin{align}\label{model-full}
Y_j \sim \text{Bernoulli}(p_j),\quad \log\left(\frac{p_j}{1-p_j}\right)= \alpha_j X + \beta_j Z,
\end{align}
with $X \sim N(\rho Z^2, 1)$ and $Z \sim N(0,1)$. We implement the MS-1dFDR, RV-1dFDR, MS-2dFDR+ and RV-2dFDR+, and report the results in Figure \ref{fig:14}.
In MS-2dFDR+, $T^C_j$ is the statistic for testing $\alpha_j=0$ under the full model (\ref{model-full}) and $T^M_j$ is the statistic for testing $\alpha_j=0$ by forcing $\beta_j=0$ in (\ref{model-full}).

\item[E.] \textit{Count response.} We consider the Poisson model
\begin{align*}
    Y_j \sim \text{Poisson}( \lambda_j), \quad
    \log \lambda_j = \alpha_j X + \beta_j Z,
\end{align*}
with $X \sim N(\rho Z, 1)$ and $Z \sim N(0,1).$ We implement the MS-1dFDR, RV-1dFDR, MS-2dFDR+ and RV-2dFDR+, and report the results in Figure \ref{fig:pois}. Additionally, we consider the negative binomial regression model
\begin{align*}
    Y_j \sim \text{Negative Binomial}(\text{size} = 3, \mu_j = e^{f_j(X,Z)}) 
\end{align*}
where $f_j(X,Z) = \alpha_j X + \beta_j Z$ for $X\sim N(\rho Z, 1)$ and $Z \sim N(0,1)$. We implement the MS-1dFDR, RV-1dFDR, MS-2dFDR+ and RV-2dFDR+, and report the results in Figure \ref{fig:nby}.




\end{itemize}

In Section \ref{sec:additional}, we report some additional numerical results under the following scenarios: (1) FWER control, (2) global null, (3) dependent errors, and (4) separating the effects of the densities of the signal of interest and the confounder signal.


\subsection{Simulation results}
We now discuss the major simulation findings under each scenario. Full simulation results are summarized in the  Figures \ref{fig:5}-\ref{fig:14} and Figures \ref{fig:6}-\ref{fig:19} in the supplementary material.   
Under Scenario A ($X$ and $Z$ are both continuous), when the underlying models between $Y$ and $(X, Z)$, and $X$ and $Z$ are both linear (see Figure \ref{fig:5}), all the methods provide tight FDR control except for the 2dFDR which has slight FDR inflation in some instances when the confounding effect is strong. In contrast, the proposed RV-2dFDR+, which is equivalent to MS-2dFDR+, controls the FDR at the target level across all cases, indicating more robustness of the proposed method than the original 2dFDR. In terms of power, we observe that the power decreases as the confounding effect becomes stronger for all procedures. The 2d procedure is comparable to the 1d counterpart when the confounding effect is weak but is substantially more powerful when the confounding effect is strong. We also observe that RV-2dFDR+ is comparable to 2dFDR and is more powerful than HSIC-2dFDR+.  When the underlying model is nonlinear, 2dFDR suffers from severe FDR inflation (see, e.g., Figure \ref{fig:3a}). In contrast, 2dFDR+ controls the FDR at the target level across different cases. Among the 2dFDR+ variants, RV-2dFDR+ delivers the highest power in most cases.

Under Scenario B, where X is discrete while Z is continuous (see, e.g., Figure \ref{fig:9}), the empirical FDR is well controlled for 2dFDR+ even when the confounding effect is strong. 2dFDR suffers from moderate FDR inflation (e.g Figure \ref{fig:10a}) in some instances, e.g., in the case of strong confounding. Not surprisingly, the 2d procedure is significantly more powerful than the corresponding 1d version. Moreover, the RV-based methods generally make more true rejections compared to the HSIC-based methods. 

Under Scenario C, where X and Z are both Bernoulli, as seen from Figure \ref{fig:12}, all the approaches have empirical FDR under control. When the degree of confounding is high, 2dFDR+ delivers higher power than 1dFDR does. 

Under Scenarios D and E, the original 2dFDR is not applicable, and hence only 1dFDR and 2dFDR+ have been compared in the simulations.
As seen from Figures \ref{fig:14}, \ref{fig:pois} and \ref{fig:nby}, for Scenarios D-E (binary and count response), all the methods provide reliable FDR control. 2dFDR+ produces significant power improvement over the 1dFDR methods.

To sum up, the proposed 2dFDR+ provides reliable FDR control for all the simulation settings even when the degree of confounding is strong because 2dFDR+ explicitly models the relationship between $X$ and $Z$. The 2d procedure delivers more rejections compared to the 1d counterpart, and the larger number of rejections typically translates into a higher detection power for the 2d methods. We also see that RV-2dFDR+ provides the best power in many simulation settings. As the (conditional) RV coefficients are calculated based on spline transformed covariates and confounding factors, RV-2dFDR+ can capture the nonlinearity between $Y$ and $(X, Z)$ and $X$ and $Z$ in many cases.

\begin{figure}[H]
\centering
\begin{subfigure}{0.9\textwidth}
  \centering
  \includegraphics[width=0.8\linewidth]{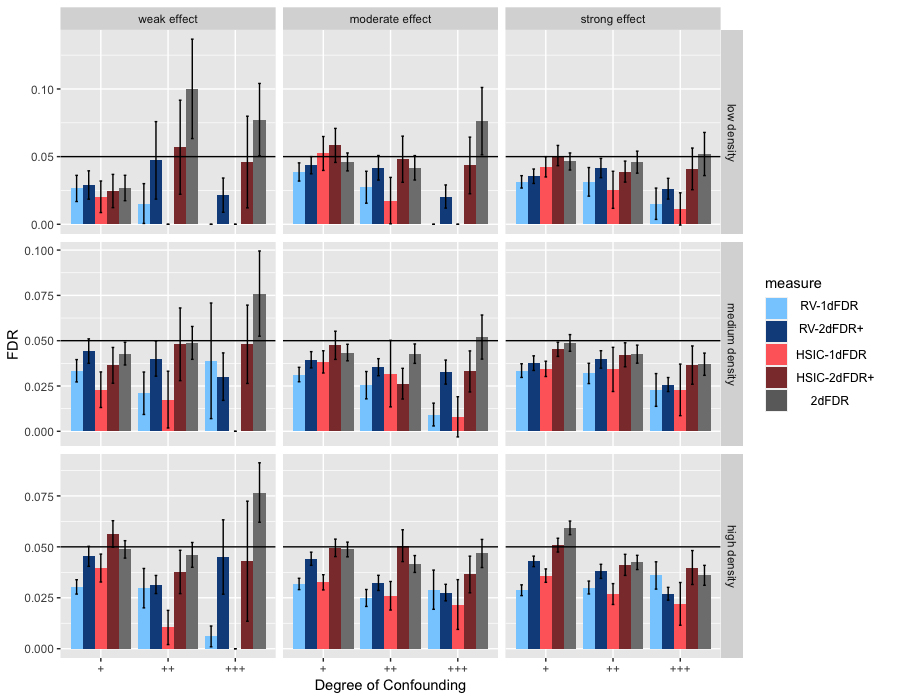}
  \caption{FDR}
  \label{fig:5a}
\end{subfigure}
\begin{subfigure}{0.9\textwidth}
  \centering
  \includegraphics[width=0.8\linewidth]{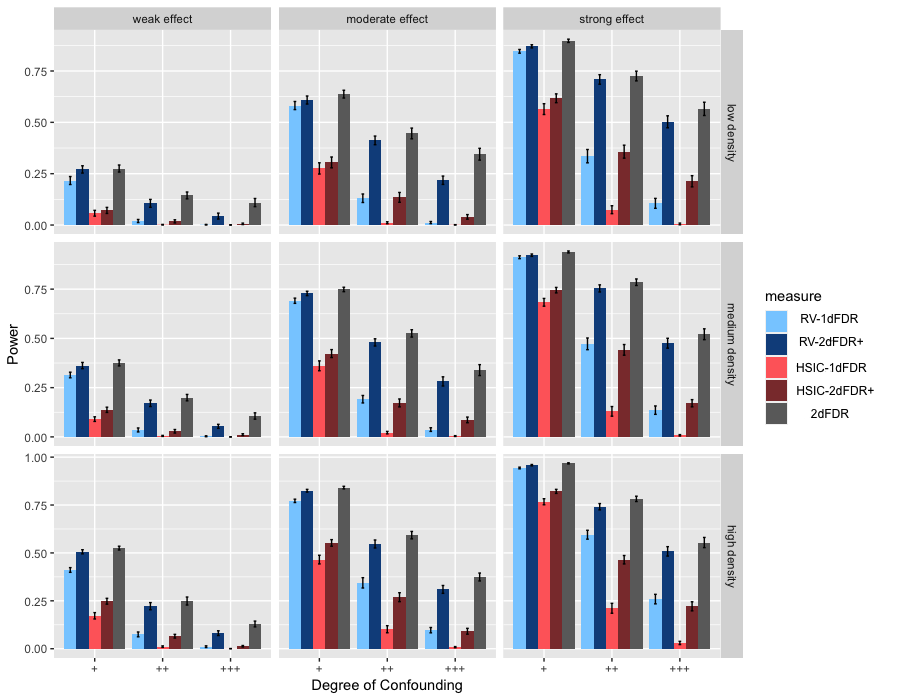}
  \caption{Power}
  \label{fig:5b}
\end{subfigure}
\caption{Empirical FDR and power for HSIC-1dFDR, RV-1dFDR, 2dFDR, HSIC-2dFDR+, RV-2dFDR+ under the model $Y_j = \alpha_j X + \beta_j Z + \epsilon_j$ and $X \sim N(\rho Z, 1)$, where $Z \sim N(0,1)$. Error bars represent the 95\% CIs and the horizontal line in (a) indicates the target FDR level of 0.05.  }
\label{fig:5}
\end{figure}

\begin{figure}[H]
\centering
\begin{subfigure}{0.9\textwidth}
  \centering
  \includegraphics[width=0.8\linewidth]{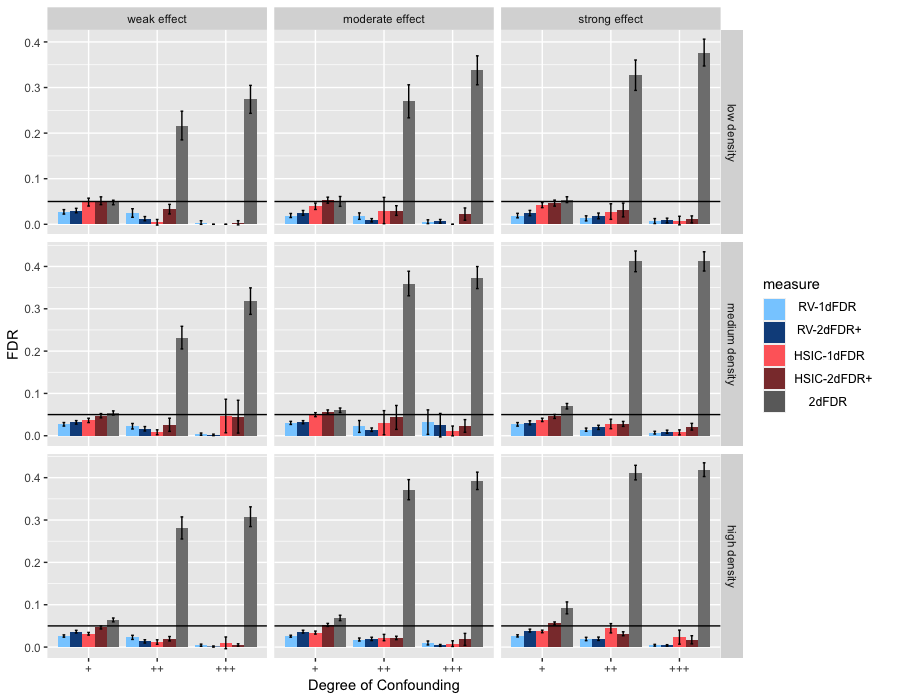}
  \caption{FDR}
  \label{fig:3a}
\end{subfigure}
\begin{subfigure}{0.9\textwidth}
  \centering
  \includegraphics[width=0.8\linewidth]{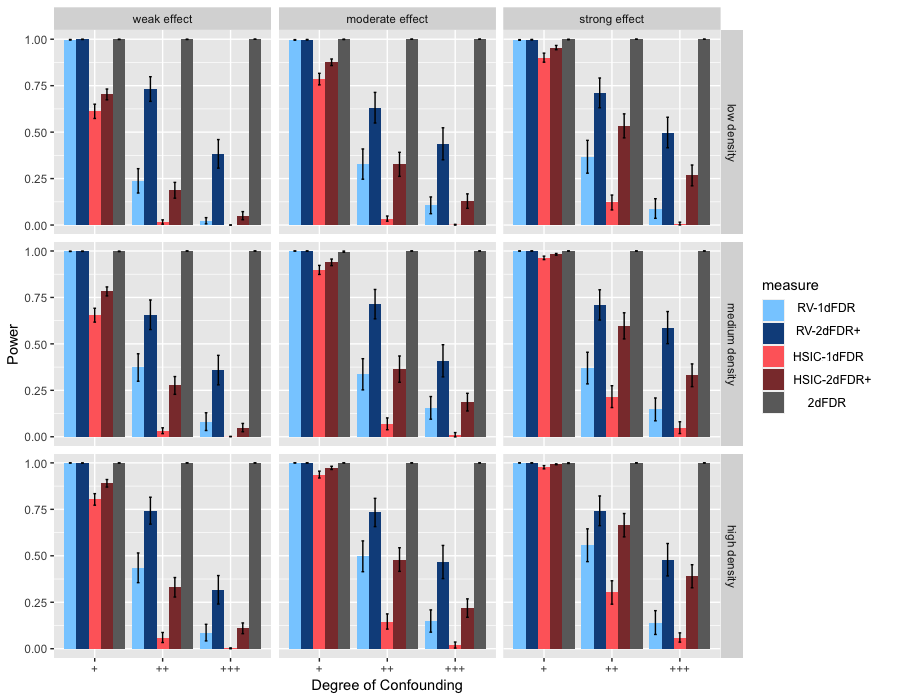}
  \caption{Power}
  \label{fig:3b}
\end{subfigure}
\caption{Empirical FDR and power for HSIC-1dFDR, RV-1dFDR, 2dFDR, HSIC-2dFDR+, RV-2dFDR+ under the model $Y_j = \alpha_j X^3 + \beta_j e^Z + \epsilon_j$ and $X\sim N(\rho Z^2, 1)$, where $Z \sim N(0,1)$. Error bars represent the 95\% CIs and the horizontal line in (a) indicates the target FDR level of 0.05. 
}
\label{fig:3}
\end{figure}

\begin{figure}[H]
\centering
\begin{subfigure}{0.9\textwidth}
  \centering
  \includegraphics[width=0.8\linewidth]{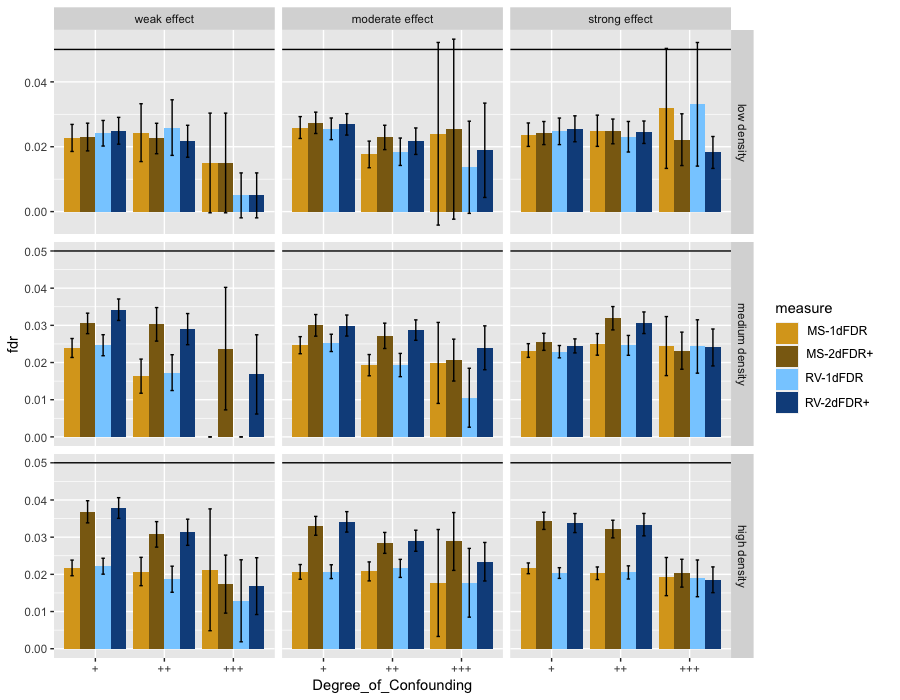}
  \caption{FDR}
  \label{fig:14a}
\end{subfigure}
\begin{subfigure}{0.9\textwidth}
  \centering
  \includegraphics[width=0.8\linewidth]{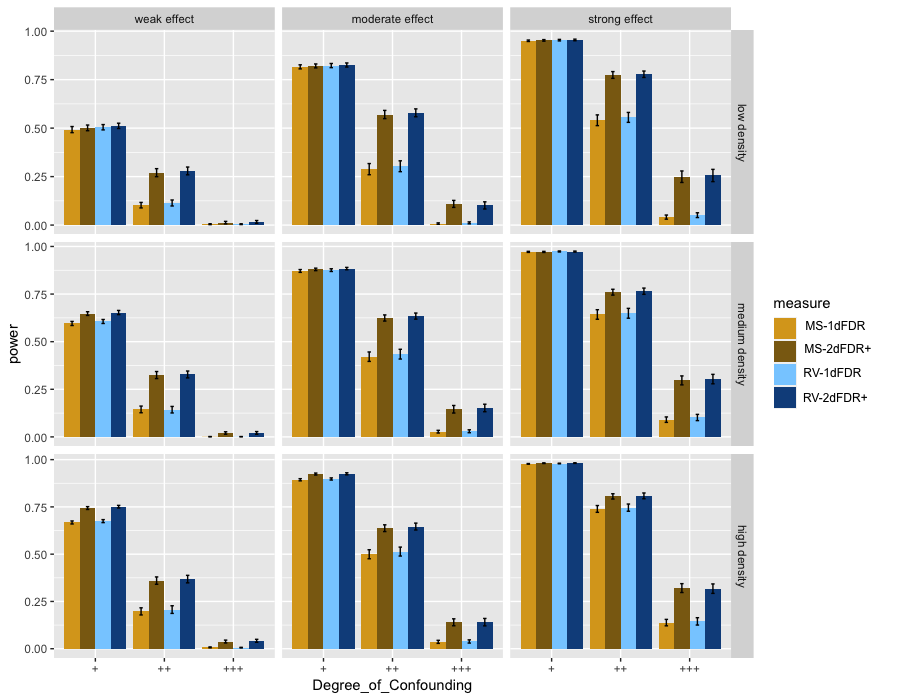}
  \caption{Power}
  \label{fig:14b}
\end{subfigure}
\caption{Empirical FDR and power for MS-1dFDR, RV-1dFDR, MS-2dFDR+, RV-2dFDR+ under the model $Y_j \sim \text{Bernoulli}((1 + e^{-f_j(X,Z)})^{-1})$, where $f_j(X,Z) = \alpha_j X + \beta_j Z ,$ $X\sim N(\rho Z, 1)$ and $Z \sim N(0,1)$. Error bars represent the 95\% CIs and the horizontal line in (a) indicates the target FDR level of 0.05. 
}
\label{fig:14}
\end{figure}



\section{Real Data Analysis}\label{sec:8}
\subsection{Microbiome data}
In the first example, we analyze a microbiome dataset in the R package \textit{GUniFrac}, which is part of a microbiome data set for studying the smoking effect on the upper respiratory tract microbiome \citep{gunifrac, charlson10}. The original data set contains samples from the right and left nasopharynx and oropharynx. Here we use the data from the left oropharynx of 32 nonsmokers and 28 smokers ($n=60$). The microbiome composition was profiled using  16S rRNA gene-targeted sequencing and processed using the QIIME bioinformatics pipeline \citep{qiime}, resulting in a count table recording the frequencies of 856 detected OTUs (operational taxonomic units). Sex is a confounding factor in this data set, with more smokers in males (odds ratio equals 2.3). The aim here is to identify smoking-associated OTUs while adjusting sex.

For illustration purposes, the OTU abundances were treated as both continuous and binary outcomes. The results for the binary outcomes are given in supplement. We first filtered out the OTUs occurring in less than 10\% of the subjects, which resulted in a total of 174 OTUs. The OTU abundance data were then transformed using a center log-ratio transformation, adding a pseudo-count of 0.5. The numbers of rejections for varying levels of FDR (ranging from 0 to 0.2) were calculated for the following methods: Benjamini-Hochberg (BH, \cite{Benjamini1995}) procedure, 2dFDR, MS-2dFDR+, RV-2dFDR+, MS-1dFDR, RV-1dFDR. The BH procedure was applied to the p-values corresponding to the tests of significance of the coefficients of IR in a linear regression model with the IR and BMI being the predictors. The numbers of rejections at different FDR levels are shown in Figure \ref{fig:combined}a.
The trend is consistent with the simulations, where we have observed that the 2dFDR+ procedure is more powerful than the corresponding 1dFDR procedure and RV-2dFDR+ makes the highest number of true rejections in most simulation setups. In addition, we produced a Venn diagram (Figure \ref{fig:smoking2}) of the rejected features for each method at the FDR level 0.10 to visualize the degree to which the rejected features in various methods overlap. We find that at level 0.1, MS-2dFDR+ successfully identifies all the seven features identified by the 2dFDR procedure and five additional features.

\subsection{Metabolomics data}
Next, we consider an Insulin Resistance dataset \citep{Pedersen2016, Pedersen2018} where the goal is to identify serum metabolites associated with insulin resistance (IR) while controlling the effect of the Body Mass Index (BMI) of the individual. 289 non-diabetic Danish adults were recruited for the study, where their IR was estimated by homeostatic model assessment (HOMA-IR) \citep{Matthews1985}. Untargeted metabolome profiles were generated on fasting serum samples, producing measurements on 325 polar metabolites and 876 molecular lipids (collectively called serum metabolites, $m = 1201$). The BMI of a subject is a confounding factor as the IR of a subject is largely influenced by the BMI (correlation coefficient $=0.57$). In this example, 2dFDR discovers the largest number of metabolites (481 at $5\%$ FDR), followed by RV-2dFDR+ (432 metabolites at $5\%$ FDR). Both are a significant improvement over RV-1dFDR (333), HSIC-1dFDR (323), and the BH procedure (377). The comparison of the number of rejections versus FDR level for all methods is displayed in Figure \ref{fig:combined}b.

Again, the result generally agrees with the findings from the simulation studies. While 2dFDR is the most powerful in this example, its inflated type I error rate observed in many non-linear simulation setups raises some concern about the reliability of the rejections solely found by itself. 

Figure \ref{fig:metabolome2} shows the Venn diagram of the serum metabolites detected by the different methods and their degree of overlap at FDR $= 0.05$ is provided. It is interesting to note that while RV-2dFDR+ and 2dFDR have detected 403 metabolites in common, the BH procedure has significantly fewer overlapping metabolites with either of these methods.

\begin{figure}[H]
        \centering
        \includegraphics[scale = 0.55]{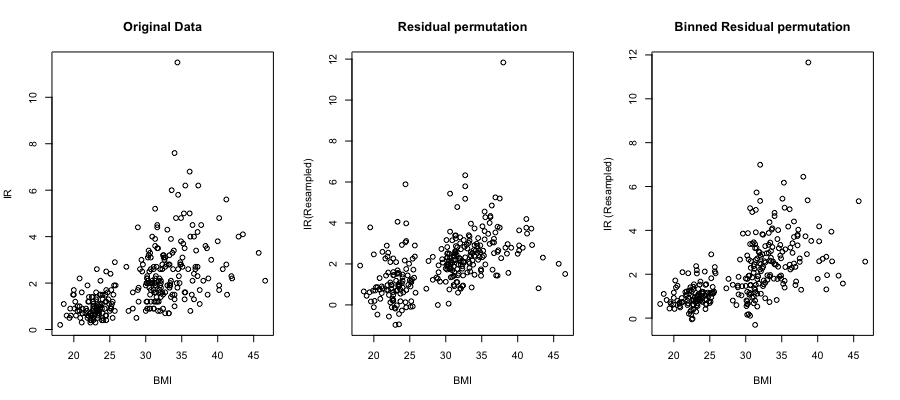}
        \caption{Scatterplots of IR versus BMI for 289 subjects. Left panel: the original data; Middle panel: Resampled data using the traditional residual permutation; Right panel: Resampled data using binned residual permutation}
        \label{fig:metabolome3}
    \end{figure}

An additional challenge we faced while analyzing the metabolomics data was generating samples from the conditional distribution $P_{\X|\Z}.$ As observed in Figure \ref{fig:metabolome3}, there is distinct heteroscedasticity in the conditional distribution of IR given BMI. Traditional resampling methods such as residual permutation \citep{winkler2014permutation} may fail as homoscedasticity is one of the underlying assumptions. To combat this, the data set was binned into two parts, namely BMI $\leq 26$ and BMI $> 26$, and two separate regressions were fitted to these two subsamples, and the residuals were permuted within each segment. The right panel of Figure \ref{fig:metabolome3}, which plots the resampled IR versus BMI using binned residual permutation, shows that the heteroscedastic structure has been preserved in the resampled data. The middle panel shows resampling using the traditional residual permutation and we can see that the original shape of the data has not been maintained in this case.

\subsection{Gene expression data}
\begin{figure}
\centering
    \includegraphics[scale = 0.65]{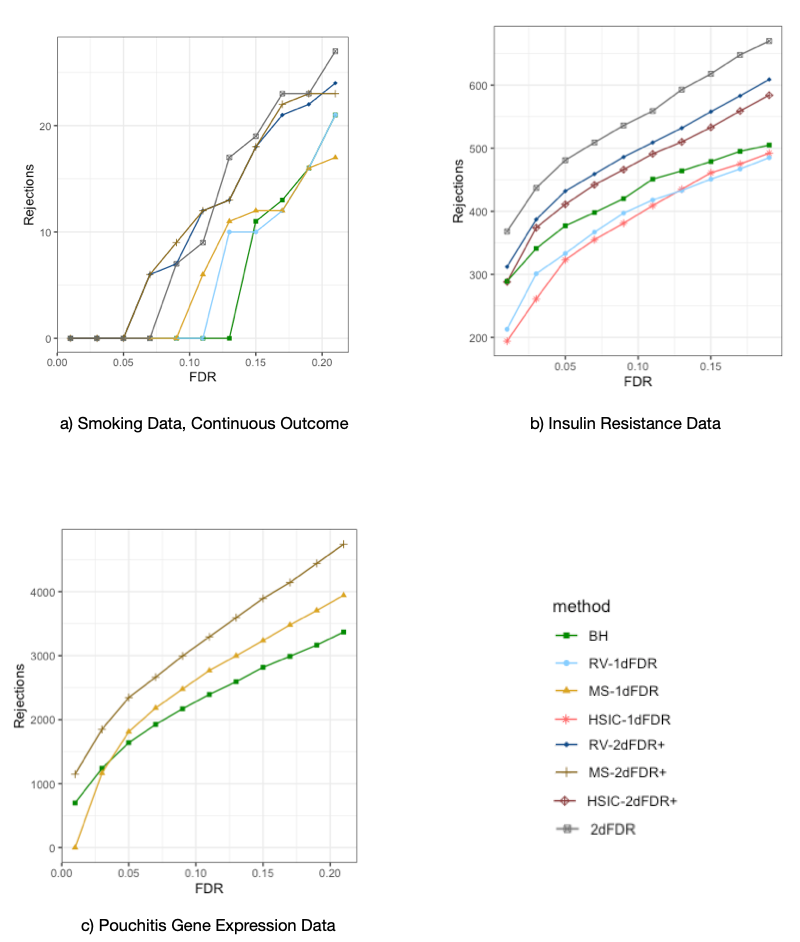}
     \caption{Number of Rejections versus FDR for different methods in the Smoking (continuous outcomes), Insulin Resistance and Pouchitis gene expression dataset}
      \label{fig:combined}
\end{figure}

Finally, we consider a Pouchitis dataset \citep{Morgan2015}, where the goal is to identify gene expressions associated with patient outcomes in a cohort with ileal pouch-anal anastomosis (IPAA) surgery in the past one year, adjusting for potential confounders such as antibiotics use and sex. This dataset considered a large population of patients having undergone IPAA at Mount Sinai Hospital, Canada. The expression levels of 19,908 genes were measured in two sites, the J-pouch and the pre-pouch ileum (PPI), using the procedure described in \citep{Morgan2015}. We considered the biopsies collected only from the pouch ($n = 74$) in this example. The conditioning variables were sex, smoking status, and antibiotic use in the previous month. The variable of interest is the disease outcome, including FAP (Familial Adenomatous Polyposis), No Pouchitis, Acute Pouchitis, Chronic Pouchitis, and Crohn's Disease like Inflammation. 
As the variable of interest is nominal, we did not use the RV coefficients in this case. Figure \ref{fig:combined}c  shows the number of genes identified as associated with the disease outcome conditioning on sex, smoking status, and antibiotic usage. At the FDR level of 0.05, the 2dFDR+ identifies the maximum number of genes (2345), followed by MS-1dFDR (1811) and BH procedure (1640),  respectively. 

\section{Conclusion}\label{sec:9}
We have proposed a general framework (2dFDR+) for performing multiple hypothesis testing while adjusting for confounding effects. Within this new framework, the conditional distribution of the omics features given the variable of interest and confounders can be arbitrary and completely unknown. The framework is flexible by allowing the joint use of any conditional and marginal independence tests, continuous/binary/count/multivariate responses, and various ways of modeling the conditional distribution of the variable of interest given the confounders. As a general methodology, 2dFDR+ can be applied to multiple types of omics data. 
In view of the numerical results, we recommend using RV-2dFDR+ (based on the spline-transformed variables) under most scenarios due to its robustness and efficiency. In cases where the RV-based statistics are not applicable, for instance, when either of $\X, \Y$ or $\Z$ are categorical, or when $\Y$ is discrete (e.g., originating from a Poisson or Negative Binomial distribution), the model-based statistics are recommended. The statistics will differ depending on the types of the variable of interest and the confounding variable. For example, when all the three variables $(\X, \Y, \Z)$ are categorical, as in the binary outcome case for the smoking microbiome data, the Pearson's chi-square statistic and the Cochran–Mantel–Haenszel (CMH) statistic are recommended for testing the marginal and the conditional dependence, respectively. Table \ref{tab:statistics} summarizes the statistics that we recommend using under different scenarios.


\begin{center}
\begin{table}[H]
\begin{tabular}{c  c  c | c  } 
 \hline
 $\Y$ & $\X$ & $\Z$ & $T^M$ and $T^C$ \\ [0.5ex] 
 \hline
 Continuous & Continuous & Continuous & RV and cRV \\ 
 \hline
 Categorical/Discrete & Continuous & Continuous & Model-based statistics (GLM) \\
 \hline
 Categorical/Discrete & Categorical & Continuous & Model-based statistics (GLM) \\
 \hline
  Categorical/Discrete & Continuous & Categorical & Model-based statistics (GLM) \\
 \hline
 Continuous & Categorical & Categorical & Model-based statistics (ANOVA) \\
 \hline
 Continuous & Categorical & Continuous & Model-based statistics (ANCOVA) \\
 \hline
 Categorical & Categorical & Categorical & $\chi^2$ and CMH-statistics \\ [1ex] 
 \hline
\end{tabular}
\caption{\label{tab:statistics} Recommended statistics under various scenarios.}
\end{table}
\end{center}

\bibliographystyle{biom}
\bibliography{reference.bib}

\begin{thebibliography}{}

\bibitem[\protect\citeauthoryear{Alter, Brown, and Botstein}{Alter
  et~al.}{2000}]{alter2000singular}
Alter, O., Brown, P.~O., and Botstein, D. (2000).
\newblock Singular value decomposition for genome-wide expression data
  processing and modeling.
\newblock {\em Proceedings of the National Academy of Sciences} {\bf 97,}
  10101--10106.

\bibitem[\protect\citeauthoryear{Benjamini and Hochberg}{Benjamini and
  Hochberg}{1995a}]{benjamini1995controlling}
Benjamini, Y. and Hochberg, Y. (1995a).
\newblock Controlling the false discovery rate: a practical and powerful
  approach to multiple testing.
\newblock {\em Journal of the Royal statistical society: series B
  (Methodological)} {\bf 57,} 289--300.

\bibitem[\protect\citeauthoryear{Benjamini and Hochberg}{Benjamini and
  Hochberg}{1995b}]{Benjamini1995}
Benjamini, Y. and Hochberg, Y. (1995b).
\newblock Controlling the false discovery rate: A practical and powerful
  approach to multiple testing.
\newblock {\em Journal of the Royal Statistical Society: Series B
  (Methodological)} {\bf 57,} 289--300.

\bibitem[\protect\citeauthoryear{Bergsma and Dassios}{Bergsma and
  Dassios}{2014}]{bergsma2014consistent}
Bergsma, W. and Dassios, A. (2014).
\newblock A consistent test of independence based on a sign covariance related
  to kendall’s tau.
\newblock {\em Bernoulli} {\bf 20,} 1006--1028.

\bibitem[\protect\citeauthoryear{Cao, Chen, and Zhang}{Cao
  et~al.}{2022}]{Cao:2020}
Cao, H., Chen, J., and Zhang, X. (2022).
\newblock Optimal false discovery rate control for large scale multiple testing
  with auxiliary information.
\newblock {\em The Annals of Statistics} {\bf 50,} 807--857.

\bibitem[\protect\citeauthoryear{Charlson, Chen, Custers-Allen, Bittinger, Li,
  Sinha, Hwang, Bushman, and Collman}{Charlson et~al.}{2010}]{charlson10}
Charlson, E.~S., Chen, J., Custers-Allen, R., Bittinger, K., Li, H., Sinha, R.,
  Hwang, J., Bushman, F.~D., and Collman, R.~G. (2010).
\newblock Disordered microbial communities in the upper respiratory tract of
  cigarette smokers.
\newblock {\em PLOS ONE} {\bf 5,} e15216.
\newblock
  https://journals.plos.org/plosone/article/file?id=10.1371/journal.pone.0015216\&type=printable.

\bibitem[\protect\citeauthoryear{Chen, Zhang, and Zhou}{Chen
  et~al.}{2021}]{gunifrac}
Chen, J., Zhang, X., and Zhou, H. (2021).
\newblock Gunifrac r package version 1.4.

\bibitem[\protect\citeauthoryear{Davison and Hinkley}{Davison and
  Hinkley}{1997}]{davison1997bootstrap}
Davison, A.~C. and Hinkley, D.~V. (1997).
\newblock {\em Bootstrap methods and their application}.
\newblock Number~1. Cambridge university press.

\bibitem[\protect\citeauthoryear{D’Argenio, Casaburi, Precone, and
  Salvatore}{D’Argenio et~al.}{2014}]{qiime}
D’Argenio, V., Casaburi, G., Precone, V., and Salvatore, F. (2014).
\newblock Comparative metagenomic analysis of human gut microbiome composition
  using two different bioinformatic pipelines.
\newblock {\em BioMed Research International} {\bf 2014,} 1--10.

\bibitem[\protect\citeauthoryear{Efron}{Efron}{2011}]{efron2011tweedie}
Efron, B. (2011).
\newblock Tweedie’s formula and selection bias.
\newblock {\em Journal of the American Statistical Association} {\bf 106,}
  1602--1614.

\bibitem[\protect\citeauthoryear{Fan and Lv}{Fan and Lv}{2008}]{fan2008sure}
Fan, J. and Lv, J. (2008).
\newblock Sure independence screening for ultrahigh dimensional feature space.
\newblock {\em Journal of the Royal Statistical Society: Series B (Statistical
  Methodology)} {\bf 70,} 849--911.

\bibitem[\protect\citeauthoryear{Fare, Coffey, Dai, He, Kessler, Kilian, Koch,
  LeProust, Marton, Meyer, et~al\mbox{.}}{Fare et~al.}{2003}]{fare2003effects}
Fare, T.~L., Coffey, E.~M., Dai, H., He, Y.~D., Kessler, D.~A., Kilian, K.~A.,
  Koch, J.~E., LeProust, E., Marton, M.~J., Meyer, M.~R., et~al. (2003).
\newblock Effects of atmospheric ozone on microarray data quality.
\newblock {\em Analytical chemistry} {\bf 75,} 4672--4675.

\bibitem[\protect\citeauthoryear{Fithian, Sun, and Taylor}{Fithian
  et~al.}{2014}]{fithian2014optimal}
Fithian, W., Sun, D., and Taylor, J. (2014).
\newblock Optimal inference after model selection.
\newblock {\em arXiv preprint arXiv:1410.2597} .

\bibitem[\protect\citeauthoryear{Friguet, Kloareg, and Causeur}{Friguet
  et~al.}{2009}]{friguet2009factor}
Friguet, C., Kloareg, M., and Causeur, D. (2009).
\newblock A factor model approach to multiple testing under dependence.
\newblock {\em Journal of the American Statistical Association} {\bf 104,}
  1406--1415.

\bibitem[\protect\citeauthoryear{Gagnon-Bartsch and Speed}{Gagnon-Bartsch and
  Speed}{2012}]{gagnon2012using}
Gagnon-Bartsch, J.~A. and Speed, T.~P. (2012).
\newblock Using control genes to correct for unwanted variation in microarray
  data.
\newblock {\em Biostatistics} {\bf 13,} 539--552.

\bibitem[\protect\citeauthoryear{Garreau, Jitkrittum, and Kanagawa}{Garreau
  et~al.}{2017}]{Garreau2017}
Garreau, D., Jitkrittum, W., and Kanagawa, M. (2017).
\newblock Large sample analysis of the median heuristic.

\bibitem[\protect\citeauthoryear{Gasch, Spellman, Kao, Carmel-Harel, Eisen,
  Storz, Botstein, and Brown}{Gasch et~al.}{2000}]{gasch2000genomic}
Gasch, A.~P., Spellman, P.~T., Kao, C.~M., Carmel-Harel, O., Eisen, M.~B.,
  Storz, G., Botstein, D., and Brown, P.~O. (2000).
\newblock Genomic expression programs in the response of yeast cells to
  environmental changes.
\newblock {\em Molecular biology of the cell} {\bf 11,} 4241--4257.

\bibitem[\protect\citeauthoryear{Gershoni and Pietrokovski}{Gershoni and
  Pietrokovski}{2017}]{gershoni2017landscape}
Gershoni, M. and Pietrokovski, S. (2017).
\newblock The landscape of sex-differential transcriptome and its consequent
  selection in human adults.
\newblock {\em BMC biology} {\bf 15,} 1--15.

\bibitem[\protect\citeauthoryear{Glass, Vi{\~n}uela, Davies, Ramasamy, Parts,
  Knowles, Brown, Hedman, Small, Buil, et~al\mbox{.}}{Glass
  et~al.}{2013}]{glass2013gene}
Glass, D., Vi{\~n}uela, A., Davies, M.~N., Ramasamy, A., Parts, L., Knowles,
  D., Brown, A.~A., Hedman, {\AA}.~K., Small, K.~S., Buil, A., et~al. (2013).
\newblock Gene expression changes with age in skin, adipose tissue, blood and
  brain.
\newblock {\em Genome biology} {\bf 14,} 1--12.

\bibitem[\protect\citeauthoryear{Gretton, Bousquet, Smola, and
  Sch{\"o}lkopf}{Gretton et~al.}{2005}]{gretton2005measuring}
Gretton, A., Bousquet, O., Smola, A., and Sch{\"o}lkopf, B. (2005).
\newblock Measuring statistical dependence with hilbert-schmidt norms.
\newblock In {\em International conference on algorithmic learning theory},
  pages 63--77. Springer.

\bibitem[\protect\citeauthoryear{Gretton, Fukumizu, Teo, Song, Sch{\"o}lkopf,
  Smola, et~al\mbox{.}}{Gretton et~al.}{2007}]{gretton2007kernel}
Gretton, A., Fukumizu, K., Teo, C.~H., Song, L., Sch{\"o}lkopf, B., Smola,
  A.~J., et~al. (2007).
\newblock A kernel statistical test of independence.
\newblock In {\em Nips}, volume~20, pages 585--592. Citeseer.

\bibitem[\protect\citeauthoryear{Lazar, Meganck, Taminau, Steenhoff, Coletta,
  Molter, Weiss-Sol{\'\i}s, Duque, Bersini, and Now{\'e}}{Lazar
  et~al.}{2013}]{lazar2013batch}
Lazar, C., Meganck, S., Taminau, J., Steenhoff, D., Coletta, A., Molter, C.,
  Weiss-Sol{\'\i}s, D.~Y., Duque, R., Bersini, H., and Now{\'e}, A. (2013).
\newblock Batch effect removal methods for microarray gene expression data
  integration: a survey.
\newblock {\em Briefings in bioinformatics} {\bf 14,} 469--490.

\bibitem[\protect\citeauthoryear{Leek, Scharpf, Bravo, Simcha, Langmead,
  Johnson, Geman, Baggerly, and Irizarry}{Leek et~al.}{2010}]{leek2010tackling}
Leek, J.~T., Scharpf, R.~B., Bravo, H.~C., Simcha, D., Langmead, B., Johnson,
  W.~E., Geman, D., Baggerly, K., and Irizarry, R.~A. (2010).
\newblock Tackling the widespread and critical impact of batch effects in
  high-throughput data.
\newblock {\em Nature Reviews Genetics} {\bf 11,} 733--739.

\bibitem[\protect\citeauthoryear{Leek and Storey}{Leek and
  Storey}{2007}]{leek2007capturing}
Leek, J.~T. and Storey, J.~D. (2007).
\newblock Capturing heterogeneity in gene expression studies by surrogate
  variable analysis.
\newblock {\em PLoS genetics} {\bf 3,} e161.

\bibitem[\protect\citeauthoryear{Leek and Storey}{Leek and
  Storey}{2008}]{leek2008general}
Leek, J.~T. and Storey, J.~D. (2008).
\newblock A general framework for multiple testing dependence.
\newblock {\em Proceedings of the National Academy of Sciences} {\bf 105,}
  18718--18723.

\bibitem[\protect\citeauthoryear{Liang and Cookson}{Liang and
  Cookson}{2014}]{liang2014grasping}
Liang, L. and Cookson, W.~O. (2014).
\newblock Grasping nettles: cellular heterogeneity and other confounders in
  epigenome-wide association studies.
\newblock {\em Human molecular genetics} {\bf 23,} R83--R88.

\bibitem[\protect\citeauthoryear{Lin, Coleman, Hawley, Huang, Dumpit, Gifford,
  Kezele, Hung, Knudsen, Kristal, et~al\mbox{.}}{Lin
  et~al.}{2006}]{lin2006influence}
Lin, D.~W., Coleman, I.~M., Hawley, S., Huang, C.~Y., Dumpit, R., Gifford, D.,
  Kezele, P., Hung, H., Knudsen, B.~S., Kristal, A.~R., et~al. (2006).
\newblock Influence of surgical manipulation on prostate gene expression:
  implications for molecular correlates of treatment effects and disease
  prognosis.
\newblock {\em Journal of clinical oncology} {\bf 24,} 3763--3770.

\bibitem[\protect\citeauthoryear{Lu, Pan, Kao, Li, Kohane, Chan, and
  Yankner}{Lu et~al.}{2004}]{lu2004gene}
Lu, T., Pan, Y., Kao, S.-Y., Li, C., Kohane, I., Chan, J., and Yankner, B.~A.
  (2004).
\newblock Gene regulation and dna damage in the ageing human brain.
\newblock {\em Nature} {\bf 429,} 883--891.

\bibitem[\protect\citeauthoryear{Majewski and Bernards}{Majewski and
  Bernards}{2011}]{majewski2011taming}
Majewski, I.~J. and Bernards, R. (2011).
\newblock Taming the dragon: genomic biomarkers to individualize the treatment
  of cancer.
\newblock {\em Nature medicine} {\bf 17,} 304--312.

\bibitem[\protect\citeauthoryear{Matthews, Hosker, Rudenski, Naylor, Treacher,
  and Turner}{Matthews et~al.}{1985}]{Matthews1985}
Matthews, D.~R., Hosker, J.~P., Rudenski, A.~S., Naylor, B.~A., Treacher,
  D.~F., and Turner, R.~C. (1985).
\newblock Homeostasis model assessment: insulin resistance and beta-cell
  function from fasting plasma glucose and insulin concentrations in man.
\newblock {\em Diabetologia} {\bf 28,} 412--419.

\bibitem[\protect\citeauthoryear{Mirza and Osindero}{Mirza and
  Osindero}{2014}]{mirza2014conditional}
Mirza, M. and Osindero, S. (2014).
\newblock Conditional generative adversarial nets.
\newblock {\em arXiv preprint arXiv:1411.1784} .

\bibitem[\protect\citeauthoryear{Morgan, Kabakchiev, Waldron, Tyler, Tickle,
  Milgrom, Stempak, Gevers, Xavier, Silverberg, and Huttenhower}{Morgan
  et~al.}{2015}]{Morgan2015}
Morgan, X.~C., Kabakchiev, B., Waldron, L., Tyler, A.~D., Tickle, T.~L.,
  Milgrom, R., Stempak, J.~M., Gevers, D., Xavier, R.~J., Silverberg, M.~S.,
  and Huttenhower, C. (2015).
\newblock Associations between host gene expression, the mucosal microbiome,
  and clinical outcome in the pelvic pouch of patients with inflammatory bowel
  disease.
\newblock {\em Genome Biology} {\bf 16,} 67.

\bibitem[\protect\citeauthoryear{Naaman}{Naaman}{2021}]{naaman2021tight}
Naaman, M. (2021).
\newblock On the tight constant in the multivariate
  dvoretzky--kiefer--wolfowitz inequality.
\newblock {\em Statistics \& Probability Letters} {\bf 173,} 109088.

\bibitem[\protect\citeauthoryear{Pedersen, Forslund, Gudmundsdottir,
  Østergaard Petersen, Hildebrand, Hyötyläinen, Nielsen, Hansen, Bork,
  Ehrlich, Brunak, Oresic, Pedersen, and Nielsen}{Pedersen
  et~al.}{2018}]{Pedersen2018}
Pedersen, H.~K., Forslund, S.~K., Gudmundsdottir, V., Østergaard Petersen, A.,
  Hildebrand, F., Hyötyläinen, T., Nielsen, T., Hansen, T., Bork, P.,
  Ehrlich, S.~D., Brunak, S., Oresic, M., Pedersen, O., and Nielsen, H.~B.
  (2018).
\newblock A computational framework to integrate high-throughput ‘-omics’
  datasets for the identification of potential mechanistic links.
\newblock {\em Nature Protocols} {\bf 13,} 2781--2800.

\bibitem[\protect\citeauthoryear{Pedersen, Gudmundsdottir, Nielsen,
  Hyotylainen, Nielsen, Jensen, Forslund, Hildebrand, Prifti, Falony,
  Chatelier, Levenez, Doré, Mattila, Plichta, Pöhö, Hellgren, Arumugam,
  Sunagawa, Vieira-Silva, Jørgensen, Holm, Trošt, Consortium, Kristiansen,
  Brix, Raes, Wang, Hansen, Bork, Brunak, Oresic, Ehrlich, and
  Pedersen}{Pedersen et~al.}{2016}]{Pedersen2016}
Pedersen, H.~K., Gudmundsdottir, V., Nielsen, H.~B., Hyotylainen, T., Nielsen,
  T., Jensen, B. A.~H., Forslund, K., Hildebrand, F., Prifti, E., Falony, G.,
  Chatelier, E.~L., Levenez, F., Doré, J., Mattila, I., Plichta, D.~R.,
  Pöhö, P., Hellgren, L.~I., Arumugam, M., Sunagawa, S., Vieira-Silva, S.,
  Jørgensen, T., Holm, J.~B., Trošt, K., Consortium, M., Kristiansen, K.,
  Brix, S., Raes, J., Wang, J., Hansen, T., Bork, P., Brunak, S., Oresic, M.,
  Ehrlich, S.~D., and Pedersen, O. (2016).
\newblock Human gut microbes impact host serum metabolome and insulin
  sensitivity.
\newblock {\em Nature} {\bf 535,} 376--381.

\bibitem[\protect\citeauthoryear{Sriperumbudur, Fukumizu, and
  Lanckriet}{Sriperumbudur et~al.}{2011}]{sriperumbudur2011universality}
Sriperumbudur, B.~K., Fukumizu, K., and Lanckriet, G.~R. (2011).
\newblock Universality, characteristic kernels and rkhs embedding of measures.
\newblock {\em Journal of Machine Learning Research} {\bf 12,}.

\bibitem[\protect\citeauthoryear{Storey}{Storey}{2002}]{storey2002direct}
Storey, J.~D. (2002).
\newblock A direct approach to false discovery rates.
\newblock {\em Journal of the Royal Statistical Society: Series B (Statistical
  Methodology)} {\bf 64,} 479--498.

\bibitem[\protect\citeauthoryear{Storey, Taylor, and Siegmund}{Storey
  et~al.}{2004}]{storey2004strong}
Storey, J.~D., Taylor, J.~E., and Siegmund, D. (2004).
\newblock Strong control, conservative point estimation and simultaneous
  conservative consistency of false discovery rates: a unified approach.
\newblock {\em Journal of the Royal Statistical Society: Series B (Statistical
  Methodology)} {\bf 66,} 187--205.

\bibitem[\protect\citeauthoryear{Sz{\'e}kely, Rizzo, and Bakirov}{Sz{\'e}kely
  et~al.}{2007}]{szekely2007measuring}
Sz{\'e}kely, G.~J., Rizzo, M.~L., and Bakirov, N.~K. (2007).
\newblock Measuring and testing dependence by correlation of distances.
\newblock {\em The annals of statistics} {\bf 35,} 2769--2794.

\bibitem[\protect\citeauthoryear{Walsh, Hokenstad, Chen, Sung, Jenkins, Chia,
  Nelson, Mariani, and Walther-Antonio}{Walsh
  et~al.}{2019}]{walsh2019postmenopause}
Walsh, D.~M., Hokenstad, A.~N., Chen, J., Sung, J., Jenkins, G.~D., Chia, N.,
  Nelson, H., Mariani, A., and Walther-Antonio, M.~R. (2019).
\newblock Postmenopause as a key factor in the composition of the endometrial
  cancer microbiome (ecbiome).
\newblock {\em Scientific reports} {\bf 9,} 1--16.

\bibitem[\protect\citeauthoryear{Wang, Zhao, Hastie, and Owen}{Wang
  et~al.}{2017}]{wang2017confounder}
Wang, J., Zhao, Q., Hastie, T., and Owen, A.~B. (2017).
\newblock Confounder adjustment in multiple hypothesis testing.
\newblock {\em Annals of statistics} {\bf 45,} 1863.

\bibitem[\protect\citeauthoryear{Winkler, Ridgway, Webster, Smith, and
  Nichols}{Winkler et~al.}{2014}]{winkler2014permutation}
Winkler, A.~M., Ridgway, G.~R., Webster, M.~A., Smith, S.~M., and Nichols,
  T.~E. (2014).
\newblock Permutation inference for the general linear model.
\newblock {\em Neuroimage} {\bf 92,} 381--397.

\bibitem[\protect\citeauthoryear{Yi, Zhang, Yang, Huang, Liu, Wang, Schaid, and
  Chen}{Yi et~al.}{2021}]{yi20212dfdr}
Yi, S., Zhang, X., Yang, L., Huang, J., Liu, Y., Wang, C., Schaid, D.~J., and
  Chen, J. (2021).
\newblock 2dfdr: a new approach to confounder adjustment substantially
  increases detection power in omics association studies.
\newblock {\em Genome biology} {\bf 22,} 1--18.

\bibitem[\protect\citeauthoryear{Zhang, Peters, Janzing, and
  Sch{\"o}lkopf}{Zhang et~al.}{2012}]{zhang2012kernel}
Zhang, K., Peters, J., Janzing, D., and Sch{\"o}lkopf, B. (2012).
\newblock Kernel-based conditional independence test and application in causal
  discovery.
\newblock {\em arXiv preprint arXiv:1202.3775} .

\bibitem[\protect\citeauthoryear{Zhou, Jiao, Liu, and Huang}{Zhou
  et~al.}{2022}]{zhou2022deep}
Zhou, X., Jiao, Y., Liu, J., and Huang, J. (2022).
\newblock A deep generative approach to conditional sampling.
\newblock {\em Journal of the American Statistical Association} pages 1--12.

\bibitem[\protect\citeauthoryear{Ziegler, Koch, Krockenberger, and
  Gro{\ss}hennig}{Ziegler et~al.}{2012}]{ziegler2012personalized}
Ziegler, A., Koch, A., Krockenberger, K., and Gro{\ss}hennig, A. (2012).
\newblock Personalized medicine using dna biomarkers: a review.
\newblock {\em Human genetics} {\bf 131,} 1627--1638.

\end{thebibliography}

\appendix
\section{Appendix}
The supplement contains more discussions on Assumption 2, the 2d FWER-controlling procedure, the technical details, asymptotic power analysis, the DGPs used in the simulation studies, and additional numerical results.

\subsection{Discussions on Assumption \ref{as2}}\label{as2-example}
We provide further discussions on Assumption \ref{as2} and justify it under model (\ref{eq-model}) with 
\begin{align*}
u_j(\X)=\sum^{J_1}_{k=1}\alpha_{k,j}B_{k,\X}(\X),\quad v_j(\Z)=\sum^{J_2}_{k=1}\beta_{k,j}B_{k,\Z}(\Z), \quad 
\boldsymbol{\epsilon}_j=(\epsilon_{1,j},\dots,\epsilon_{n,j})^\top \sim N(0,\sigma_j^2\mathbf{I}),
\end{align*}
where $B_{k,\X}(\cdot):\mathbb{R}^p\rightarrow \mathbb{R}$ and $B_{k,\Z}(\cdot):\mathbb{R}^d\rightarrow \mathbb{R}$
are some known basis functions. Define $\B_{\X}=(B_{k,\X}(\X_i))_{1\leq i\leq n,1\leq k\leq J_1}\in\mathbb{R}^{n\times J_1}$ and $\B_{\Z}=(B_{k,\Z}(\Z_i))_{1\leq i\leq n,1\leq k\leq J_2}\in\mathbb{R}^{n\times J_2}$.
Let $\P_\Z^\perp$ be the orthogonal projection onto the column space of $\B_{\Z}$. We consider the statistics
\begin{align*}
& T_{j}^M=\hat{\sigma}_j^{-2}\|(\B_{\X}^{\top}\B_{\X})^{-1/2} \B_{\X}^\top \widetilde{\Y}_j\|^2, \\
& T_{j}^C=\hat{\sigma}_j^{-2}\|(\B_{\X}^{\top}\P_\Z^\perp\B_{\X})^{-1/2} \B_{\X}^\top \P_\Z^\perp\widetilde{\Y}_j\|^2,
\end{align*} 
where $\hat{\sigma}_j^2$ is a consistent variance estimator of $\sigma_j^2$ such that $\hat{\sigma}_j^2\rightarrow^p \sigma_j^2.$ Conditional on $(\widetilde{\X},\widetilde{\Z})$, 
$(\B_{\X}^{\top}\B_{\X})^{-1/2} \B_{\X}^\top \widetilde{\Y}_j$ and $(\B_{\X}^{\top}\P_\Z^\perp\B_{\X})^{-1/2} \B_{\X}\P_\Z^\perp\widetilde{\Y}_j$
jointly follow the multivariate normal distribution with the mean 
\begin{align*}
\begin{pmatrix}
(\B_{\X}^{\top}\B_{\X})^{1/2}\balpha_j + 
(\B_{\X}^{\top}\B_{\X})^{-1/2}(\B_{\X}^{\top}\B_{\Z})\bbeta_j\\
(\B_{\X}^{\top}\P_\Z^\perp\B_{\X})^{1/2}\balpha_j
\end{pmatrix}    
\end{align*}
and the covariance matrix
\begin{align*}
\sigma_j^2
\begin{pmatrix}
\mathbf{I} & (\B_{\X}^{\top}\B_{\X})^{-1/2}(\B_{\X}^{\top}\P_\Z^\perp\B_{\X})^{1/2} \\
(\B_{\X}^{\top}\P_\Z^\perp\B_{\X})^{1/2}(\B_{\X}^{\top}\B_{\X})^{-1/2}  & \mathbf{I}
\end{pmatrix}.    
\end{align*}
Define $\boldsymbol{\Sigma}_\X=\text{cov}(\widetilde{\B}_\X)$, $\boldsymbol{\Sigma}_{\X\Z}=\text{cov}(\widetilde{\B}_\X,\widetilde{\B}_\Z)$ 
and $\boldsymbol{\Sigma}_{\X|\Z}=\boldsymbol{\Sigma}_\X-\boldsymbol{\Sigma}_{\X\Z}\boldsymbol{\Sigma}_{\Z}^{-1}\boldsymbol{\Sigma}_{\Z\X}$, where $\widetilde{\B}_\X=(B_{1,\X}(\X),\dots,B_{J_1,\X}(\X))^\top$ and $\widetilde{\B}_\Z=(B_{1,\Z}(\Z),\dots,B_{J_2,\X}(\Z))^\top$. By the law of large numbers, we have 
\begin{align*}
&n^{-1}\B_{\X}^{\top}\B_{\X}\rightarrow^p  \boldsymbol{\Sigma}_\X,\\
&n^{-1/2}(\B_{\X}^{\top}\B_{\X})^{-1/2}(\B_{\X}^{\top}\B_{\Z})\rightarrow^p \boldsymbol{\Sigma}_\X^{-1/2}\boldsymbol{\Sigma}_{\X\Z},\\
& n^{-1}\B_{\X}^{\top}\P_\Z^\perp\B_{\X}\rightarrow^p \boldsymbol{\Sigma}_{\X|\Z}.
\end{align*}
In this case, we have
\begin{align*}
&\widetilde{V}(t_1,t_2)=\lim_{m,n\rightarrow+\infty}\frac{1}{m_0}\sum_{j\in \mathcal{M}_0}F(t_1,t_2;0,\sqrt{n}\boldsymbol{\beta}_j/\sigma_j),\\
&\widetilde{S}(t_1,t_2)=\lim_{m,n\rightarrow+\infty}\frac{1}{m}\sum_{j=1}^m F(t_1,t_2;\sqrt{n}\boldsymbol{\alpha}_j/\sigma_j,\sqrt{n}\boldsymbol{\beta}_j/\sigma_j),
\end{align*}
with $F(t_1,t_2;\mathbf{a},\mathbf{b})=P(\|\mathbf{V}_{1,j}\|^2>t_1,\|\mathbf{V}_{2,j}\|^2>t_2)$, where $(\mathbf{V}_{1,j},\mathbf{V}_{2,j})$ follow the multivariate normal distribution with the mean 
\begin{align*}
\begin{pmatrix}
\boldsymbol{\Sigma}^{1/2}_\X\mathbf{a} + 
\boldsymbol{\Sigma}^{-1/2}_\X\boldsymbol{\Sigma}_{\X\Z}\mathbf{b}\\
\boldsymbol{\Sigma}^{1/2}_{\X|\Z}\mathbf{a}
\end{pmatrix}    
\end{align*}
and the covariance matrix
\begin{align*}
\begin{pmatrix}
\mathbf{I} & \boldsymbol{\Sigma}^{-1/2}_\X\boldsymbol{\Sigma}^{1/2}_{\X|\Z} \\
\boldsymbol{\Sigma}^{1/2}_{\X|\Z}\boldsymbol{\Sigma}^{-1/2}_\X & \mathbf{I}
\end{pmatrix}.    
\end{align*}
If $(\sqrt{n}\boldsymbol{\alpha}_j/\sigma_j,\sqrt{n}\boldsymbol{\beta}_j/\sigma_j)$ follows some distribution $\mathcal{F}$ independently across $j$ and conditional on $\boldsymbol{\alpha}_j=0$, $\sqrt{n}\boldsymbol{\beta}_j/\sigma_j$ follows the distribution $\mathcal{F}_0$ independently for $j\in\mathcal{M}_0$, then we have 
\begin{align*}
&\widetilde{V}(t_1,t_2)=\int F(t_1,t_2,(0,\bbeta))d\mathcal{F}_0(\bbeta),\\
&\widetilde{S}(t_1,t_2)=\int F(t_1,t_2,(\balpha,\bbeta))d\mathcal{F}(\balpha,\bbeta).
\end{align*}

\subsection{Family-wise error rate control}\label{sec:2d-FWER}
Family-wise error rate (FWER), referring to the probability of making one false discovery, provides more stringent type I error rate control. It is preferable to the FDR if the overall conclusion from various individual inferences is likely to be erroneous when at least one of them is, or the existence of a single false claim would cause significant loss. It is natural to ask whether our method can be modified to control other error measures such as FWER. Here we describe such a procedure to control the FWER. Given the rejection rule $\mathbf{1}\{T_j^M\geq t_1,T_j^C\geq t_2\}$, we let 
$\widetilde{\text{FWER}}(t_1,t_2):=\sum_{j=1}^m \bar{F}_{j,B}(t_1,t_2)$ be an estimate of the FWER. We choose the optimal cut-off value as the one that maximizes the number of rejections while controls the FWER estimate at a prespecified level $q$:
$$(\breve{t}_1,\breve{t}_2)=\argmax_{(t_1,t_2)\in\mathcal{G}_q}\sum^{m}_{j=1}\mathbf{1}\{T_j^{M}\geq t_{1},T_j^{C}\geq t_{2}\},$$
where $\mathcal{G}_q=\{(t_1,t_2)\in \mathbb{R}^+\times\mathbb{R}^+:\widetilde{\text{FWER}}(t_1,t_2) \leq q\}$.
Then we reject $H_{0,j}$ whenever 
$T_j^M\geq \breve{t}_{1}$ and $T_j^C\geq \breve{t}_{2}.$ We name the above procedure 2dFWER+. In Section \ref{sec:additional}, we investigate its finite sample performance and report the empirical FWER and power for 2dFWER+ and its corresponding 1d version (1dFWER) in Figures \ref{fig:18} and \ref{fig:19}.

\subsection{Technical details}\label{sec:tech}
In this section, we prove the main theoretical results in the paper.
We first present the following lemma which was recently proved in \cite{naaman2021tight}.
\begin{lemma}[Dvoretzky–Kiefer–Wolfowitz inequality]\label{lem-1}
Let $\boldsymbol{\xi}_1,\cdots,\boldsymbol{\xi}_n$ be independent $d$-dimensional random vectors with the distribution function $F(\mathbf{t}) =P(\boldsymbol{\xi}_i\leq \mathbf{t})$, where $\boldsymbol{\xi}_i\leq \mathbf{t}$ means that $\xi_{ij}\leq t_j$ for $\boldsymbol{\xi}_i=(\xi_{i1},\dots,\xi_{id})$ and $1\leq j\leq d$. Denote the standard empirical distribution function by
$F_n(\mathbf{t})=n^{-1}\sum^{n}_{i=1}\mathbf{1}\{\boldsymbol{\xi}_i\leq \mathbf{t}\}$. Then we have 
\begin{align*}
P\left(\sup_{\mathbf{t}\in\mathbb{R}^d}|F_n(\mathbf{t})-F(\mathbf{t})|>\epsilon\right)\leq d(n+1) \exp\left(-2n\epsilon^2\right).
\end{align*}
\end{lemma}

We now present the proof of the main theoretical results.

\begin{proof}[Proof of Theorem \ref{thm:finite}]
Define the filtration
\begin{align*}
\mathcal{F}_s=\sigma\left(\left\{\mathbf{1}\{T^M_{j,b}\geq t_1(a)\},\mathbf{1}\{T_{j,b}^C\geq t_2(a)\}\right\}_{1\leq j\leq m,0\leq b\leq B}: 1\leq a \leq s\right)   
\end{align*}
for $1\leq s\leq \mathcal{S}$ and the process
$U(s)=\widetilde{V}^0(s)/\{\sum^{B}_{b=0}\widetilde{V}^b(s)\}$,
which is adapted to the filtration $\mathcal{F}_s.$ The conditional distribution of $\widetilde{V}^b(t)$ given the sigma-field $\sigma(\{\mathbf{1}\{T^M_{j,b}\geq t_1(a)\},\mathbf{1}\{T^M_{j,b}\geq t_2(a)\}\}_{1\leq j\leq m}: 1\leq a \leq s)$ with $s<t$ are the same across all $b=0,1,\dots,B.$ By the symmetry, we must have for $s<t$,
$\bE[U(t)|\mathcal{F}_s]=(B+1)^{-1}.$
Thus $U(t)-1/(B+1)$ is a martingale difference sequence. Also, we have
$\{s^*\leq t\} \in\mathcal{F}_t.$   
Therefore, $s^*$ is a stopping time. 
By the optional stopping time theorem, 
\begin{align}\label{eq-fin-1}
\bE[U(s^*)]=\frac{1}{B+1}.
\end{align}
Recall from the definition of $s^*$ that
\begin{align}\label{eq-fin-2}
\frac{(B+1)^{-1}\sum^{B}_{b=0}V^b(s^*)}{1\vee V^0(s^*)}\leq q.    
\end{align}
Using (\ref{eq-fin-1}) and (\ref{eq-fin-2}), we obtain
\begin{align*}
\bE\left[\frac{\widetilde{V}^0(s^*)}{1\vee V^0(s^*)}\right]
\leq (B+1)q \bE\left[\frac{\widetilde{V}^0(s^*)}{\sum^{B}_{b=0}V^b(s^*)}\right] 
= (B+1)q \bE\left[U(s^*)\right]=q. 
\end{align*}
\end{proof}

\begin{proof}[Proof of Theorem \ref{thm-main}]
For $(t_1,t_2)\in \mathbb{R}^+\times \mathbb{R}^+$, define the following processes
\begin{align*}
&S_{n,m}(t_1,t_2)=m^{-1}\sum^{m}_{j=1}\mathbf{1}\{T_{j}^M\geq t_1,T_j^C\geq t_2\},\\ &V_{n,m}(t_1,t_2)=m^{-1}_0\sum_{j\in \mathcal{M}_0}\mathbf{1}\{T_{j}^M\geq t_1,T^C_j\geq t_2\},\\
&Q_{n,m}(t_1,t_2)=m_0^{-1}\sum_{j\in\mathcal{M}_0} P_0(T_j^{M}\geq t_1,T_j^{C}\geq t_2|\widetilde{\Y}_j,\widetilde{\Z}).
\end{align*}

We divide the proof into two steps. In Step 1, we obtain some uniform convergence results while in Step 2, we apply these results to show the FDR control.
\\ \noindent \textbf{Step 1.} Conditional on $(\widetilde{\X},\widetilde{\Z})$, $\mathbf{1}\{T_j^M\geq t_1,T_j^C\geq t_2\}$ are independent across $j\in\mathcal{M}_0$. By Lemma \ref{lem-1}, we have
\begin{align}\label{eq-11}
\sup_{t_1\leq t_{0,1},t_{2}\leq t_{0,2}}\left|
\frac{1}{m_0}\sum_{j\in\mathcal{M}_0}\left[\mathbf{1}\{T_j^M\geq t_1,T_j^C\geq t_2\}-P(T_j^M\geq t_1,T_j^C\geq t_2|\widetilde{X},\widetilde{\Z})\right]
\right|\rightarrow^p 0.    
\end{align}
By Assumption \ref{as1}, conditional on $\widetilde{\Z}$ and for any fixed $t_1$ and $t_2$, $P(T_j^M\geq t_1,T_j^C\geq t_2|\widetilde{\Y}_j,\widetilde{\Z})$ are independent across $j\in\mathcal{M}_0$. Therefore, by the law of large numbers,
\begin{align*}
\frac{1}{m_0}\sum_{j\in\mathcal{M}_0}\Bigg\{P(T_j^M\geq t_1,T_j^C\geq t_2|\widetilde{\Y}_j,\widetilde{\Z})-P(T_j^M\geq t_1,T_j^C\geq t_2|\widetilde{\Z})\Bigg\}\rightarrow^p 0.
\end{align*}
Following the proof of the Glivenko-Cantelli Theorem, we can strengthen the point-wise convergence to the uniform convergence, i.e., 
\begin{align}\label{eq-12}
&\sup_{t_1\leq t_{0,1},t_{2}\leq t_{0,2}}\Bigg|
\frac{1}{m_0}\sum_{j\in\mathcal{M}_0}\Bigg\{P(T_j^M\geq t_1,T_j^C\geq t_2|\widetilde{\Y}_j,\widetilde{\Z})
-P(T_j^M\geq t_1,T_j^C\geq t_2|\widetilde{\Z})\Bigg\}
\Bigg|\rightarrow^p 0.
\end{align}
Similarly, the result in Assumption \ref{as2} can also be strengthened to the uniform convergence, i.e.,
\begin{align}\label{eq-13}
&\sup_{t_1\leq t_{0,1},t_{2}\leq t_{0,2}}\left|\frac{1}{m_0}\sum_{j\in\mathcal{M}_0} P(T_j^M\geq t_1,T_j^C\geq t_2|\widetilde{X},\widetilde{\Z})-\widetilde{V}(t_1,t_2)\right|\rightarrow^p 0.
\end{align}
It implies that
\begin{equation}\label{eq-14}
\begin{split}
&\sup_{t_1\leq t_{0,1},t_{2}\leq t_{0,2}}\left|\frac{1}{m_0}\sum_{j\in\mathcal{M}_0} P(T_j^M\geq t_1,T_j^C\geq t_2|\widetilde{\Z})-\widetilde{V}(t_1,t_2)\right|
\\ = &\sup_{t_1\leq t_{0,1},t_{2}\leq t_{0,2}}\left|\bE\left[\frac{1}{m_0}\sum_{j\in\mathcal{M}_0} P(T_j^M\geq t_1,T_j^C\geq t_2|\widetilde{\X},\widetilde{\Z})-\widetilde{V}(t_1,t_2)\Bigg| \widetilde{\Z}\right]\right|
\\ \leq & \bE\left[\sup_{t_1\leq t_{0,1},t_{2}\leq t_{0,2}}\left|\frac{1}{m_0}\sum_{j\in\mathcal{M}_0} P(T_j^M\geq t_1,T_j^C\geq t_2|\widetilde{\Z})-\widetilde{V}(t_1,t_2)\right|\Bigg|\widetilde{\Z}\right]\rightarrow^p 0,
\end{split}
\end{equation}
by Lebesgue's dominated convergence theorem. Combining (\ref{eq-11}), (\ref{eq-12}), (\ref{eq-13}) and (\ref{eq-14}) together, we get
\begin{align}\label{eq-17}
&\sup_{t_1\leq t_{0,1},t_{2}\leq t_{0,2}}\left|V_{n,m}(t_1,t_2)-\widetilde{V}(t_1,t_2)\right|\rightarrow^{p} 0,\\
&\sup_{t_1\leq t_{0,1},t_{2}\leq t_{0,2}}\left|Q_{n,m}(t_1,t_2)-\widetilde{V}(t_1,t_2)\right|\rightarrow^{p} 0.
\end{align}
Using similar arguments by conditioning on $(\widetilde{\X},\widetilde{\Z})$, we can show that
\begin{align} \label{eq-18}
\sup_{t_1\leq t_{0,1},t_{2}\leq t_{0,2}}\left|S_{n,m}(t_1,t_2)-\widetilde{S}(t_1,t_2)\right|\rightarrow^{p} 0.
\end{align}
Following the arguments in the proof of Lemma 8.2 of Cao et al. (2020), we have under Assumptions \ref{as1}-\ref{as3} that
\begin{align}
& \sup_{t_1\leq t_{0,1},t_2\leq t_{0,2}}\left|\text{FDP}(t_1,t_2)-\frac{\pi_0\widetilde{V}(t_1,t_2)}{\widetilde{S}(t_1,t_2)}\right|\rightarrow^p 0, \nonumber \\ 
& \sup_{t_1\leq t_{0,1},t_2\leq t_{0,2}}\left|\frac{m_0Q_{n,m}(t_1,t_2)}{1\vee m S_{n,m}(t_1,t_2)}-\frac{\pi_0\widetilde{V}(t_1,t_2)}{\widetilde{S}(t_1,t_2)}\right|\rightarrow^p 0. \label{eq-Q-con}
\end{align}

Moreover, under the null, we have $\bE[\mathbf{1}\{T_{j,b}^M\geq t_1,T_{j,b}^C\geq t_2\}|\widetilde{\Y}_j,\widetilde{\Z}]=P(T_j^M\geq t_1,T_j^C\geq t_2|\widetilde{\Y}_j,\widetilde{\Z})$ by the way we generate $\X_{i,b}$. Thus
$m_0^{-1}\sum_{j\in\mathcal{M}_0}\bE[\bar{F}_{j,B}(t_1,t_2)-Q_{n,m}(t_1,t_2)|\widetilde{\Z},\widetilde{\Y}_j,j\in\mathcal{M}_0]=0$
and 
\begin{align*}
&\text{var}\left(\frac{1}{m_0}\sum_{j\in\mathcal{M}_0}\left\{\bar{F}_{j,B}(t_1,t_2)-Q_{n,m}(t_1,t_2)\right\}\Bigg|\widetilde{\Z},\widetilde{\Y}_j,j\in\mathcal{M}_0\right)
\\=&\frac{1}{B+1}\text{var}\left(\frac{1}{m_0}\sum_{j\in\mathcal{M}_0}\left(\mathbf{1}\{T_{j,1}^M\geq t_1,T_{j,1}^C\geq t_2\}-Q_{n,m}(t_1,t_2)\right)\Bigg|\widetilde{\Z},\widetilde{\Y}_j,j\in\mathcal{M}_0\right)\leq \frac{1}{4(B+1)},
\end{align*}
where we have used the fact that $\text{var}(X)\leq 1/4$ for $X\in[0,1]$. Therefore, 
\begin{align*}
\frac{1}{m_0}\sum_{j\in\mathcal{M}_0}\left\{\bar{F}_{j,B}(t_1,t_2)-Q_{n,m}(t_1,t_2)\right\}\rightarrow^p 0, 
\end{align*}
which can be strengthened to the uniform convergence
\begin{align*}
\sup_{t_1\leq t_{0,1},t_2\leq t_{0,2}}\left|\frac{1}{m_0}\sum_{j\in\mathcal{M}_0}\left\{\bar{F}_{j,B}(t_1,t_2)-Q_{n,m}(t_1,t_2)\right\}\right|\rightarrow^p 0. 
\end{align*}
Together with (\ref{eq-Q-con}), we obtain
\begin{align}\label{eq-F-con}
\sup_{t_1\leq t_{0,1},t_2\leq t_{0,2}}\left|\frac{\sum_{j\in\mathcal{M}_0}\bar{F}_{j,B}(t_1,t_2)}{1\vee m S_{n,m}(t_1,t_2)}-\frac{\pi_0\widetilde{V}(t_1,t_2)}{\widetilde{S}(t_1,t_2)}\right|\rightarrow^p 0.    
\end{align}
In view of Assumption \ref{as3}, (\ref{eq-F-con}) implies that
\begin{align*}
&P\left(\widetilde{\text{FDP}}(t_{0,1},0)<q, \widetilde{\text{FDP}}(0,t_{0,2})<q\right)
\\ = &P\left(\frac{\sum_{j\in\mathcal{M}_0}\bar{F}_{j,B}(t_{0,1},0)+m_1U_{n,m}(t_{0,1},0)}{1\vee m S_{n,m}(t_{0,1},0)}<q, \frac{\sum_{j\in\mathcal{M}_0}\bar{F}_{j,B}(0,t_{0,2})+m_1U_{n,m}(0,t_{0,2})}{1\vee m S_{n,m}(0,t_{0,2})}<q\right)
\rightarrow 1,
\end{align*}
where $U_{n,m}(t_1,t_2)=m_1^{-1}\sum_{j\in\mathcal{M}_1} P_0(T_j^{M}\geq t_1,T_j^{C}\geq t_2|\widetilde{\Y}_j,\widetilde{\Z}).$
Thus we must have 
$$P(t_{1}^*\leq t_{0,1},t_{2}^*\leq t_{0,2})\rightarrow 1.$$ 

\noindent \textbf{Step 2.} Note that $\widetilde{\text{FDP}}(t_1,t_2)\geq \sum_{j\in\mathcal{M}_0}\bar{F}_{j,B}(t_1,t_2)/\{1\vee m S_{n,m}(t_1,t_2)\}$. On the event $t_{1}^*\leq t_{0,1}$ and $t_{2}^*\leq t_{0,2}$ which has probability converging to one, we have
\begin{align*}
&\text{FDP}(t_1^*,t_2^*)-\widetilde{\text{FDP}}(t_1^*,t_2^*) 
\\ \leq& 
\text{FDP}(t_1^*,t_2^*)-\frac{\sum_{j\in\mathcal{M}_0}\bar{F}_{j,B}(t_1^*,t_2^*)}{1\vee m S_{n,m}(t_1^*,t_2^*)}
\\=&\text{FDP}(t_1^*,t_2^*)-\frac{\pi_0\widetilde{V}(t_1^*,t_2^*)}{\widetilde{S}(t_1^*,t_2^*)}+\frac{\pi_0\widetilde{V}(t_1^*,t_2^*)}{\widetilde{S}(t_1^*,t_2^*)}-\frac{\sum_{j\in\mathcal{M}_0}\bar{F}_{j,B}(t_1^*,t_2^*)}{1\vee m S_{n,m}(t_1^*,t_2^*)}
\\ \leq & \sup_{t_1\leq t_{0,1},t_2\leq t_{0,2}}\left|\text{FDP}(t_1,t_2)-\frac{\pi_0\widetilde{V}(t_1,t_2)}{\widetilde{S}(t_1,t_2)}\right|
\\&+\sup_{t_1\leq t_{0,1},t_2\leq t_{0,2}}\left|\frac{\sum_{j\in\mathcal{M}_0}\bar{F}_{j,B}(t_1,t_2)}{1\vee m S_{n,m}(t_1,t_2)}-\frac{\pi_0\widetilde{V}(t_1,t_2)}{\widetilde{S}(t_1,t_2)}\right|=o_p(1).
\end{align*}
Thus we have 
\begin{equation}\label{eq-S}
\text{FDP}(t_1^*,t_2^*)\leq \widetilde{\text{FDP}}(t_1^*,t_2^*) +o_p(1)=q+o_p(1).
\end{equation}
By Lemma 8.3 of \cite{Cao:2020}, we get
\begin{align*}
\limsup_{n,m\rightarrow +\infty}\bE\left[\text{FDP}(t_1^*,t_2^*)\right]\leq q.
\end{align*}
\end{proof}

\begin{proof}[Proof of Corollary \ref{cor-1}]
Recall that 
$\text{TP}_{\text{2d}} \geq    \frac{1-q_2}{1-q_1}\text{TP}_{\text{1d}}.$
By (\ref{eq-S}) in the proof of Theorem \ref{thm-main}, $P(q_2\leq q+\epsilon)\rightarrow 1$. As $q_1\geq 0$, the conclusion follows.
\end{proof}

\subsection{Asymptotic Power Analysis}\label{sec:s-power}
We perform an asymptotic power analysis by comparing the asymptotic power of 2dFDR+ with that of the associated 1d procedure. For $(t_1,t_2)\in \mathbb{R}^+\times \mathbb{R}^+$, define $\widetilde{K}(t_1, t_2)$ as
\begin{equation} \label{eq-20}
   \widetilde{K}(t_1, t_2) = \frac{\widetilde{S}(t_1, t_2) - \pi_0 \widetilde{V}(t_1, t_2)}{(1-\pi_0)},
\end{equation}
which can be considered as the limiting power process. Assume that
\begin{align*}
\sup_{t_1\leq t_{0,1},t_2\leq t_{0,2}}\left|\frac{1}{m_1}\sum_{j\in\mathcal{M}_1}\bar{F}_{j,B}(t_1,t_2)-\widetilde{U}(t_1,t_2)\right|\rightarrow^p 0
\end{align*}
for some non-negative function $\widetilde{U}$. Let 
\begin{align*}
\widetilde{\text{FDP}^{\infty}}(t_1,t_2)=\frac{\pi_0 \widetilde{V}(t_1, t_2)+(1-\pi_0)\widetilde{U}(t_1,t_2)}{\widetilde{S}(t_1, t_2)},
\end{align*}
which is the limiting process for $\widetilde{\text{FDP}}(t_1,t_2)$ in view of the derivations in Section \ref{sec:tech}. To understand the power behavior of 2dFDR+ and the associated 1d procedure, we consider the following (infeasible) procedures based on the above limiting processes:
\begin{align*}
&\text{Limiting 2dFDR+:}\quad (t_{1,\text{2d}}^*,t_{2,\text{2d}}^*) = \argmax_{(t_1,t_2)\in \mathbb{R}^+\times\mathbb{R}^+}\widetilde{S}(t_1,t_2)\quad \text{subject to}\quad \widetilde{\text{FDP}^{\infty}}(t_1,t_2) \leq q,\\
&\text{Limiting 1dFDR:}\quad t_{\text{1d}}^*= \argmax_{t\in \mathbb{R}^+}\widetilde{S}(0,t)\quad \text{subject to}\quad \widetilde{\text{FDP}^{\infty}}(0,t) \leq q.
\end{align*}
As $(0,t_{\text{1d}}^*)$ is a feasible point of the optimization problem in limiting 2dFDR+, we must have
\begin{align}\label{ineq-S}
\widetilde{S}(t_{1,\text{2d}}^*,t_{2,\text{2d}}^*)\geq \widetilde{S}(0,t_{\text{1d}}^*).
\end{align}
Assume that $\widetilde{\text{FDP}^{\infty}}(t_1,t_2)$ is a continuous function of $(t_1,t_2)$. Then we have
$$\widetilde{\text{FDP}^{\infty}}(t_{1,\text{2d}}^*,t_{2,\text{2d}}^*)=\widetilde{\text{FDP}^{\infty}}(0,t_{\text{1d}}^*)=q,$$
as otherwise one can lower the values of $(t_1,t_2)$ to increase the value of the objective function $\widetilde{S}$.
Some algebra yields that
\begin{align*}
\frac{(1-\pi_0)\{\widetilde{K}(t_{1,\text{2d}}^*,t_{2,\text{2d}}^*) -\widetilde{U}(t_{1,\text{2d}}^*,t_{2,\text{2d}}^*) \}}{\widetilde{S}(t_{1,\text{2d}}^*,t_{2,\text{2d}}^*) }=\frac{(1-\pi_0)\{\widetilde{K}(0,t_{\text{1d}}^*)-\widetilde{U}(0,t_{\text{1d}}^*)\}}{\widetilde{S}(0,t_{\text{1d}}^*)}=1-q.    
\end{align*}
By (\ref{ineq-S}), we have
\begin{align*}
\widetilde{K}(t_{1,\text{2d}}^*,t_{2,\text{2d}}^*)\geq  \frac{1-q+(1-\pi_0)\widetilde{U}(t_{1,\text{2d}}^*,t_{2,\text{2d}}^*)/\widetilde{S}(t_{1,\text{2d}}^*,t_{2,\text{2d}}^*)}{1-q+(1-\pi_0)\widetilde{U}(0,t_{\text{1d}}^*)/\widetilde{S}(0,t_{\text{1d}}^*)}\widetilde{K}(0,t_{\text{1d}}^*)\geq (1-q)\widetilde{K}(0,t_{\text{1d}}^*).  
\end{align*}
Comparing to the result in Corollary \ref{cor-1}, we derive two terms $(1-\pi_0)\widetilde{U}(t_{1,\text{2d}}^*,t_{2,\text{2d}}^*)/\widetilde{S}(t_{1,\text{2d}}^*,t_{2,\text{2d}}^*)$ and $(1-\pi_0)\widetilde{U}(0,t_{\text{1d}}^*)/\widetilde{S}(0,t_{\text{1d}}^*)$ that determine the power improvement. In the worst-case scenario, $\widetilde{K}(t_{1,\text{2d}}^*,t_{2,\text{2d}}^*)\geq (1-q)\widetilde{K}(0,t_{\text{1d}}^*)$, which again suggests that the power loss is at most $q$.

\subsection{DGPs in the simulation studies}\label{sec:sim-model}
We provide the specific data generating processes (DGPs) considered in Section \ref{sec:sim-model-1}:
\begin{enumerate}
    \item $Y_j = \alpha_j X + \beta_j Z + \epsilon_j$ and $X \sim N(\rho Z, 1)$, where $Z \sim N(0,1)$;
    
   \item $Y_j = \alpha_j X^3 + \beta_j e^Z + \epsilon_j$ and $X\sim N(\rho Z^2 , 1)$, where $Z \sim N(0,1)$;
   
    
    \item $Y_j = \alpha_j X^3 + \beta_j Z^3 + \epsilon_j$ and $X \sim N(\rho(Z + Z^2) , 1)$, where $Z \sim N(0,1)$;
    
    \item $Y_j = \alpha_j (X + |X^3|) + \beta_j e^Z + \epsilon$ and $X\sim N(\rho(Z + Z^2) , 1)$, where $Z\sim N(0,1)$;
    
    
      \item $Y_j = \alpha_j e^X + \beta_j Z + \epsilon_j$ and $X \sim \text{Bernoulli}((1 + e^{-\rho Z})^{-1})$,  where $Z \sim N(0,1)$;
      
    \item $Y_j = \alpha_j e^X + \beta_j e^Z + \epsilon_j$ and $X \sim \text{Bernoulli}((1 + e^{-\rho Z})^{-1})$, where $Z \sim N(0,1)$;
    
    \item $Y_j = \alpha_j e^X + \beta_j Z^2 + \epsilon_j$ and $X \sim \text{Bernoulli}((1 + e^{-\rho Z})^{-1})$, where $Z \sim N(0,1)$;
    
    \item $Y_j = \alpha_j X + \beta_j Z + \epsilon_j$ and $X \sim \text{Bernoulli}((1 + e^{-\rho Z})^{-1})$, where $Z \sim \text{Bernoulli}(0.7)$;

    \item $Y_j \sim \text{Bernoulli}((1 + e^{-f_j(X,Z)})^{-1})$, where $f_j(X,Z) = \alpha_j X + \beta_j Z ,$ $X\sim N(\rho Z, 1)$ and $Z \sim N(0,1)$;
    
    \item $Y_j \sim \text{Poisson}(\lambda_j)$, where
    $\log \lambda_j = \alpha_j X + \beta_j Z$ with $X \sim N(\rho Z, 1)$ and $Z \sim N(0,1)$;
    
    \item $Y_j \sim \text{Negative Binomial}(\text{size} = 3, \mu_j = e^{f_j(X,Z)})$, where $f_j(X,Z) = \alpha_j X + \beta_j Z ,$ $X\sim N(\rho Z, 1)$ and $Z \sim N(0,1)$.
    
\end{enumerate}

\subsection{Additional simulation results}\label{sec:additional}
\begin{figure}[H]
\centering
\begin{subfigure}{0.9\textwidth}
  \centering
  \includegraphics[width=0.8\linewidth]{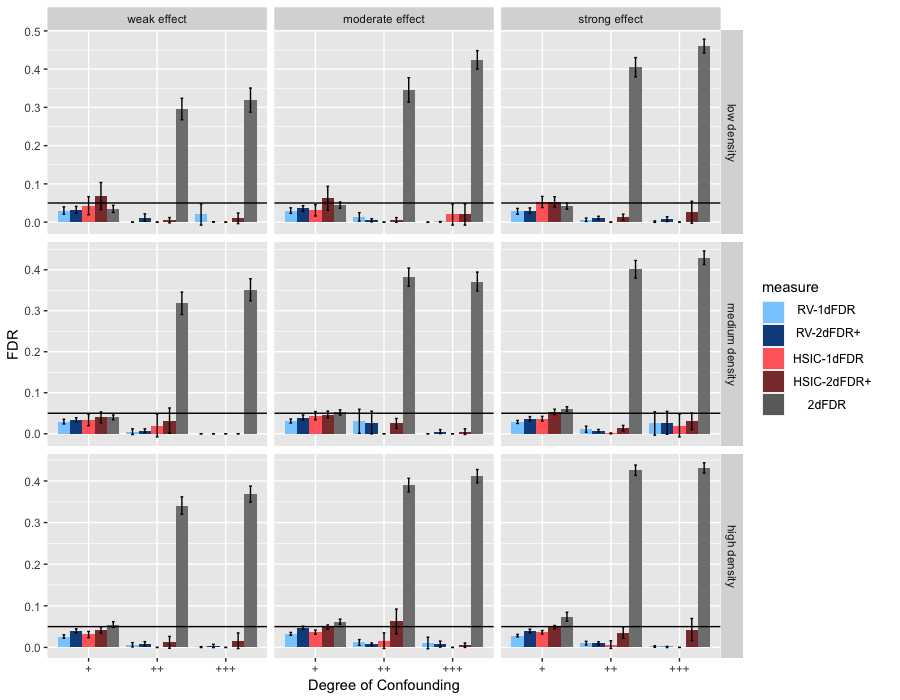}
  \caption{FDR}
  \label{fig:15a}
\end{subfigure}
\begin{subfigure}{0.9\textwidth}
  \centering
  \includegraphics[width=0.8\linewidth]{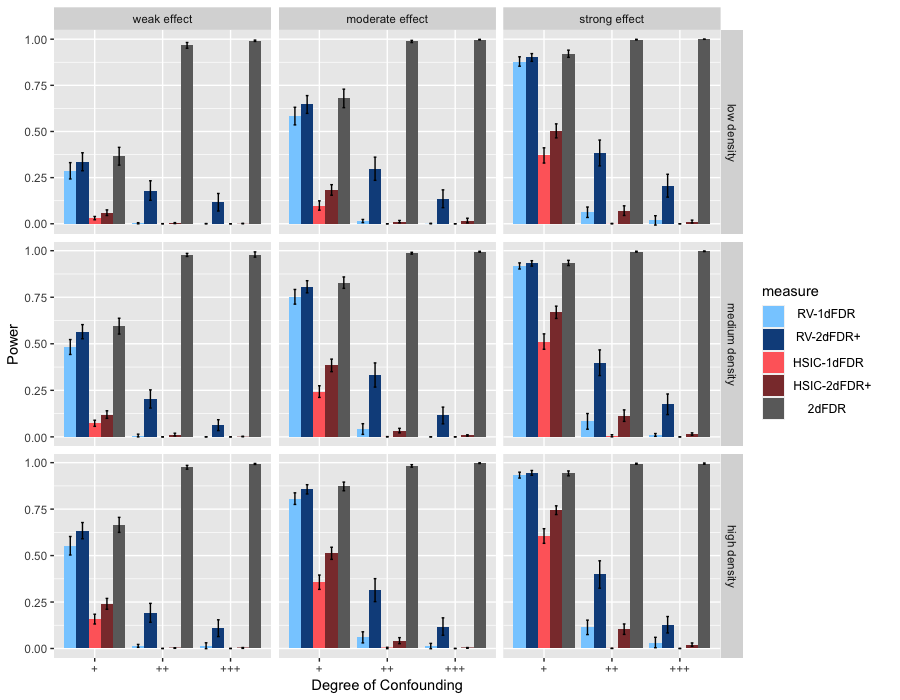}
  \caption{Power}
  \label{fig:15b}
\end{subfigure}
\caption{Empirical FDR and power for HSIC-1dFDR, RV-1dFDR, 2dFDR, HSIC-2dFDR+, RV-2dFDR+ under the model $Y_j = \alpha_j e^X + \beta_j e^Z + \epsilon_j$, where $\epsilon_j$ follows an AR(1) model with the AR(1) coefficient being 0.7, $X\sim N(\rho(Z + Z^2) , 1)$ and $Z \sim N( 0 , 1)$ Error bars represent the 95\% CIs and the horizontal line in (a) indicates the target FDR level of 0.05.}
\label{fig:15}
\end{figure}

\begin{figure}[H]
\centering
\begin{subfigure}{0.9\textwidth}
  \centering
  \includegraphics[width=0.8\linewidth]{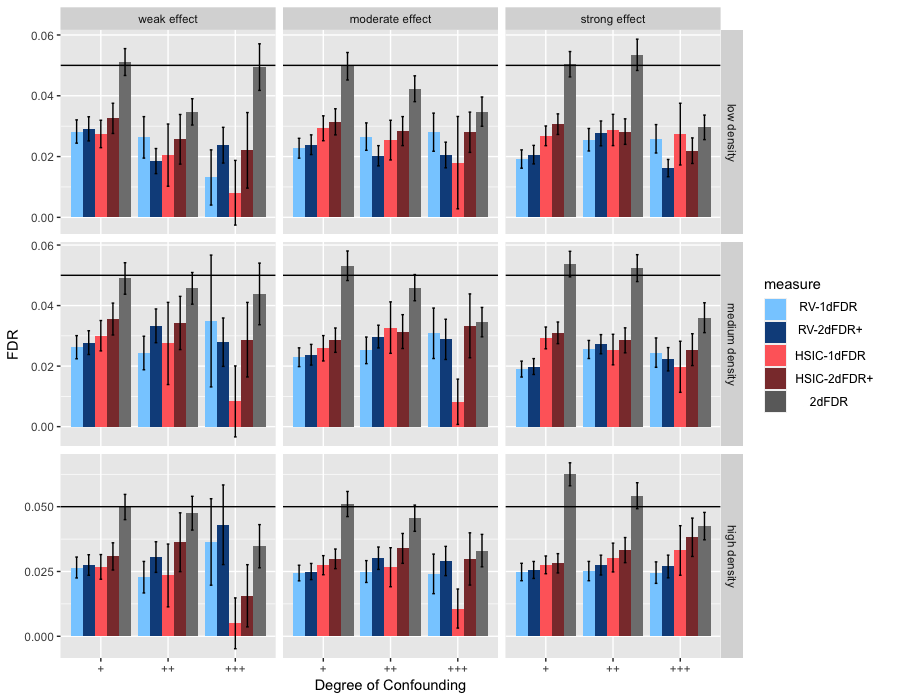}
  \caption{FDR}
  \label{fig:20a}
\end{subfigure}
\begin{subfigure}{0.9\textwidth}
  \centering
  \includegraphics[width=0.8\linewidth]{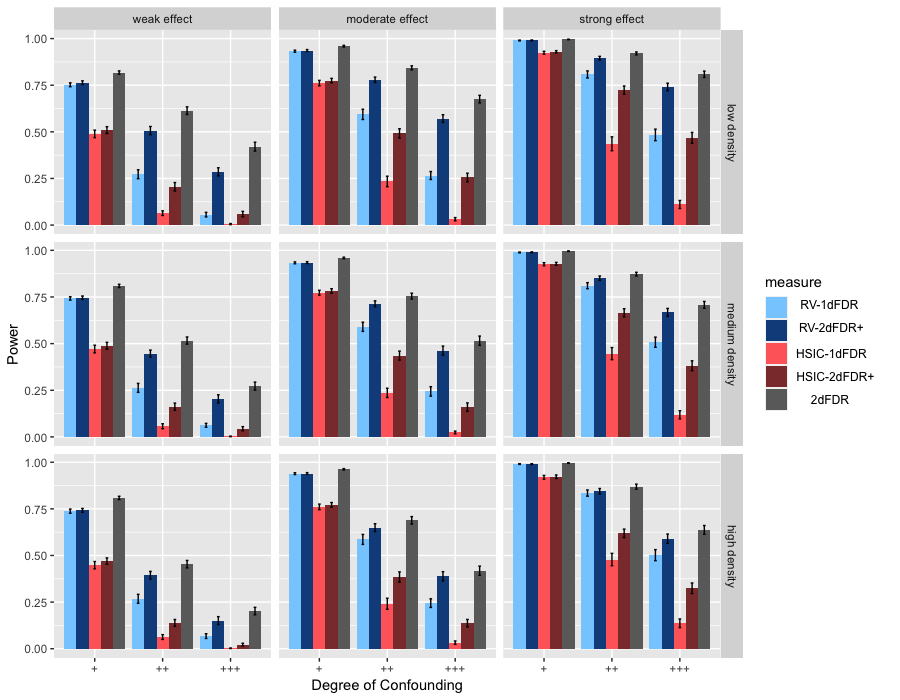}
  \caption{Power}
  \label{fig:20b}
\end{subfigure}
\caption{Empirical FDR and power for HSIC-1dFDR, RV-1dFDR, 2dFDR, HSIC-2dFDR+, RV-2dFDR+ under the model $Y_j = \alpha_j X + \beta_j Z + \epsilon_j$ where $X\sim N(\rho Z , 1)$ and $Z \sim N( 0 , 1).$ The signal density of $\alpha_j$ has been fixed at 10 \% while the signal density of $\beta_j$ has been varied through 1\%, 5\% and 10\%. Error bars represent the 95\% CIs and the horizontal line in (a) indicates the target FDR level of 0.05.}
\label{fig:20}
\end{figure}

\begin{figure}[H]
\centering
\begin{subfigure}{0.9\textwidth}
  \centering
  \includegraphics[width=0.8\linewidth]{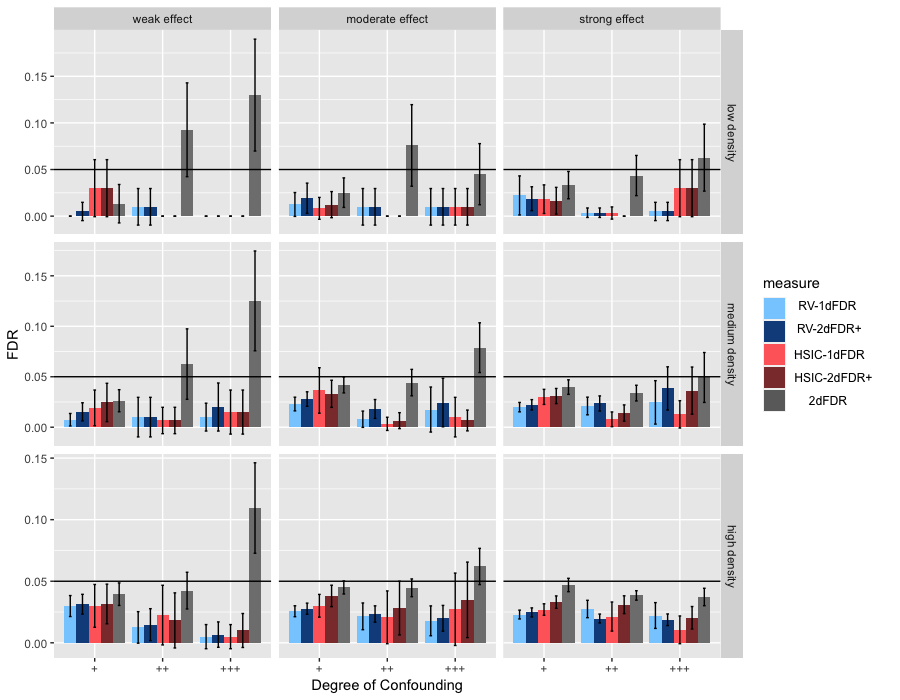}
  \caption{FDR}
  \label{fig:21a}
\end{subfigure}
\begin{subfigure}{0.9\textwidth}
  \centering
  \includegraphics[width=0.8\linewidth]{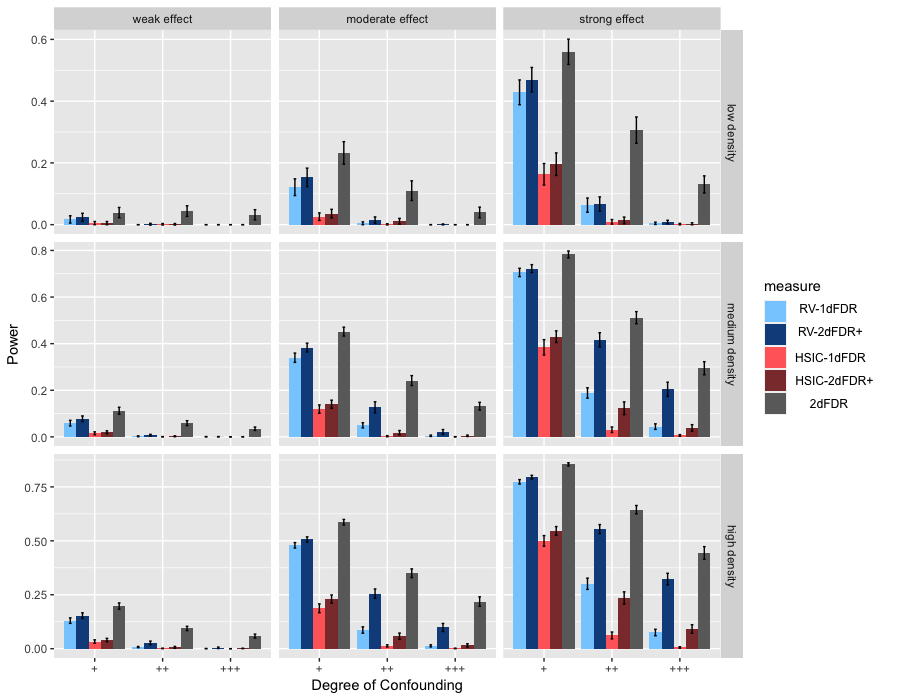}
  \caption{Power}
  \label{fig:21b}
\end{subfigure}
\caption{Empirical FDR and power for HSIC-1dFDR, RV-1dFDR, 2dFDR, HSIC-2dFDR+, RV-2dFDR+ under the model $Y_j = \alpha_j X + \beta_j Z + \epsilon_j$ where $X\sim N(\rho Z , 1)$ and $Z \sim N( 0 , 1).$ The signal density of $\beta_j$ has been fixed at 10 \% while the signal density of $\alpha_j$ has been varied through 1\%, 5\% and 10\%. Error bars represent the 95\% CIs and the horizontal line in (a) indicates the target FDR level of 0.05.}
\label{fig:21}
\end{figure}

\begin{figure}[H]
\centering
\begin{subfigure}{0.9\textwidth}
  \centering
  \includegraphics[width=0.8\linewidth]{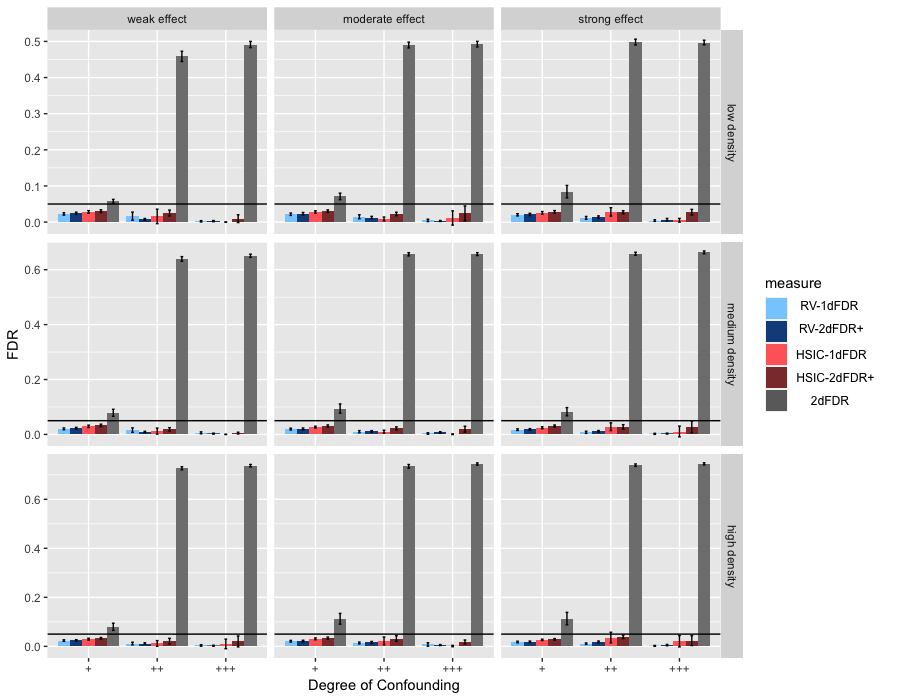}
  \caption{FDR}
  \label{fig:16a}
\end{subfigure}
\begin{subfigure}{0.9\textwidth}
  \centering
  \includegraphics[width=0.8\linewidth]{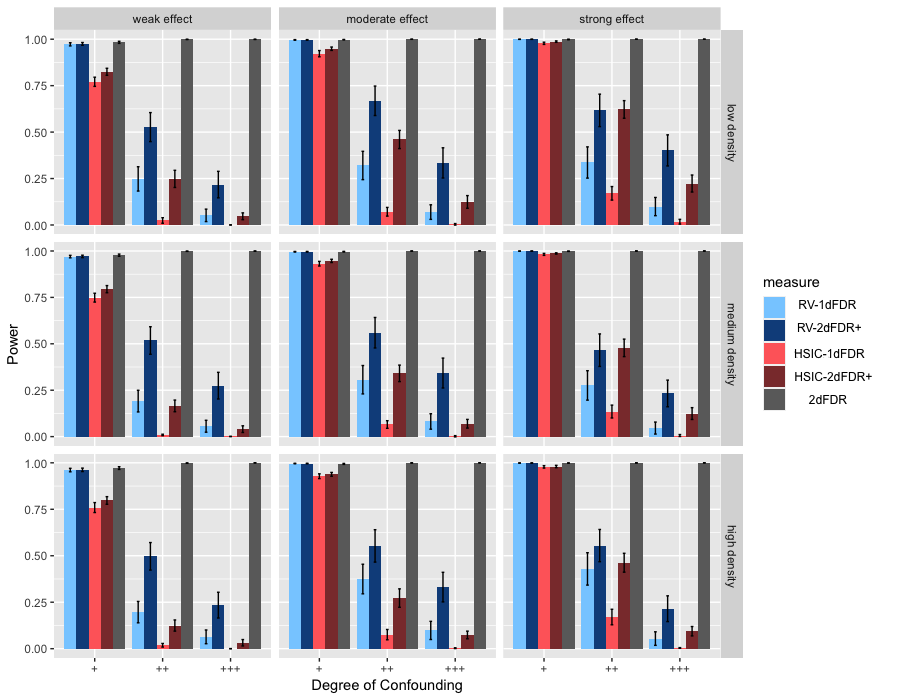}
  \caption{Power}
  \label{fig:16b}
\end{subfigure}
\caption{Empirical FDR and power for HSIC-1dFDR, RV-1dFDR, 2dFDR, HSIC-2dFDR+, RV-2dFDR+ under the model $Y_j = \alpha_j e^X + \beta_j Z^2 + \epsilon_j$ where $X\sim N(\rho Z^2 , 1)$ and $Z \sim N( 0 , 1).$ The signal density of $\alpha_j$ has been fixed at 10 \% while the signal density of $\beta_j$ has been varied through 1\%, 5\% and 10\%. Error bars represent the 95\% CIs and the horizontal line in (a) indicates the target FDR level of 0.05.}
\label{fig:16}
\end{figure}

\begin{figure}[H]
\centering
\begin{subfigure}{0.9\textwidth}
  \centering
  \includegraphics[width=0.8\linewidth]{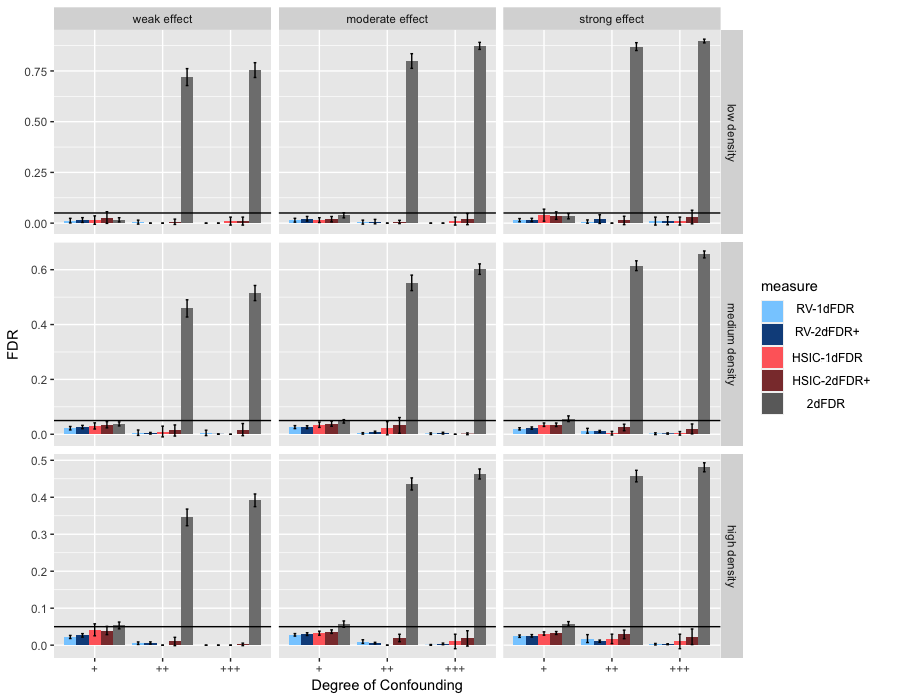}
  \caption{FDR}
  \label{fig:17a}
\end{subfigure}
\begin{subfigure}{0.9\textwidth}
  \centering
  \includegraphics[width=0.8\linewidth]{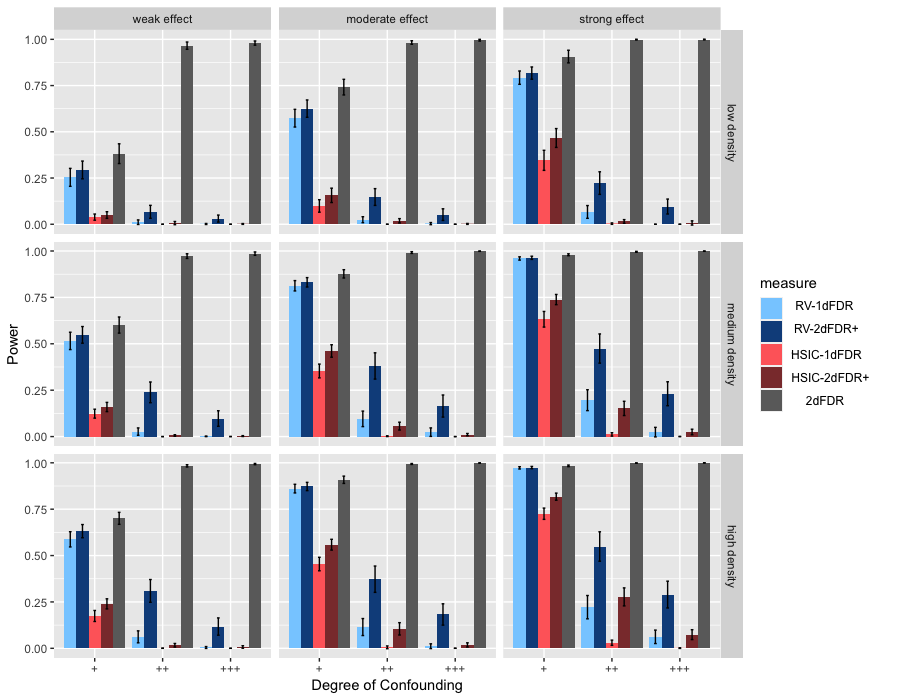}
  \caption{Power}
  \label{fig:17b}
\end{subfigure}
\caption{Empirical FDR and power for HSIC-1dFDR, RV-1dFDR, 2dFDR, HSIC-2dFDR+, RV-2dFDR+ under the model $Y_j = \alpha_j e^X + \beta_j Z^2 + \epsilon_j$ where $X\sim N(\rho Z^2 , 1)$ and $Z \sim N( 0 , 1).$ The signal density of $\beta_j$ has been fixed at 10 \% while the signal density of $\alpha_j$ has been varied through 1\%, 5\% and 10\%. Error bars represent the 95\% CIs and the horizontal line in (a) indicates the target FDR level of 0.05.}
\label{fig:17}
\end{figure}

\begin{figure}[H]
\centering
\begin{subfigure}{0.9\textwidth}
  \centering
  \includegraphics[width=0.8\linewidth]{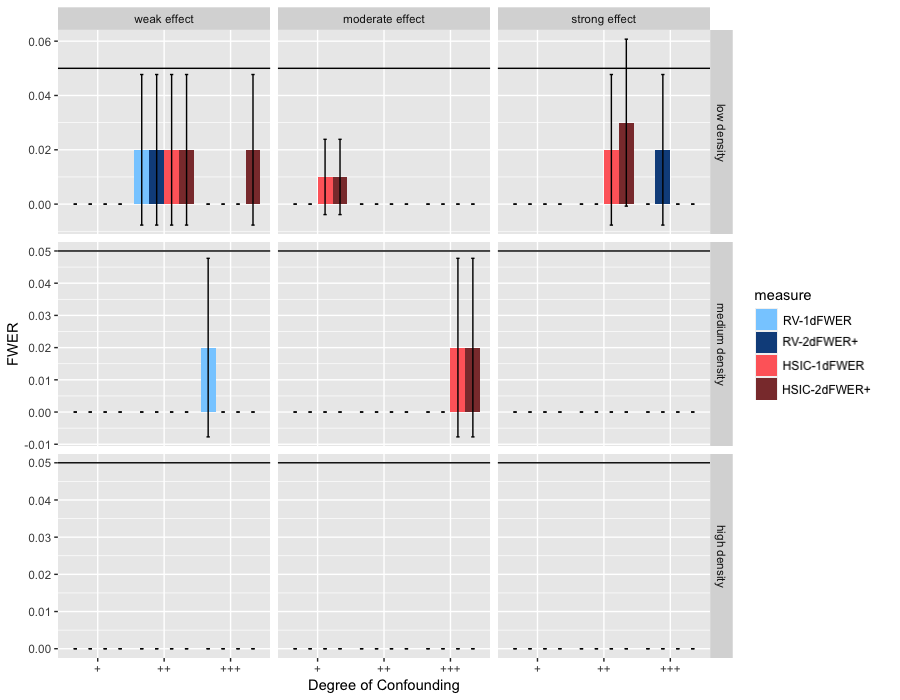}
  \caption{FWER}
  \label{fig:18a}
\end{subfigure}
\begin{subfigure}{0.9\textwidth}
  \centering
  \includegraphics[width=0.8\linewidth]{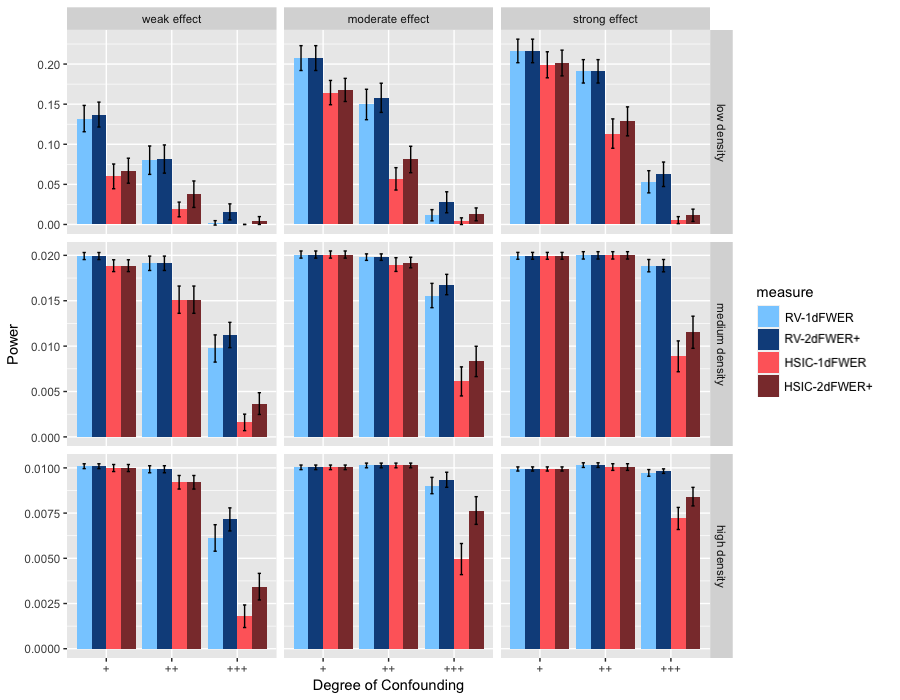}
  \caption{Power}
  \label{fig:18b}
\end{subfigure}
\caption{Empirical FWER and power for HSIC-1dFWER, RV-1dFWER, HSIC-2dFWER+, RV-2dFWER+ under the model $Y_j = \alpha_j X + \beta_j Z + \epsilon_j$, where $X\sim N(\rho Z , 1)$ and $Z \sim N( 0 , 1).$ Error bars represent the 95\% CIs and the horizontal line in (a) indicates the target FWER level of 0.05.}
\label{fig:18}
\end{figure}

\begin{figure}[H]
\centering
\begin{subfigure}{0.9\textwidth}
  \centering
  \includegraphics[width=0.8\linewidth]{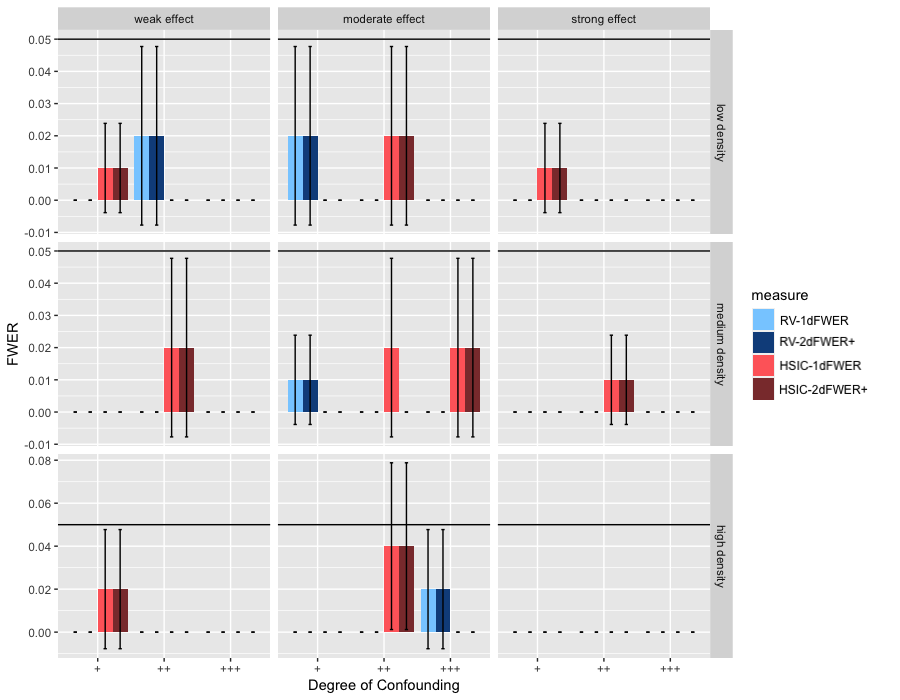}
  \caption{FWER}
  \label{fig:19a}
\end{subfigure}
\begin{subfigure}{0.9\textwidth}
  \centering
  \includegraphics[width=0.8\linewidth]{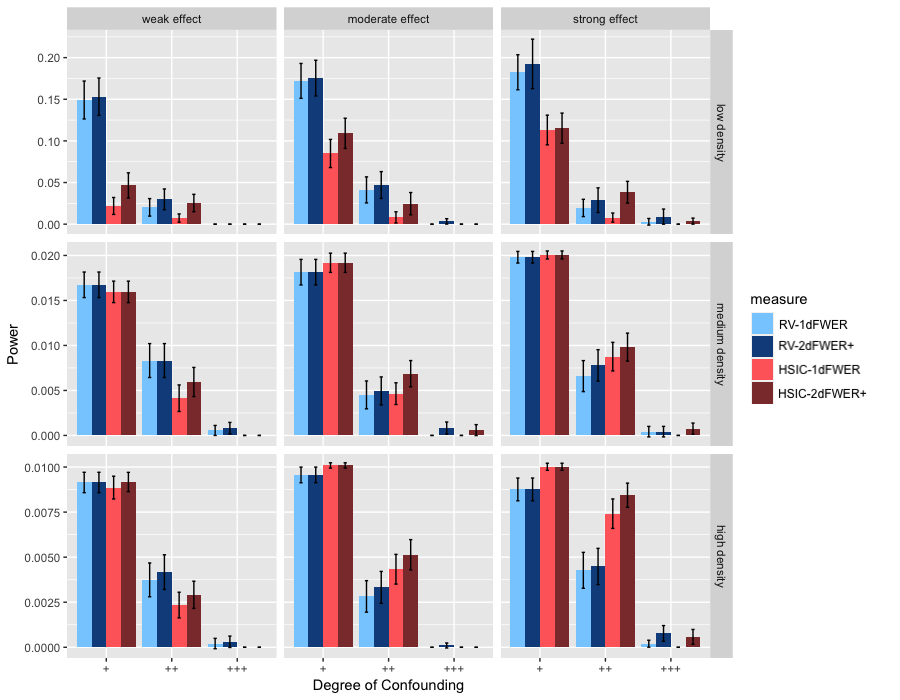}
  \caption{Power}
  \label{fig:19b}
\end{subfigure}
\caption{Empirical FWER and power for HSIC-1dFWER, RV-1dFWER, HSIC-2dFWER+, RV-2dFWER+ under the model $Y_j = \alpha_j e^X + \beta_j Z^2 + \epsilon_j$, where $X\sim N(\rho Z^2 , 1)$ and $Z \sim N( 0 , 1).$ Error bars represent the 95\% CIs and the horizontal line in (a) indicates the target FWER level of 0.05.}
\label{fig:19}
\end{figure}

\begin{itemize}
    \item FWER control: We investigate the finite sample performance of 2dFWER+ and its corresponding 1d version. In Figures \ref{fig:18}-\ref{fig:19}, we report the empirical FWER and power of 2dFWER+ and 1dFWER for both the linear and nonlinear models. In either case, the empirical FWER is well controlled for both methods. The 2d procedure again produces higher power than the 1d version, especially for stronger confounders.

    \item Global null: We examine the performance of 2dFDR, RV-2dFDR+, HSIC-2dFDR+, RV-1dFDR and HSIC-1dFDR under the global null. Specifically, we consider the model $Y_j = \beta_j Z$, where $X \sim N(\rho Z , 1)$ and $Z \sim N(0,1)$. None of the methods produced any rejections for all degrees of confounding.

    \item Dependent errors: To evaluate the impact of dependence on the methods' performance, we consider the model: $Y_j = \alpha_j e^X + \beta_j e^Z + \epsilon_j$ where $\epsilon_j = 0.7\epsilon_{j-1} + e_j$ and $X\sim N(\rho(Z + Z^2), 1)$ with $Z\sim N(0,1) $ and $\{e_j\}_{j = 1} ^m$ being a white noise process. The results are summarized in Figure \ref{fig:15}. Overall, 2dFDR+ is robust to the AR(1) type dependence with reliable FDR control and reasonable power.
    
    \item Separating the effects of densities of the signal of interest and the confounder signal: In all preceding simulations, the density of the signal of interest and the confounding signal had been kept at the same level---weak, moderate or strong. In this simulation setup, we attempt to tease apart the effects of the two types of signals.
    \begin{enumerate}
        \item  First, we fix the density of the signal of interest at the $10\%$ level and vary the density of the confounding signal through weak, moderate, and strong. The associated plots are given in Figure \ref{fig:20} and Figure \ref{fig:16}, corresponding to linear and non-linear DGPs respectively. 
        \item Next, we fix the density of the confounding signal to $10\%$ and vary the density of the signal of interest through weak, moderate, and strong. The associated plots are in Figure \ref{fig:21} and Figure \ref{fig:17}, corresponding to linear and non-linear DGPs respectively. 
    \end{enumerate}
    In both the linear and the non-linear DGPs, we find that varying the density of the signal of interest while keeping the density of the confounding variable constant is displaying a starker difference (increase) in the power as the densities are increased.
\end{itemize}

\subsection{Microbiome data: Binary Outcomes}
\begin{figure}[H]
        \centering
        \includegraphics[scale = 0.8]{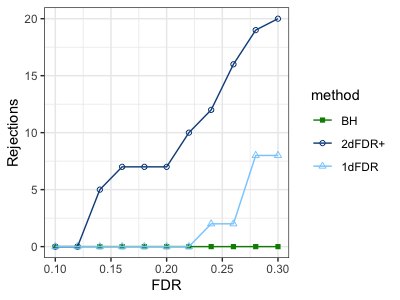}
        \caption{Number of Rejections versus FDR for different methods for smoking microbiome data, where the continuous abundance data were transformed into presence/absence (binary) data.}
        \label{fig:smokingbin}
    \end{figure}

We consider the microbiome data analyzed in Section \ref{sec:8} of the main paper. The abundance data of the 174 OTUs were converted into presence/absence data after rarefaction to the minimal sequencing depth (since presence/absence depends on the sequencing depth strongly, rarefaction removes the confounding effect due to sequence depth). Because $X$, $Y$, and $Z$ are all categorical (specifically, binary) in this case, for the conditional statistic, i.e., $T^C$, the Mantel Haenszel statistic was used. For the marginal statistic, i.e., $T^M$, the Pearson's chi-square statistic was used. Note that the original 2dFDR in \cite{yi20212dfdr} is not applicable in this case as the outcomes are binary. As shown in Figure \ref{fig:smokingbin}, for all levels of FDR, the BH procedure makes no rejections, and overall, the 2dFDR+ procedure makes a higher number of rejections compared to the corresponding 1dFDR procedure. 
    
\newpage

\begin{figure}[H]
\centering
\begin{subfigure}{0.9\textwidth}
  \centering
  \includegraphics[width=0.8\linewidth]{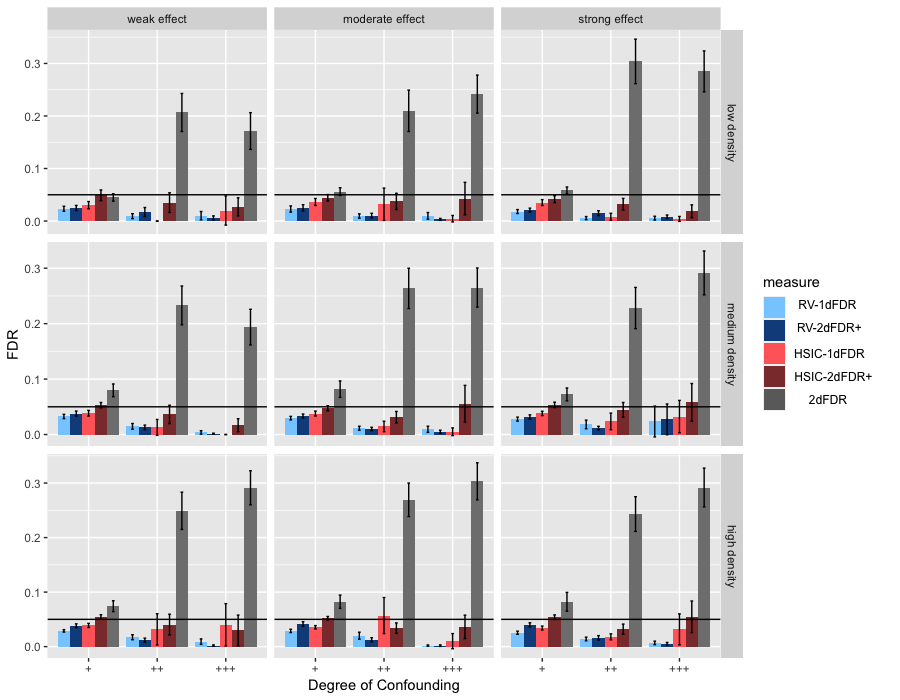}
  \caption{FDR}
  \label{fig:6a}
\end{subfigure}
\begin{subfigure}{0.9\textwidth}
  \centering
  \includegraphics[width=0.8\linewidth]{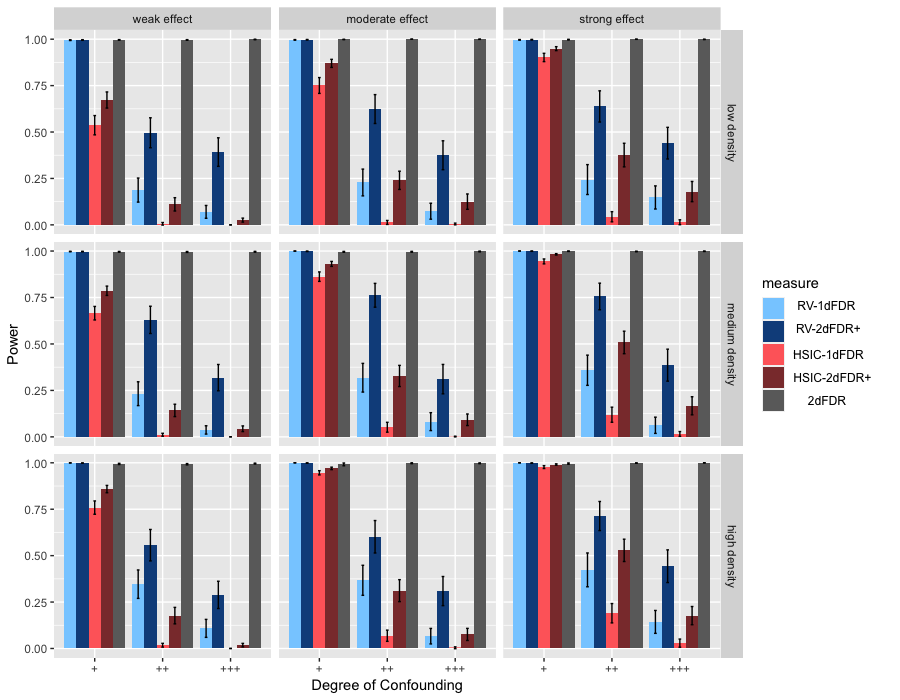}
  \caption{Power}
  \label{fig:6b}
\end{subfigure}
\caption{Empirical FDR and power for HSIC-1dFDR, RV-1dFDR, 2dFDR, HSIC-2dFDR+, RV-2dFDR+ under the model $Y_j = \alpha_j X^3 + \beta_j Z^3 + \epsilon_j$ and $X \sim N(\rho(Z + Z^2) , 1)$, where $Z \sim N(0,1)$. Error bars represent the 95\% CIs and the horizontal line in (a) indicates the target FDR level of 0.05.}
\label{fig:6}
\end{figure}

\newpage

\begin{figure}[H]
\centering
\begin{subfigure}{0.9\textwidth}
  \centering
  \includegraphics[width=0.8\linewidth]{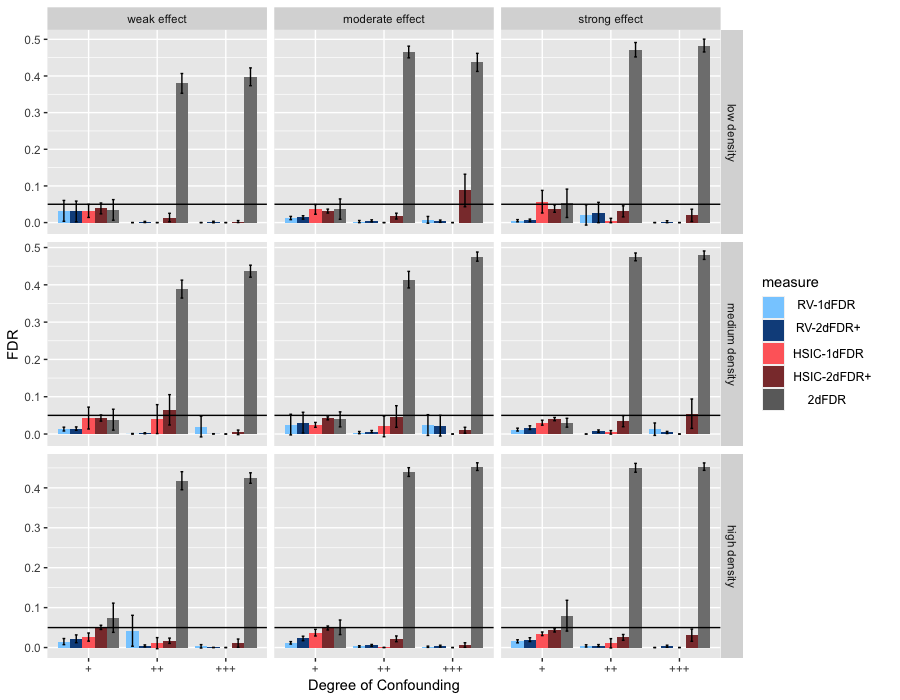}
  \caption{FDR}
  \label{fig:7a}
\end{subfigure}
\begin{subfigure}{0.9\textwidth}
  \centering
  \includegraphics[width=0.8\linewidth]{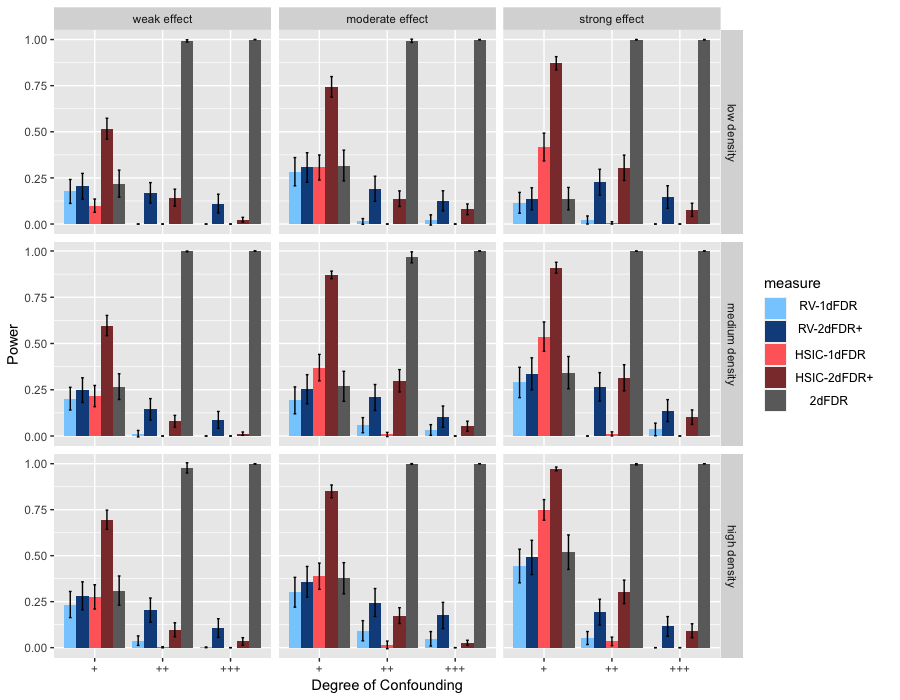}
  \caption{Power}
  \label{fig:7b}
\end{subfigure}
\caption{Empirical FDR and power for HSIC-1dFDR, RV-1dFDR, 2dFDR, HSIC-2dFDR+, RV-2dFDR+ under the model $Y_j = \alpha_j (X + |X^3|) + \beta_j e^Z + \epsilon$ and $X\sim N(\rho(Z + Z^2) , 1)$, where $Z\sim N(0,1)$. Error bars represent the 95\% CIs and the horizontal line in (a) indicates the target FDR level of 0.05.}
\label{fig:7}
\end{figure}

\newpage

\begin{figure}[H]
\centering
\begin{subfigure}{0.9\textwidth}
  \centering
  \includegraphics[width=0.8\linewidth]{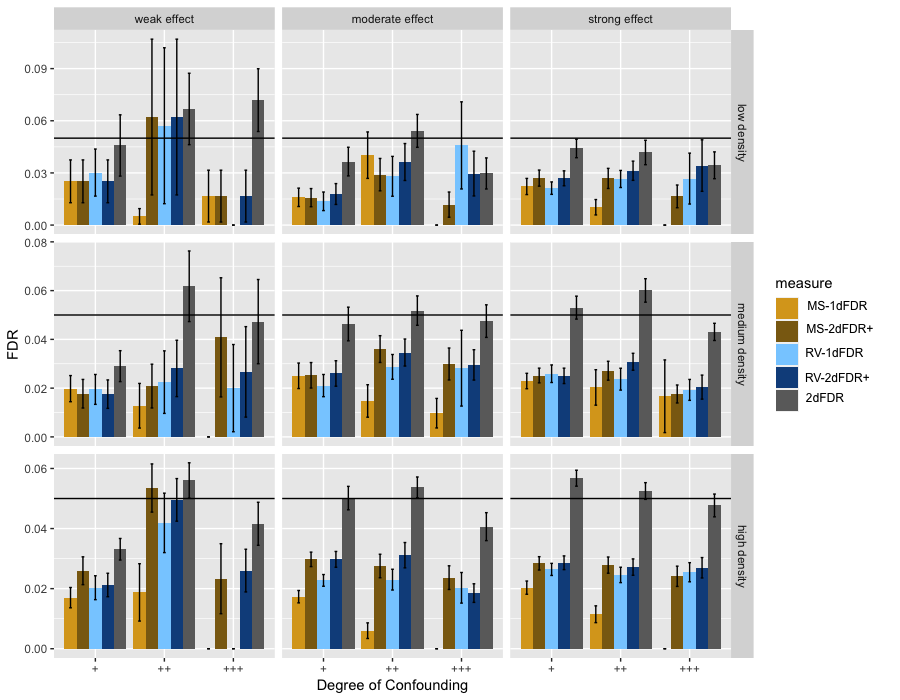}
  \caption{FDR}
  \label{fig:9a}
\end{subfigure}
\begin{subfigure}{0.9\textwidth}
  \centering
  \includegraphics[width=0.8\linewidth]{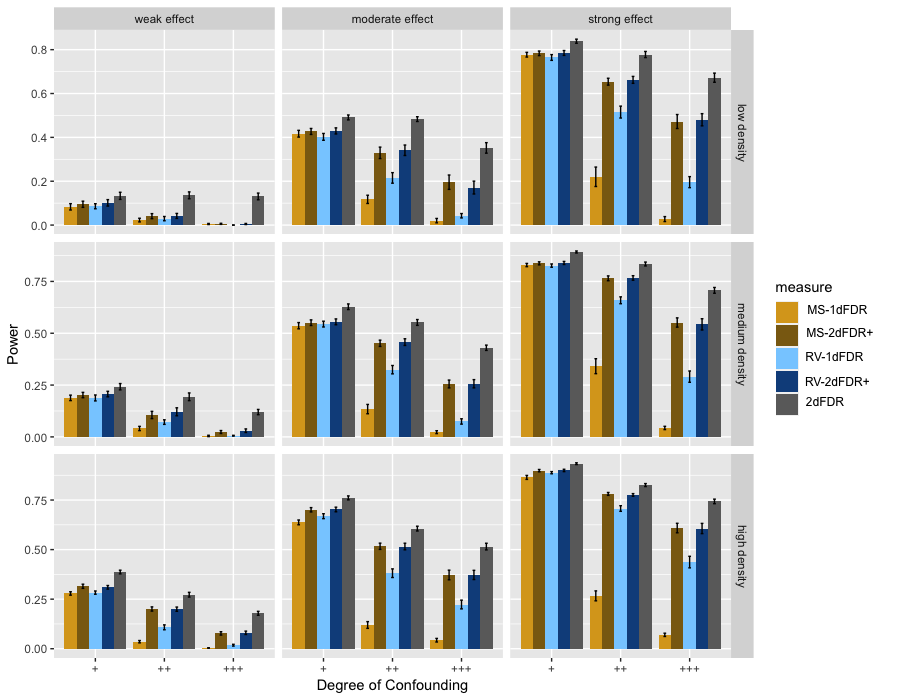}
  \caption{Power}
  \label{fig:9b}
\end{subfigure}
\caption{Empirical FDR and power for MS-1dFDR, RV-1dFDR, 2dFDR, MS-2dFDR+, RV-2dFDR+ under the model $Y_j = \alpha_j e^X + \beta_j Z + \epsilon_j$ and $X \sim \text{Bernoulli}((1 + e^{-\rho Z})^{-1})$,  where $Z \sim N(0,1)$. Error bars represent the 95\% CIs and the horizontal line in (a) indicates the target FDR level of 0.05. }
\label{fig:9}
\end{figure}

\newpage

\begin{figure}[H]
\centering
\begin{subfigure}{0.9\textwidth}
  \centering
  \includegraphics[width=0.8\linewidth]{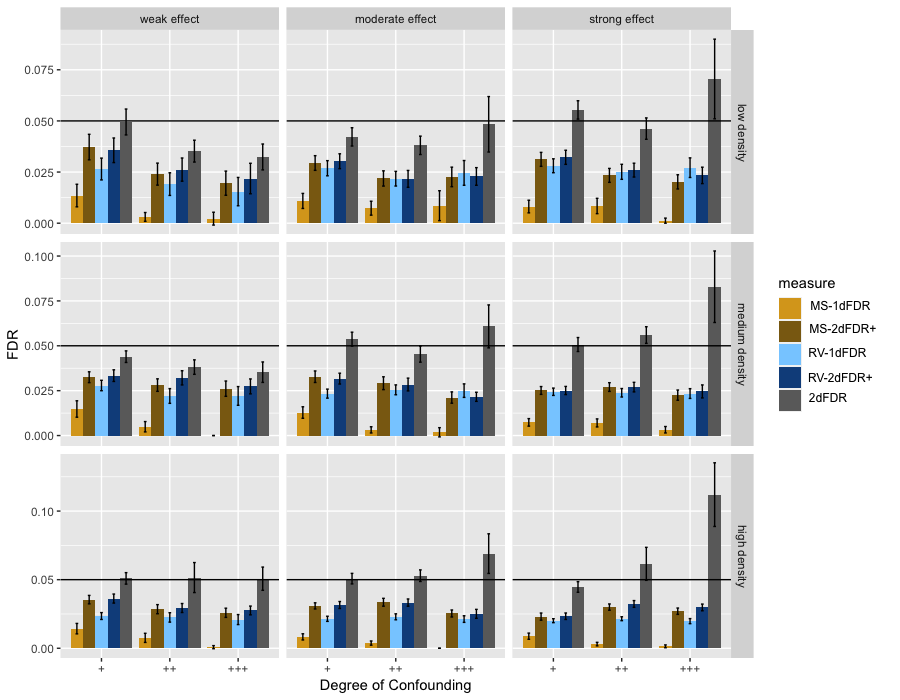}
  \caption{FDR}
  \label{fig:10a}
\end{subfigure}
\begin{subfigure}{0.9\textwidth}
  \centering
  \includegraphics[width=0.8\linewidth]{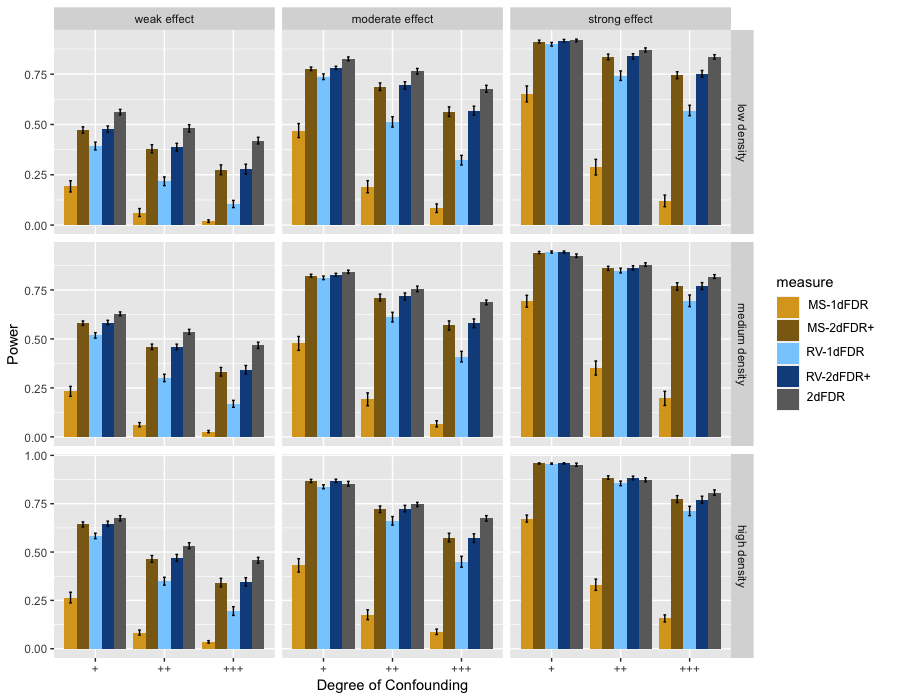}
  \caption{Power}
  \label{fig:10b}
\end{subfigure}
\caption{Empirical FDR and power for MS-1dFDR, RV-1dFDR, 2dFDR, MS-2dFDR+, RV-2dFDR+ under the model $Y_j = \alpha_j e^X + \beta_j e^Z + \epsilon_j$ and $X \sim \text{Bernoulli}((1 + e^{-\rho Z})^{-1})$, where $Z \sim N(0,1)$. Error bars represent the 95\% CIs and the horizontal line in (a) indicates the target FDR level of 0.05.}
\label{fig:10}
\end{figure}

\newpage

\begin{figure}[H]
\centering
\begin{subfigure}{0.9\textwidth}
  \centering
  \includegraphics[width=0.8\linewidth]{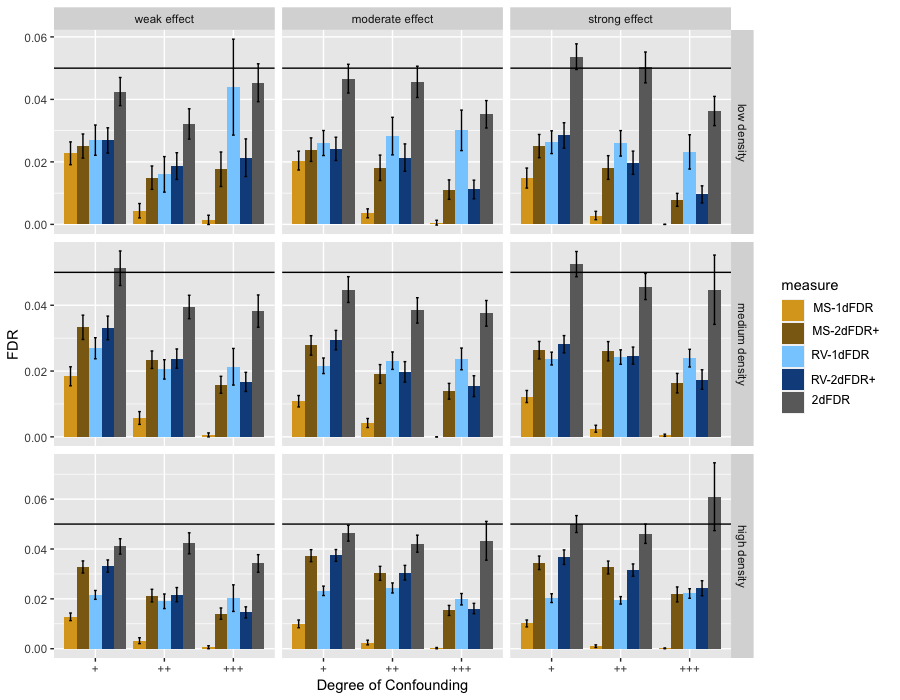}
  \caption{FDR}
  \label{fig:11a}
\end{subfigure}
\begin{subfigure}{0.9\textwidth}
  \centering
  \includegraphics[width=0.8\linewidth]{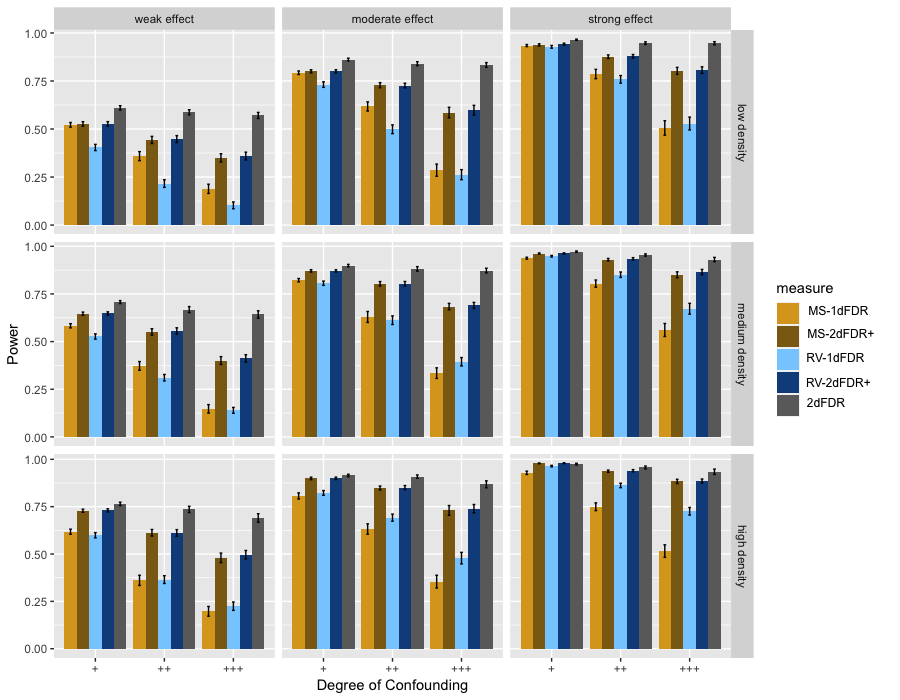}
  \caption{Power}
  \label{fig:11b}
\end{subfigure}
\caption{Empirical FDR and power for MS-1dFDR, RV-1dFDR, 2dFDR, MS-2dFDR+, RV-2dFDR+ under the model $Y_j = \alpha_j e^X + \beta_j Z^2 + \epsilon_j$ and $X \sim \text{Bernoulli}((1 + e^{-\rho Z})^{-1})$,  where $Z \sim N(0,1)$. Error bars represent the 95\% CIs and the horizontal line in (a) indicates the target FDR level of 0.05. }
\label{fig:11}
\end{figure}

\newpage

\begin{figure}[H]
\centering
\begin{subfigure}{0.9\textwidth}
  \centering
  \includegraphics[width=0.8\linewidth]{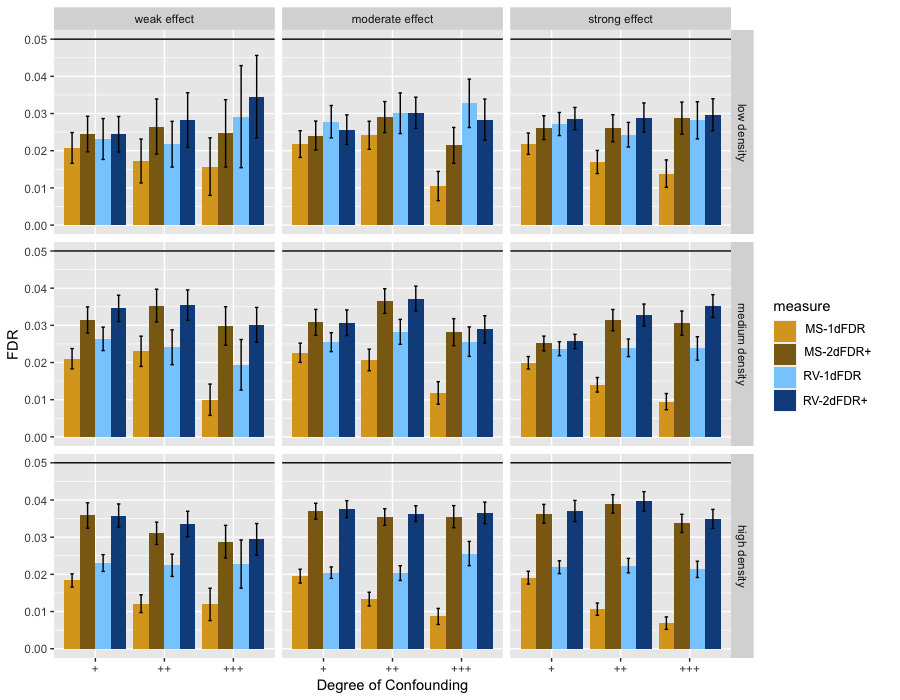}
  \caption{FDR}
  \label{fig:12a}
\end{subfigure}
\begin{subfigure}{0.9\textwidth}
  \centering
  \includegraphics[width=0.8\linewidth]{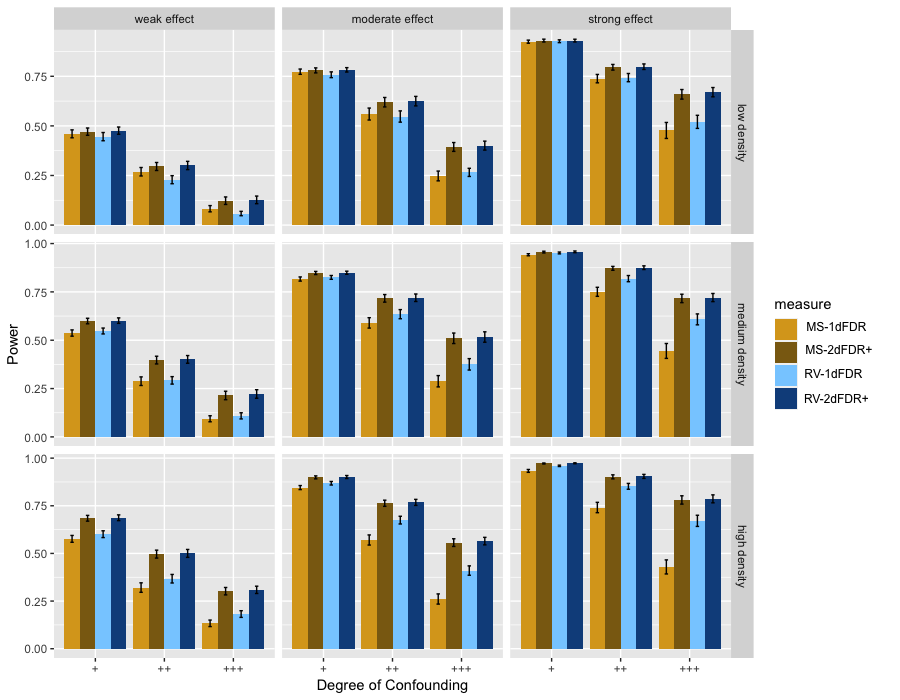}
  \caption{Power}
  \label{fig:12b}
\end{subfigure}
\caption{Empirical FDR and power for MS-1dFDR, RV-1dFDR, MS-2dFDR+, RV-2dFDR+ under the model $Y_j = \alpha_j X + \beta_j Z + \epsilon_j$ and $X \sim \text{Bernoulli}((1 + e^{-\rho Z})^{-1})$,  where $Z \sim \text{Bernoulli}(0.7)$. Error bars represent the 95\% CIs and the horizontal line in (a) indicates the target FDR level of 0.05.}
\label{fig:12}
\end{figure}

\begin{figure}[H]
\centering
\begin{subfigure}{0.9\textwidth}
  \centering
  \includegraphics[width=0.8\linewidth]{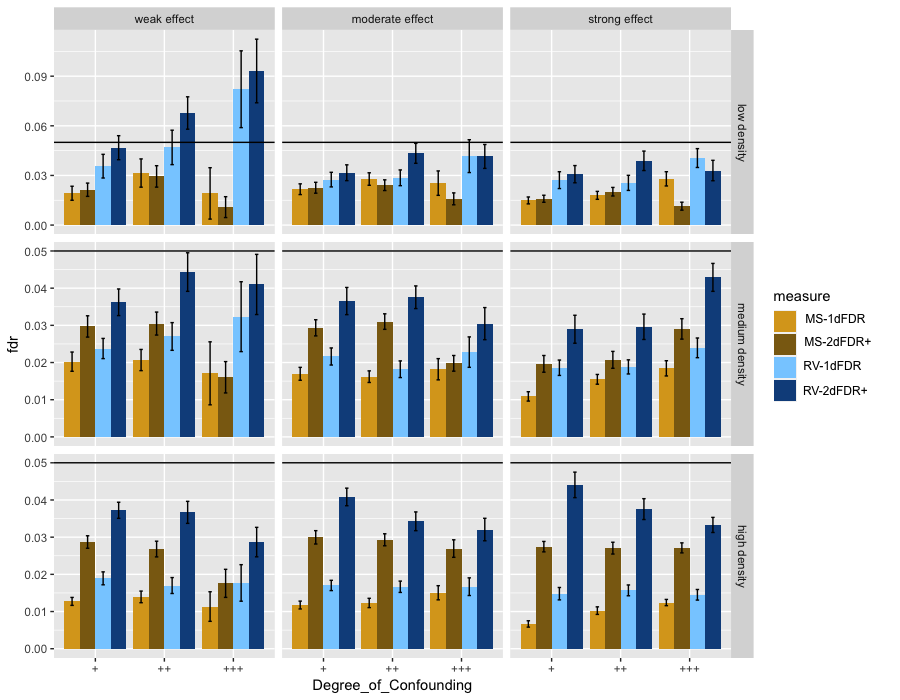}
  \caption{FDR}
  \label{fig:pois_fdr}
\end{subfigure}
\begin{subfigure}{0.9\textwidth}
  \centering
  \includegraphics[width=0.8\linewidth]{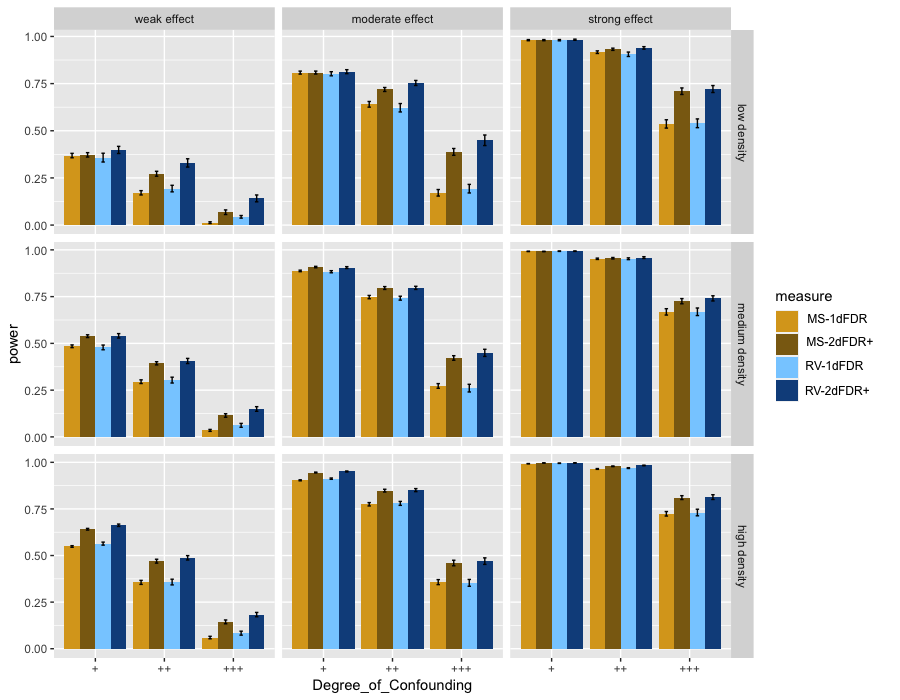}
  \caption{Power}
  \label{fig:pois_pow}
\end{subfigure}
\caption{Empirical FDR and power for MS-1dFDR, RV-1dFDR, MS-2dFDR+, RV-2dFDR+ under the model $Y_j \sim \text{Poisson}(\lambda_j)$, where $
    \log \lambda_j = \alpha_j X + \beta_j Z$ with $X \sim N(\rho Z, 1)$ and $Z \sim N(0,1).$ Error bars represent the 95\% CIs and the horizontal line in (a) indicates the target FDR level of 0.05. 
    }
\label{fig:pois}
\end{figure}

\begin{figure}[H]
\centering
\begin{subfigure}{0.9\textwidth}
  \centering
  \includegraphics[width=0.8\linewidth]{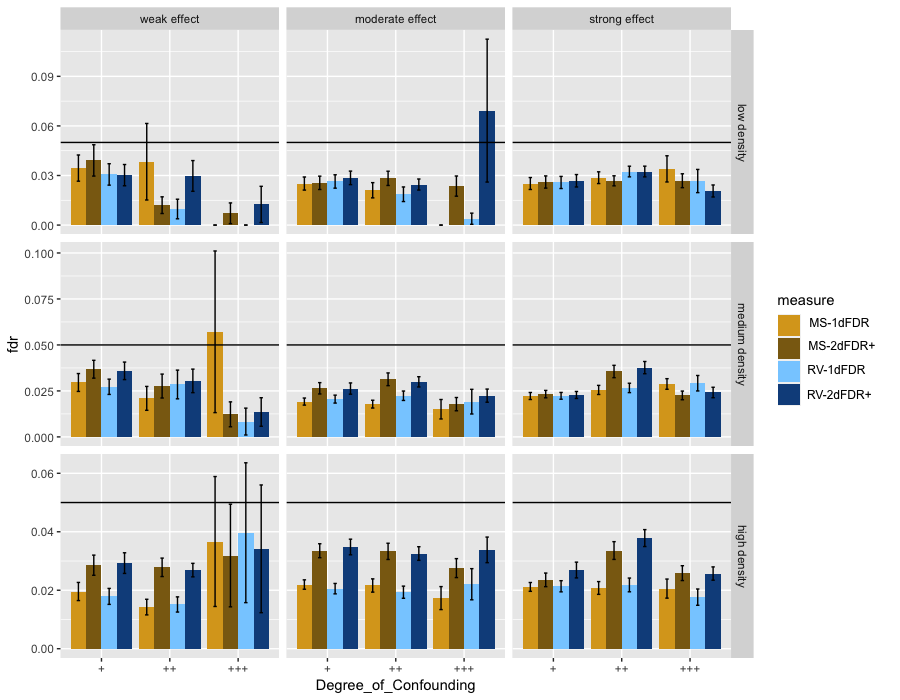}
  \caption{FDR}
  \label{fig:nbya}
\end{subfigure}
\begin{subfigure}{0.9\textwidth}
  \centering
  \includegraphics[width=0.8\linewidth]{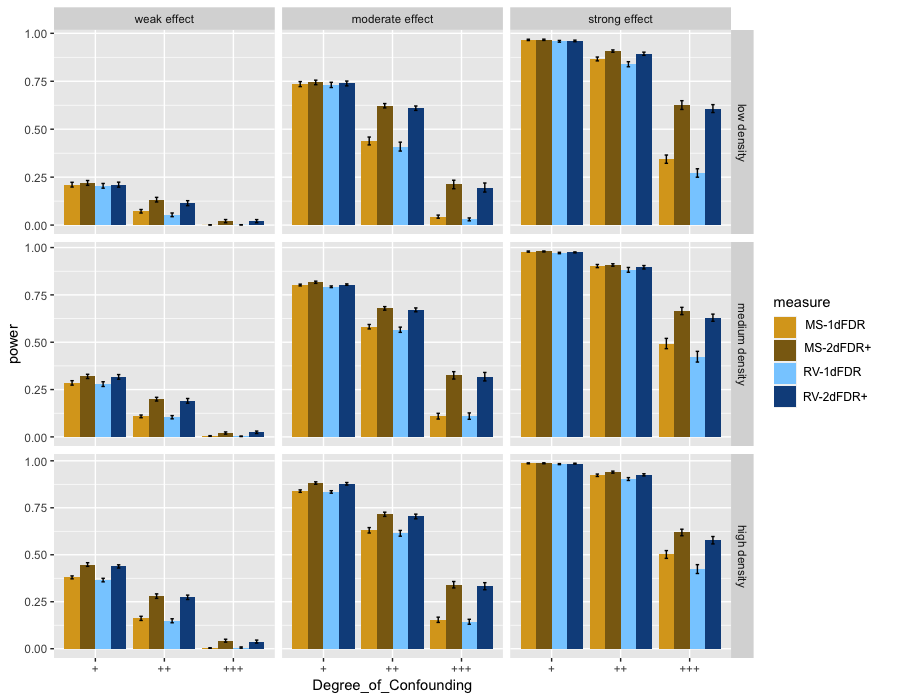}
  \caption{Power}
  \label{fig:nbyb}
\end{subfigure}
\caption{Empirical FDR and power for MS-1dFDR, RV-1dFDR, MS-2dFDR+, RV-2dFDR+ under the model $Y_j \sim \text{Negative Binomial}(\text{size} = 3, \mu_j = e^{f_j(X,Z)})$, where $f_j(X,Z) = \alpha_j X + \beta_j Z ,$ $X\sim N(\rho Z, 1)$ and $Z \sim N(0,1)$. Error bars represent the 95\% CIs and the horizontal line in (a) indicates the target FDR level of 0.05.}
\label{fig:nby}
\end{figure}

\newpage

\newpage

\newpage

\newpage

\newpage

\newpage

\newpage

  \begin{figure}[H]
        \centering
        \includegraphics[scale = 0.3]{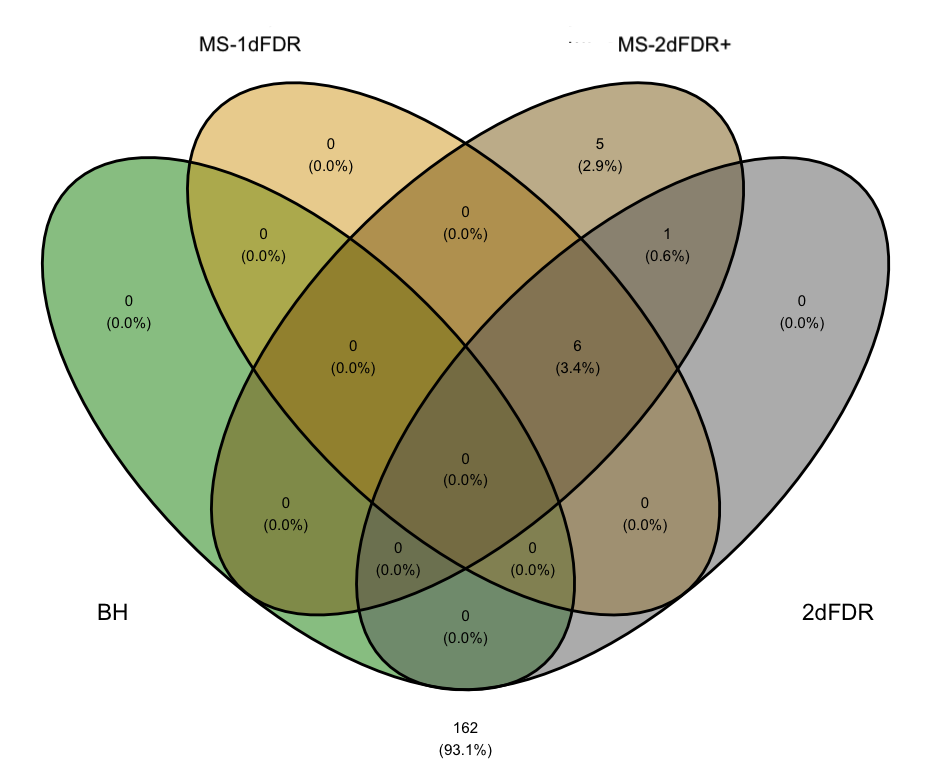}
        \caption{Venn diagram of features identified by different methods for smoking microbiome data}
        \label{fig:smoking2}
    \end{figure}

     \begin{figure}[H]
        \centering
        \includegraphics[scale = 0.3]{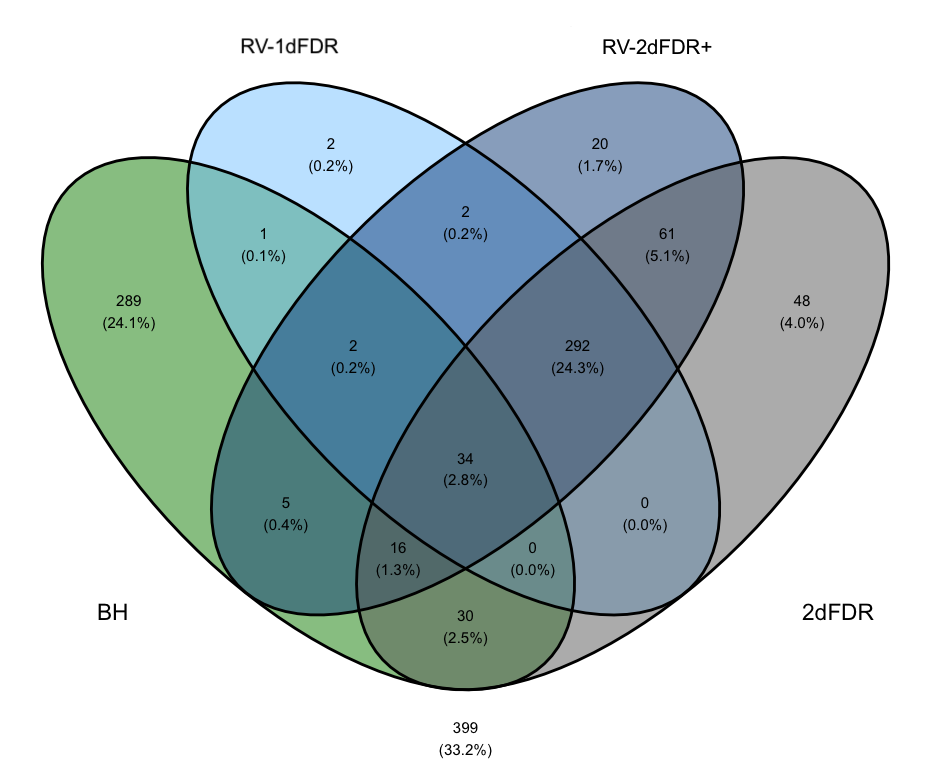}
        \caption{Venn diagram of features identified by different methods for  metabolomics data}
        \label{fig:metabolome2}
    \end{figure}

\singlespace

\end{document}